\documentclass[11pt]{article}
\usepackage{amsfonts}
\usepackage{amsthm}
\usepackage{amscd}
\usepackage{amsmath,amsbsy}
\usepackage{url}
\usepackage{graphicx}
\usepackage{color}
\usepackage{graphics}
\usepackage{amsmath}
\usepackage{rotating}
\usepackage{float}

\newcommand\bb {\mathbf b}
\newcommand\bc {\mathbf c}

\newcommand\be {\mathbf e}

\newcommand\bg {\mathbf g}

\newcommand\Bell{\ensuremath{\boldsymbol\ell}}

\newcommand\bese {\mathbf s}

\newcommand\bx {\mathbf x}

\newcommand\bB {\mathbf B}

\newcommand\bI {\mathbf I}

\newcommand\bM {\mathbf M}

\newcommand\bV {\mathbf V}

\newcommand\bX {\mathbf X}

\newcommand\indica {\mathbb{I}}

\newcommand\wa {\widehat{{a}}}
 
\newcommand\wb {\widehat{{b}}}
 
\newcommand\wc {\widehat{{c}}}

\newcommand\wh {\widehat{h}}

\newcommand\wese {\widehat{s}}

\newcommand\wy {\widehat{y}}

\newcommand\wtbB {\widetilde{\bB}}

\newcommand\itB {{\mathcal{B}}}

\newcommand\itF {{\mathcal{F}}}

\newcommand\itH {{\mathcal{H}}}
\newcommand\itI {{\mathcal{I}}}

\newcommand\itK {{\mathcal{K}}}

\newcommand\itS {{\mathcal{S}}}

\newcommand\itU {{\mathcal{U}}}

\newcommand\itX {{\mathcal{X}}}

\newcommand\bbe {\mbox{\boldmath $\beta$}}

\newcommand\bbech {\mbox{\scriptsize${\bbe}$}}

\newcommand\bla {\mbox{\boldmath $\lambda$}}

\newcommand\bnu {\mbox{\boldmath $\nu$}}

\newcommand\bthe {\mbox{\boldmath $\theta$}}
\newcommand\bthech {\mbox{\scriptsize${\bthe}$}}

\newcommand\bxi {\mbox{\boldmath $\xi$}}

\newcommand\bSi {\mbox{\boldmath $\Sigma$}}

\newcommand\walfa {\widehat{\alpha}}
\newcommand\walfach {\mbox{\scriptsize${\walfa}$}}

\newcommand\wbeta {\widehat{\beta}}
\newcommand\wbbe {\widehat{\bbe}}
\newcommand\wbbech {\mbox{\scriptsize$\wbbe$}}

\newcommand\weta {\widehat{\eta}}
\newcommand\wetach {\mbox{\scriptsize$\weta$}}

\newcommand\wgamma {\widehat{\gamma}}

\newcommand\wbnu {\widehat{\bnu}}
\newcommand\wsigma {\widehat{\sigma}}

\newcommand\wbthe {\widehat{\bthe}}
\newcommand\wbthech {\mbox{\scriptsize$\wbthe$}}

\newcommand\wtbbe {\widetilde{\bbe}}
\newcommand\wtbbech {\mbox{\scriptsize$\wtbbe$}}

\newcommand\wteta {\widetilde{\eta}}

\newcommand\wtsigma {\widetilde{\sigma}}

\newcommand\wtbSi {\widetilde{\bSi}}

\def\real{\mathbb{R}}

\def\qu{\mathbb{Q}}

\newcommand{\esp}{\mathbb{E}}
\newcommand{\prob}{\mathbb{P}}

\newcommand{\convpp}{ \buildrel{a.s.}\over\longrightarrow}

\newcommand{\convprob  }{ \buildrel{p}\over\longrightarrow}

\newcommand{\convdist}{ \buildrel{D}\over\longrightarrow}

\newcommand{\trasp}{^{\mbox{\footnotesize \sc t}}}

\newcommand\bcero {{\bf{0}}}

\def\dst{\displaystyle}

\def\argmin{\mathop{\mbox{argmin}}}

\def\EIF{\mathop{\rm EIF}}

\newcommand{\identidad}{\mbox{\bf I}}

\newcommand\new{\newline}
\newcommand\noi{\noindent}
\def\dst{\displaystyle}

\def\square{\ifmmode\sqr\else{$\sqr$}\fi}
\def\sqr{\vcenter{
         \hrule height.1mm
         \hbox{\vrule width.1mm height2.2mm\kern2.18mm
\vrule width.1mm}
         \hrule height.1mm}}

\newcommand{\tuk}{\mbox{\scriptsize \sc t}}

\newcommand{\rob}{\mbox{\footnotesize \sc r}}

\newcommand{\cl}{\mbox{\scriptsize \sc cl}}

\begin{document}

\title{Robust estimation in single index models  when the errors have a unimodal density with unknown nuisance parameter}
\author{Claudio Agostinelli\\
{\small  Universit\`a di Trento, Italy}\\
Ana M. Bianco\\
{\small  Universidad de Buenos Aires and  CONICET, Buenos Aires, Argentina} \\
 Graciela Boente\\
 {\small Universidad de Buenos Aires and IMAS, CONICET,  Buenos Aires, Argentina} 
 }
 \date{}
 
 \maketitle 


\begin{abstract}
In this paper, we propose a robust profile estimation method for the parametric and nonparametric components of a single index model when the errors   have a strongly unimodal density with  unknown nuisance parameter. Under regularity conditions, we derive consistency results for the link function  estimators as well as consistency and asymptotic distribution results for the single index parameter estimators.  Under a log--Gamma model, the sensitivity to anomalous observations is studied by means of the  empirical influence curve.  We also discuss a robust $K-$fold procedure to select the smoothing parameters involved. A numerical study is conducted to evaluate  the small sample performance of the robust proposal  with that of their classical relatives, both for errors following a log--Gamma model and   for contaminated schemes. The numerical experiment shows the good robustness properties of the proposed estimators and the advantages of considering a robust approach instead of the classical one. 
\end{abstract}

\date{}
\maketitle

\section{Introduction}{\label{intro}}
 
Semiparametric models are an appealing compromise between  
parametric  and nonparametric pa\-ra\-digms. 
These models represent an intermediate point  between a fully parametric model, which  is usually of easy interpretation  but vulnerable to poor specification, and a fully nonparametric model, which is more flexible but suffers from the well--known curse of dimensionality.  Semiparametric modeling combines parametric components with nonparametric ones, retaining the advantages of both types of approaches and avoiding their drawbacks.

Single index models  are a relevant topic within the broad class of semiparametric methods with a great potentiality when modelling data in different scientific disciplines.  These models have raised a lot of interest in part due
to the fact that they reduce the dimensionality of the covariates through a suitable projection linked to the parametric component, while at the same time they capture a possible nonlinear relationship through an unknown smooth function. 

Under a single index model, the response variable $y$ is related to the covariates $\bx$ through the equation
\begin{eqnarray}
y &=& \eta(\bbe\trasp \bx)+ \epsilon \, , \label{modelo}
\end{eqnarray}  
where the single index parameter $\bbe \in \real^q$ and the link univariate real valued function $\eta: \real \rightarrow \real$ are both unknown. 
For the sake of identifiability, it is assumed with no loss of generality that $\|\bbe\|=1$ and the last component of $\bbe$ is positive, where $\| \cdot \|$ denotes the Euclidean norm. Furthermore, in the classical setting, it is usually assumed that $\esp(\epsilon|\bx)=0$ and $\esp(\epsilon^2|\bx)< \infty$.

As noted above, in our framework $\|\bbe\|=1$, so we may assume that $\beta_q \neq 0$, without loss of generality. However, some authors consider a different  parametrization given by
\begin{eqnarray}
y &=& \eta^\star(\bthe\trasp \bx)+ \epsilon \, , \label{modelotheta}
\end{eqnarray}
where $\bthe=(\bthe^\star,\theta_q)$ with $\theta_q=1$ and $\bthe^\star=(\theta_1,\dots,\theta_{q-1}) \in \real^{q-1}$, which also leads to an identifiable model. One of the advantages of the parametrization (\ref{modelo}) over that given in (\ref{modelotheta})  is that the finite dimensional parameter $\bbe$ naturally belongs to a compact set. The relation between both parametrizations is given by $\bbe=\bthe/\|\bthe\|$ and $\eta(u)=\eta^\star(u \, \|\bthe\|)$, while $\bthe=\bbe/\beta_q$  and $\eta^\star(u )=\eta(u \, \beta_q)$. So, estimators in any of these two parametrizations lead to  estimators in the other one. 

Single index models  have received  an increasing amount of attention in the last years, probably because they have an appealing feature: they cope with the \textsl{curse of dimensionality} combining nonparametric and parametric--driven approaches. Beneath single index models underlies the idea that the contribution of the vector of  covariates $\bx$ to the response $y$ can be expressed in terms of a one--dimensional projection.
 In this sense, these models  can   be seen as a    dimension reduction technique since, once $\bbe$ has been estimated,  the unidimensional variable $\bbe\trasp \bx$ can be used as a univariate carrier to estimate nonparametrically the function $\eta$. 

There is an extensive literature in this area. Among the first works, we can mention Powell \textsl{et al.} (1989), H\"ardle and Stoker (1989), H\"ardle \textsl{et al.} (1993), Xia \textsl{et al.} (2002) and  Carroll \textsl{et al.} (1997). 
More recently,  Xia (2006) studies the asymptotic distribution of two classes of estimators,  Chang \textsl{et al.} (2010) consider the heteroscedastic case and  Xia \textsl{et al.} (2012) propose a  family of estimators of the nonparametric component for which it is not necessary to undersmooth in order to obtain a  $\sqrt{n}-$rate estimator of the parametric component. On the other hand,  Wu \textsl{et al.} (2010) consider
the estimation of the single index quantile regression, while Liu \textsl{et al.} (2013) propose robust estimators by means of the mode, without taking into account the estimation of a possible scale factor. Xue and  Zhu (2006) focus on the problem of looking for confidence regions and intervals and Zhang \textsl{et al.} (2010) study the problem of testing hypotheses that involve $\bbe$. Recently, Li and Patilea (2017) considered a a quadratic form criterion involving kernel smoothing and propose a resampling method to build  confidence intervals for the index parameter.  Wang \textsl{et al.} (2014) also consider the extension of these models to the situation in which there are missing responses. All the aforementioned procedures are based on classical methods and hence, they are very sensitive to the presence of outliers. 

Indeed, even when different approaches have been proposed for fitting model (\ref{modelo}), such as kernel smoothing   or  sliced inverse regression methods, in most cases it is assumed that the error distribution has finite first moment.  In the robust framework, this assumption is generally replaced by the symmetry of the error term distribution, in order to achieve Fisher--consistent estimators. However, in practice, situations arise in which the errors are asymmetric, as it is the case when the error term distribution belongs to some class of exponential families, such as the log--Gamma distribution.
In this paper, we focus on the problem of robust estimating the parametric and nonparametric components of model (\ref{modelo}) when the density of the error $\epsilon$ is of the form
\begin{eqnarray}
g(s,\gamma)=Q(\gamma) \exp^{\gamma \,t(s)} \, , \label{exponencial}
\end{eqnarray}
where $\gamma>0$ is an unknown parameter and $t$ is a continuous function with unique mode at $e_0$. An appealing feature of this family of distributions is that enables to model either symmetric or asymmetric errors, as well. A prominent member of this family is the log--Gamma distribution that is frequently used to fit asymmetric data.

A first approach to deal with outliers in the responses, was given in Delecroix \textsl{et al.} (2006) who considered $M-$type estimators for single index models with known nuisance parameter. However, in most cases, the nuisance parameter $\gamma$ is unknown and its estimation is crucial to down--weight large residuals. In fact, as in linear regression, it is necessary to determine  the size of the residuals to decide if an observation is an outlier or not and this task strongly depends on a good preliminary nuisance parameter estimator. Indeed, the most popular example of   model (\ref{modelo}) with errors having a density given by (\ref{exponencial}) is the usual regression model with symmetric errors $\epsilon \sim F(\cdot/\sigma)$, where $\sigma$ stands for the scale parameter. In this setting, the nuisance parameter $\gamma$ is usually taken as $\sigma$ to avoid an assumption on a fixed given errors distribution. 
On the other hand, as mentioned, a well--known regression model with asymmetric errors is the log--Gamma model which corresponds to the Generalized Linear Model for the Gamma distribution with $\log$ link function. In this case, $\gamma$ represents  the shape parameter. In both regression models, it is important to estimate $\gamma$ in order to calibrate the robust estimators. For this reason,  our approach   includes a nuisance parameter which needs to be robustly estimated prior to the estimation of $\eta$ and $\bbe$ and which may be $\gamma$ or a known function of it such as the constant needed to calibrate the estimators.  
 
The aim of this paper is to propose a class of robust estimators for single index models when the errors distribution has density   satisfying (\ref{exponencial}) with the   parameter $\gamma$  unknown.  For this purpose, we introduce a stepwise procedure based on robust profile estimators.   We make special emphasis in the case of errors with log--Gamma distribution, which is often employed in applications, and then we extend the proposal to the general setting.    Under mild conditions, the estimators of $\eta$ are consistent and the parametric component estimators  are consistent and asymptotically normal with  $\sqrt{n}-$rate.  We also provide a class of  initial estimators and a robust $K-$fold procedure to select the bandwidth parameters involved in our proposal.

The outline of the paper is as follows. In Section \ref{main}, the 
three--step procedure for robust estimation under a single index model is introduced first for log--Gamma errors and then, it is extended to more general situations. In Section {\ref{asymp}}, we give some asymptotic properties of the proposal, while in Section \ref{eif}, we compute the empirical influence function which may be helpful to study the sensitivity of the estimators to atypical observations.  Section \ref{KCV} presents a robust $K-$fold cross--validation method to select the smoothing parameters. The robustness and performance for finite
samples of the proposed method are studied by means of a numerical study in Section \ref{monte}. Proofs are relegated to the Appendix.

\section{The estimators}{\label{main}}

Let $(y_i,\bx_i)\in \real^{q+1}$ be independent observations that follow model (\ref{modelo}) for 
$\eta=\eta_0$ and $\bbe=\bbe_0$ and assume that the errors $\epsilon_i$ are independent, independent of $\bx_i$ and have density (\ref{exponencial})  with  $\gamma= \gamma_0$. Denote $\esp_0$ the expectation under the true model and $\alpha_0$ the true nuisance parameter which as mentioned above is a function of $\gamma_0$.

\subsection{The log--Gamma setting}{\label{proposalgam}}
In order to introduce the proposed estimators, let us first revisit the particular case of the purely parametric regression model with log--Gamma errors, that is, with density 
\begin{equation}
g(s,\gamma)=\frac{\gamma^{\gamma}}{\Gamma(\gamma)} \exp^{\gamma\, (s- \exp(s))}\, .
\label{ggamma}
\end{equation}
 Assume that the variable $z \in \real_{\ge 0}$ and the covariates $\bx\in \real ^{q}$ are such that $z|{\bx} \sim \Gamma(\gamma_0,\mu(\bx))$, where the parametrization is such that $\esp (z|\bx)=\mu(\bx)$ and $\log \mu(\bx)=\bx\trasp \bbe_0$. Hence, defining $u= z/\mu(\bx)$, we have that $u \sim  \Gamma(\gamma_0,1)$ and therefore, if $ y= \log(z)$ and $\epsilon= \log(u)$, we get that 
\begin{equation}
y=\bx\trasp\bbe_0 +\epsilon \, ,\label{logamma}
\end{equation}
where $ \epsilon \sim \log(\Gamma(\gamma_0,1))$ has a density given by  (\ref{ggamma}) with $\gamma=\gamma_0$, i.e., it belongs to the family given in (\ref{exponencial}).

In the  log--Gamma model, the classical estimators are based on the maximum likelihood method and are defined through the minimization of the deviance, whose components are given by $d(y,a)=\exp(y-a)-(y-a)-1$. A natural way to robustify these estimators is by means of an $M-$estimation procedure. Thus, if $(y_i,\bx_i)\in \real^{q+1}$, $1 \le i \le n$, are independent observations following model (\ref{logamma}), an $M-$estimator is defined as
\begin{equation} 
\wbbe=\argmin_{\bbe}\sum_{i=1}^n \phi(y_i,\bx_i\trasp\bbe,\wc) = \argmin_{\bbe}\sum_{i=1}^n \rho\left(\frac{\sqrt{d(y_i,\bx_i\trasp\bbe)}}\wc\right)\, ,
\end{equation}
where $\wc$ is a preliminary estimate of a tuning constant $c_0$ and $\rho$ is a bounded and continuous loss function such as the Tukey's biweight function  given by $\rho(s)=\rho_{\tuk}(s) =\min\left(1,3 s^2 - 3 s^4 +  s^6\right)$.  For this family of distributions, the nuisance parameter can be taken as the tuning constant $c_0$ that is related to the unknown shape parameter $\gamma_0$. Fisher--consistency for this family of estimators has been studied in Bianco \textsl{et al.} (2005), under general conditions.

With this background in mind, let us now consider the case of a single index    model with log--Gamma errors, that is, $(y_i,\bx_i)\in \real^{q+1}$, $1 \le i \le n$, is a random sample where
\begin{equation}
y_i=\eta_0(\bbe_0\trasp \bx)+ \epsilon_i  \quad \mbox{ and } \quad \epsilon_i \sim \log(\Gamma(\gamma_0,1))\, \label{simlogamma}.
\end{equation}
We will borrow some of the previous ideas to introduce a robust profile method that involves smoothing and parametric techniques. Profile likelihood procedures were  studied by  van der Vaart (1988) and  applied to generalized partially linear models by  Severini and Wong (1992) and Severini and Staniswalis (1994).  In order to introduce the smoothers, we will consider local weights.
For the sake of simplicity,  given $\bbe$ we define the kernel weights $W_{h}(u,\bbe\trasp \bx_i )$ as
\begin{eqnarray*}
W_{h}(u,\bbe\trasp \bx_i )={K_h\left(\bbe\trasp\bx_i-u\right)}\left\{\sum_{j=1}^n K_h\left(\bbe\trasp\bx_j-u\right)\right\}^{-1}\;,
\end{eqnarray*}
where  $K_h(u)=(1/h)\,K(u/h)$ with $K$   a kernel function, i.e., a nonnegative integrable function on $\real$ and $h$ is the bandwidth parameter. The weights $W_{h}(u,\bbe\trasp
\bx_i )$   depend  on the closeness between the point $u$ and the projection of $\bx_i$ on the direction $\bbe$, i.e., between $u$ and $\bbe\trasp\bx_i$. To assume that a consistent estimator of the tuning constant, $\wc$, is available, let $\wgamma_{\rob}$ stand for a preliminary robust consistent estimator of $\gamma_0$ allowing to define $\wc=\wc(\wgamma_{\rob})$. The latter estimators must be properly computed according to the underlying errors distribution whose density we assume in the family given in (\ref{exponencial}).    
In Section \ref{estalfa}, we introduce a robust consistent estimator of the nuisance parameter for the usual regression model with symmetric errors and for the log--Gamma regression model, as well.

Then, for the particular situation of model (\ref{simlogamma}) we propose the following stepwise procedure

\begin{description}
\item[\textbf{Step LG1}:] For each fixed $\bbe$, with $\|\bbe\|=1$, let
\begin{eqnarray*}
\weta_{\bbech}(u)&=&\dst\argmin_{a\in \real}
\sum_{i=1}^n \rho\left(\frac{\sqrt{d(y_i,a)}}{\wc}\right)  W_{h}(u,\bbe\trasp \bx_i ).
\end{eqnarray*}

\item[\textbf{Step LG2}:] Define the estimators $\wbbe$ of $\bbe_0$ as the minimum of
$\Delta_n(\bbe)$ among $\|\bbe\|=1$, where
$$\Delta_n(\bbe)=\frac 1n \sum_{i=1}^n \rho\left(\frac{\sqrt{d\left(y_i, \weta_{\bbech} \left(\bbe\trasp \bx_i\right)\right)}}{\wc}\right)\tau(\bx_i) \, $$
and $\tau$ is a weight function.

\item[\textbf{Step LG3}:] Define the final estimator  $\weta$ of $\eta_0$ as $\weta(u)=\wa(u)$ with
\begin{eqnarray*}
(\wa(u),\wb(u))
&=&\dst\argmin_{(a,b)\in \real^2}\sum_{i=1}^n  W_{h}(u,\wbbe\trasp \bx_i ) 
\rho\left(\frac{\sqrt{d(y_i,a+b\;(\wbbe\trasp\bx_j-u))}}{\wc}\right) .
\end{eqnarray*}
\end{description}
The robust estimators are obtained controlling large values of the deviance  with a bounded loss function $\rho$. A popular choice is the Tukey's bisquare loss function $\rho=\rho_{\tuk}$, while $\wc$ estimates the tuning constant $c_0$  selected   to attain a given efficiency. As mentioned above, $c_0$ depends on the shape parameter $\gamma_0$ (see Bianco \textsl{et al.}, 2005). Note that the three steps involve the function
$$\phi(y,a,c) = \rho\left(\frac{\sqrt{d(y,a)}}c\right)\, ,$$
where, as above, $d(y,a)= \exp(y-a)-(y-a)-1$.

\subsection{The proposal for the general setting (\ref{exponencial})}{\label{proposal}}

Let us now consider the general case in which the errors have a density $g$ in the family (\ref{exponencial}). In order to extend the proposal given in Section \ref{proposalgam} to this situation, one may consider a loss function  $\phi$ bounding the deviances. To be more precise, let us denote as
$$
 \phi(y,a,\alpha) =  \rho\left(\frac{\sqrt{d(y,a)}}{\alpha}\right)\;,
$$
where   $d(y,a)= t(e_0)-t(y-a)$, with $e_0$   the unique mode of the density $g$ and   $\alpha$ is the tuning constant related to the nuisance parameter. As in Maronna \textsl{et al.} (2006), 
$\rho : \real \to \real_+$ is a $\rho-$function, that is, an even function, non--decreasing on $|s|$, increasing for $s>0$ when $\rho(s)<\lim_{x\to +\infty}\rho(x)$ and such that $\rho(0) = 0$. 

We define for each $\bbe$ and any continuous function $v:\real\to\real$ the functions
\begin{eqnarray}
\Upsilon(\bbe,a,u,\alpha)& = & \esp_0
\left[\phi\left(y,a,\alpha\right)|\bbe\trasp
\bx=u\right] \; ,\label{objfunct1}\\
\Delta(\bbe,v,\alpha)& = & \esp_0
\left[\phi\left(y,v(\bbe\trasp\bx),\alpha\right) \tau(\bx) \right]. \;\label{objbetafunct2}
\end{eqnarray}
where  $\tau$ is a weight function as above. 
Denote as $\eta_{\bbech}(u)=\argmin_{a \in \real}\Upsilon(\bbe,a,u,\alpha_0)$. Note that since we are considering the deviance and a continuous family of distributions with strongly unimodal density, there is no need to introduce a correction term to attain Fisher--consistency (see Bianco \textsl{et al.}, 2005). More precisely, we have that  $\bbe_0=\argmin_{\bbech \in \real^{q}}\Delta(\bbe,\eta_0,\alpha_0)$ and $\eta_{\bbech_0}=\eta_0$, furthermore $\bbe_0$ is the unique minimum of $\Delta(\bbe,\eta_0,\alpha_0)$.

In order to define consistent estimators of the parametric and nonparametric components, let us consider the empirical versions of  the objective functions (\ref{objfunct1}) and (\ref{objbetafunct2}), respectively, as
\begin{eqnarray*}
\Upsilon_n(\bbe,a,u,\alpha)&=& \sum_{i=1}^n  W_{h}(u,\bbe\trasp \bx_i ) \phi\left(y_i,a,\alpha\right) \, ,
\\
\Delta_n(\bbe,v,\alpha)& = & \frac 1n \sum_{i=1}^n  \phi\left(y_i,v(\bbe\trasp \bx_i),\alpha\right) \tau(\bx_i)\, ,
\end{eqnarray*}
where $v$ is any continuous function $v:\real\to\real$.

Assume that an initial robust estimator of $\alpha$, $\walfa_{\rob}$, is available. For a general single index model, the \textsl{robustified profile method} can thus be defined as
\begin{description}
\item[\textbf{Step 1}:] For each fixed $\bbe$, with $\|\bbe\|=1$, let
$$
\weta_{\bbech}(u)=\dst\argmin_{a\in \real}\Upsilon_n(\bbe,a,u,\walfa_{\rob}).
$$
\item[\textbf{Step 2}:] Define the estimators $\wbbe$ of $\bbe_0$ as
$$
\wbbe=\argmin_{\|\bbech\|=1}\Delta_n(\bbe,\weta_{\bbech},\walfa_{\rob}).
$$
\item[\textbf{Step 3}:] Define the final estimator  $\weta$ of $\eta_0$ as $\weta(u)=\wa(u)$ with
$$
(\wa(u),\wb(u))=\dst\argmin_{(a,b)\in \real^2}\sum_{i=1}^n  W_{h}(u,\wbbe\trasp \bx_i ) \phi\left(y_i,a+b\;(\wbbe\trasp\bx_i-u),\walfa_{\rob}\right) \, .
$$
\end{description}

Note that the stepwise procedure defined by \textbf{Step LG.1}--\textbf{Step LG.3} corresponds to 
\textbf{Step 1}--\textbf{Step 3} for a particular choice of the function $\phi$.

It is worth noticing that this stepwise procedure only involves unidimensional nonparametric smoothers, preventing from the sparsity of the data induced by the dimensionality of the covariates. In the third step a local polynomial of first degree is computed in order to improve the estimation of the link function $\eta_0$. When nuisance parameters are present, they may be estimated using a preliminary $S-$estimator which will allow to define also the tuning constant as  motivated in the next section.

\subsection{Initial estimators} {\label{estalfa}}

The calibration of the robust estimators will need the computation of a preliminary estimator of the nuisance parameter $\gamma_0$. As described in the Introduction,  as for many robust estimators,  this is a crucial issue for the three-step procedure and it can be accomplished in different ways according to the underlying error distribution. We will illustrate the computation of an initial estimator  of the nuisance parameter for the log-Gamma model, which can be extended to the case of errors with density in the family given in (\ref{exponencial}). In Section \ref{modelsym} we consider  the situation in which the errors have a symmetric distribution.

The preliminary estimator of the shape parameter $\gamma_0$ under model (\ref{simlogamma})  allows to compute the tuning constant by
means of an $S-$estimator. $S-$estimators were introduced by Rousseeuw and Yohai (1984) for ordinary
regression and studied in the framework of linear regression with asymmetric errors in Bianco \textsl{et al.} (2005). 
Let $\rho_{\tuk}$ be the bisquare $\rho-$function and  consider the following $S-$estimator. 
\begin{description}
\item[\textbf{Step ILG.1}] For each value of $a$,  $u$ and $\bbe$, compute $s_{n,\bbech,u}(a)$ as the solution of
$$\sum_{i=1}^n \rho_{\tuk}\left(\frac{\sqrt{d(y_i,a)}}{s_{n,\bbech,u}(a)}\right)   W_{h}(u,\bbe\trasp \bx_i )=b\, ,$$
where, for instance, $b=1/2$ and $d(y,a)=\exp(y-a)-(y-a)-1$. Define $\wteta_{\bbech} (u)$  as the value  $\wteta_{\bbech} (u)  =\argmin_a s_{n,\bbech,u}(a)$
\item[\textbf{Step ILG.2}] For each $\bbe$, let $\wtsigma(\bbe)$ be the solution of 
$$\frac 1{{\sum_{i=1}^n \tau(\bx_i)}} \sum_{i=1}^n \rho_{\tuk}\left(\frac{\sqrt{d\left(y_i, \wteta_{\bbech} \left(\bbe\trasp \bx_i\right)\right)}}{\wtsigma(\bbe)}\right)\tau(\bx_i)=b \, .$$
Now, the estimator  of $\bbe_0$ is given by $\wtbbe=\argmin_{\|\bbech\|=1} \wtsigma(\bbe)$ and $\wese_n=\wtsigma(\wtbbe)$.
\item[\textbf{Step ILG.3}] Define the estimator of $\gamma_0$ as $\wgamma=S^{\star\; -1}(\wese_n)$ where $S^{\star}(\gamma)$ is the solution of
$$\esp_{\gamma} \rho_{\tuk}\left(\frac{\sqrt{d(\epsilon,0)}}{S^{\star}(\gamma)}\right)=\esp_{\gamma} \rho_{\tuk}\left(\frac{\sqrt{\exp(\epsilon)-1-\epsilon}}{S^{\star}(\gamma)}\right)=b$$
where  $\epsilon$ has density $g(s,\gamma)$ given in (\ref{ggamma}).
\end{description}  
This method provides an estimator of $\gamma_0$ as well as an initial estimator $\wtbbe$ of $\bbe_0$, which is robust, but may be inefficient. It also provides an estimator of the function $\eta_0$ as $\weta=\wteta_{\wtbbech}$. These estimators may be used to start the stepwise estimation procedure in \textbf{Steps LG1} to \textbf{LG3} given above. In Bianco \textsl{et al.} (2005) it is shown that $S^{\star}(\gamma)$ is a  one--to--one function and thus invertible. For this reason, they recommend to take $\wc_n \ge \wese_n= S^{\star}(\wgamma)$.

It is worth noting that  if we replace $d(y,a)=\exp(y-a)-(y-a)-1$ by $d(y,a)=t(e_0)-t(y-a)$  in the initial Steps \textbf{ILG.1} to \textbf{ILG.3}, the described procedure provides preliminary estimators when the errors have density given by (\ref{exponencial}).

\subsection{The model with symmetric errors}{\label{modelsym}}
 As it is noted above, the family of densities given in (\ref{exponencial}) also includes symmetric distributions. In this case, a suitable initial method that exploits this feature of the  errors distribution can be introduced. Thus, as a second example, we consider the symmetric setting. We set $\alpha=\sigma$ and  $\rho_0(u)=\rho_{\tuk}(u/c_0)$, where $c_0$ is the tuning constant needed to obtain a scale Fisher--consistent estimator. For instance, when dealing with Tukey's bisquare function $\rho_{\tuk}$, the choice $c_0 = 1.54764$ and $b=1/2$ leads to a scale estimator Fisher--consistent at the normal distribution with breakdown point 50\%.  Then, to provide a preliminary estimator of the true scale parameter $\alpha_0=\sigma_0$,  let us consider an $S-$estimator that can easily be computed as follows.
\begin{description}
\item[\textbf{Step IS.1}] For each value of $u$ and $\bbe$, compute $\wteta_{\bbech} (u)$ as the median of the empirical local distribution
$$F_{n,\bbech,h} (s)=\sum_{i=1}^n \indica_{(-\infty, s]}(y_i)   W_{h}(u,\bbe\trasp \bx_i )\, .$$
\item[\textbf{Step IS.2}] For each $\bbe$, let $\wtsigma(\bbe)$ be the solution of 
$$\frac 1{{\sum_{i=1}^n \tau(\bx_i)}} \sum_{i=1}^n \rho_0\left(\frac{y_i-\wteta_{\bbech} \left(\bbe\trasp \bx_i\right)}{\wtsigma(\bbe)}\right)\tau(\bx_i)=b \, ,$$
where, for instance, $b=1/2$. Now, the estimators of $\bbe_0$ and $\sigma_0$ are given as $\wtbbe=\argmin_{\|\bbech\|=1} \wtsigma(\bbe)$ and $\walfa_{\rob}=\wsigma=\wtsigma(\wtbbe)$.
\end{description} 
To improve the efficiency of the estimators of $\bbe_0$, consider $\rho_1(u)=\rho_{\tuk}(u/c_1)$, with $c_1 > c_0$, and define an $MM-$procedure as follows.
\begin{description}
\item[\textbf{Step S.1}] For each value of $u$ and $\bbe$, compute $\wteta_{\bbech} (u)$ as 
\begin{eqnarray*}
\weta_{\bbech}(u)=\dst\argmin_{a\in \real}  \sum_{i=1}^n  W_{h}(u,\bbe\trasp \bx_i ) \rho_1\left(\frac{y_i-a}{\wsigma}\right).
\end{eqnarray*}
\item[\textbf{Step S.2}] Define the estimator  $\wbbe$ as  
$$\wbbe= \argmin_{\|\bbech\|=1} \frac 1{n} \sum_{i=1}^n \rho_1\left(\frac{y_i-\weta_{\bbech} \left(\bbe\trasp \bx_i\right)}{\wsigma}\right)\tau(\bx_i) \, .$$
\item[\textbf{Step S.3}] 
For each value of $u$, define the final estimator $\weta$ of $\eta_0$ as $\weta(u)=\wa(u)$ with
\begin{eqnarray*}
(\wa(u),\wb(u))=\dst\argmin_{(a,b) \in \real^2}  \sum_{i=1}^n  W_{h}(u,\wbbe\trasp \bx_i )\rho_1\left(\frac{y_i- a -b(\wbbe\trasp\bx_i-u)}{\wsigma}\right) \;.
\end{eqnarray*}
\end{description} 

Note that the stepwise procedure defined by \textbf{Steps S.1} to \textbf{S.3} corresponds to 
\textbf{Step 1}--\textbf{Step 3} for a particular choice of the function $\phi$, that is, $\phi(y,a,\alpha)=\rho_1((y-a)/\alpha)$.

\section{Asymptotic results}{\label{asymp}}

In this section, we derive, under some regularity conditions, the consistency
of the estimators defined in Section \ref{proposal} through Steps 1 to 3. We will assume that  $\bx\in \itX\subset \real^p$. Let $\itX_0\subset \itX$ be a compact set and define the set $ \itU(\itX_0)=\{\bbe\trasp\bx: \bx\in \itX_0,\; \bbe \in \itS_1\}$, where  $\itS_1$ is the unit ball in $\real^p$, i.e.,  $\itS_1=\{\bbe \in \real ^{p}: \|\bbe\|=1 \}$.  For any continuous function  $v:\itU(\itX_0)\to \real$ denote $\|v\|_{0,\infty}=\sup_{u\in \itU(\itX_0)}|v(u)|$. We consider the following set of assumptions:

\begin{enumerate}
\item[\textbf{A1.}]   The loss function  $\rho$ and the function  $t$ defined in (\ref{exponencial}) are continuous. Moreover, $\rho$ and $\tau$ are  bounded.
\item[\textbf{A2.}] The kernel $K:\real \to \real$ is an even, nonnegative, continuous and bounded function, with bounded
variation, satisfying $ \int K(u)du=1$, $ \int u^2K(u)du<\infty$ and $|u| K(u)\to 0$ as $|u|\to \infty$.
\item[\textbf{A3.}] The bandwidth sequence $h=h_n$ is such that $h\to 0$, $nh/\log(n) \to \infty$ when ${n\to \infty}$.
\item[\textbf{A4.}] i) The marginal density $f_{\bX}$ of $\bx$ is  bounded  in $\itX$.
\new ii) {Given any compact set $\itX_0\subset \itX$,
there exists a positive constant $A_1\!\!\left(\itU(\itX_0\right))$ such that $A_1\!\!\left(\itU(\itX_0)\right)<f_{\bbech}(u)$ for all $u\in \itU(\itX_0)$ and $\|\bbe\|=1$, where $f_{\bbech}$ is the marginal density of $\bbe\trasp \bx$.}

\item[\textbf{A5.}] The function  $\Upsilon(\bbe,a,u,\alpha)$ satisfies the following equicontinuity  condition: given $\itX_0\subset \itX$ and $\itK\subset \real_{>0}$ compact sets,  for any $\epsilon>0$ there exists  $\delta>0$ such that for any $u_1, u_2\in\itU(\itX_0)$;
$\bbe_1, \bbe_2\in \itS_1$ and $\alpha_1,\alpha_2\in \itK$,
$$|u_1-u_2|<\delta \;, |\alpha_1-\alpha_2|<\delta\; \mbox{ and } \|\bbe_1-\bbe_2\|<\delta\;\Rightarrow \sup_{{a\in
\real}}|\Upsilon(\bbe_1, a,u_1, \alpha_1)-\Upsilon(\bbe_2, a,u_2, \alpha_2)|<\epsilon\;
.
$$
\item[\textbf{A6.}] The function $\Upsilon(\bbe,a,u,\alpha)$ is continuous and $\eta_{\bbech, \alpha}(u)=\argmin_{a\in \real} \Upsilon(\bbe,a,u,\alpha)$ is a continuous function on $(\bbe,u,\alpha)$.
\item[\textbf{A7.}] The initial estimator  of $\alpha$, $ \walfa_{\rob} $, is a consistent estimator.
\item[\textbf{A8.}]  The functions $\rho$   and $t$ are differentiable functions.
\end{enumerate}

\vskip0.2in
\noi \textbf{Remark  1.} Condition \textbf{A1} is fulfilled by the  loss functions commonly used in the framework of robustness such as Tukey's bisquare
function and guarantees that $\phi(y,a,\alpha)$ is a continuous and bounded function.   Assumptions \textbf{A2} and \textbf{A3} are standard in nonparametric regression. Moreover, \textbf{A2} is verified for the Epanechnikov and Gaussian kernels, while \textbf{A3} is satisfied choosing $h_n= n^{-q}$ for $q>0$. \textbf{A4} is a standard condition in semiparametric models; in particular ii) is achieved if $f_{\bx}(\bx)> B_1(\itX_0)$ for any $\bx \in \itU(\itX_0)$. Note that \textbf{A8} entails that $\phi(y,a,\alpha)$ is a continuously differentiable function with respect to $a$. We will denote as $\phi^{\,\prime}(y,a,\alpha)$ its partial derivative with respect to $a$.

The following Lemma gives the uniform convergence of $\weta_{\bbech, \alpha}$ to $\eta_{\bbech, \alpha}$.  Its proof is omitted since it follows using analogous arguments to those considered in the proof of Lemma 3.1 in Boente and Rodriguez (2012).

\vskip0.2in
\noindent \textbf{Lemma  1.} \textsl{Let $\itK\subset \real_{>0}$  and $\itX_0\subset\itX$ be compact sets and assume that there exists $\delta_0>0$ such that $\itX_{\delta_0,0}\subset\itX$, where  $\itX_{\delta_0,0}$ stands for the closure of a $\delta_0$-neighbourhood of $\itX_0$. Assume that \textbf{A1} to \textbf{A6} hold and that
the family of functions $\itF=\{f(y)=\phi\left(y,a, \alpha\right)\,, a \in \real, \alpha\in \itK \}$ has a covering number satisfying $ \sup_{\qu}N\left(\epsilon,\itF,L^1(\qu)\right)\le A \epsilon^{-W}$, for any   $0<\epsilon<1$ and some positive constants $A$ and $W$, where $\qu$ stands for any probability measure for $(y,\bx)$.
Then, we have that}
 \begin{enumerate}
\item [a)]\textsl{$\dst\sup_{a \in \real, \bbech\in\itS_1, \alpha\in \itK}\|\Upsilon_n(\bbe, a,\cdot,\alpha)-\Upsilon(\bbe, a,\cdot,\alpha)\|_{0,\infty}\convpp 0 $.}
\item [b)]\textsl{If $ \dst\inf_{\stackrel{\bbech \in \itS_1,\;\alpha \in \itK}{u\in \itU(\itX_0)}}\left[\dst \lim_{|a|\to \infty}\Upsilon(\bbe, a,u,\alpha)-\Upsilon(\bbe, \eta_{\bbech, \alpha}(u),u,\alpha)\right]>0$,  then
$\!\!\!\dst\sup_{\bbech \in \itS_1,\;\alpha \in \itK} \|\weta_{\bbech, \alpha}-\eta_{\bbech, \alpha}\|_{0,\infty}\convpp 0$, where
$\weta_{\bbech, \alpha}(u)= \argmin_{a\in \real} \Upsilon_n(\bbe,a,u,\alpha)$.}
\end{enumerate}

\noi \textbf{Remark  2.} The condition on the infimum assumed in Lemma  1b)   warranties that the infimum of function $\Upsilon$ in (\ref{objfunct1}) is not attained at infinity.  Recall that finite--dimensional families of functions are VC--classes of functions  as defined in   Pollard (1984). Hence, using that  
$$\itF=\{f(y)=\phi\left(y,a, \alpha\right)=\rho\left(\frac{\sqrt{t(e_0)-t(y-a)}}{\alpha}\right)\,, a \in \real, \alpha\in \itK \}\,,$$
we obtain that the required condition on the covering number depends on the behaviour of the function $t(s)$. In particular, for the log--Gamma regression model, this condition is satisfied for any $\rho-$function.

\vskip0.1in
From Lemma  1, the continuity of $\eta_{\bbech, \alpha}(u)$ as a function of $\bbe$ and \textbf{A7}, we obtain the following result recalling that  $\Delta(\bbe,\eta_0,\alpha_0)$ has a unique minimum at  $\bbe_0$.

\vskip0.1in
\noindent \textbf{Theorem 1.} \textsl{Let $\wbbe$ be defined  $\wbbe=\argmin_{\bbech}\Delta_n(\bbe,\weta_{\bbech,\walfach_{\rob}},\walfa_{\rob}) $,
where $\weta_{\bbech,\alpha}= \argmin_{a\in \real} \Upsilon_n(\bbe,a,u,\alpha)$ satisfies
\begin{equation}
\sup_{\bbech \in \itS_1,\;\alpha \in \itK} \|\weta_{\bbech, \alpha}-\eta_{\bbech, \alpha}\|_{0,\infty}\convpp 0\;.\label{condicion1}
\end{equation}
Assume that \textbf{A1} and \textbf{A8} hold and that $\walfa_{\rob} \convpp \alpha_0$. Then, we  have that}
\begin{enumerate}
\item [a)]\textsl{$\dst\sup_{\bbech,\bb \in \itS_1; a \in \itK}\left|\Delta_n(\bbe,\weta_{\bb,a},a)-\Delta(\bbe,\eta_{\bb,a},a)\right| \convpp 0$ for any compact set $\itK\subset \real_{>0}$.}
\item[b)] \textsl{$\wbbe\convpp \bbe_0$.}
\end{enumerate}

\vskip0.2in

The asymptotic distribution of $\wbbe$ can be derived using the consistency of $\walfa_{\rob}$. In fact,  similar arguments to those considered in the proof of Theorem 3.5.3 in  Rodriguez (2007) can be used, but taking into account the fact that the estimator of the nuisance parameter is consistent. In particular, we consider below the case of a   log--Gamma model. 

From now on, we assume  that $\rho$ is twice continuously differentiable with first and second derivatives   $\Psi(y)$ and $\Psi^{\prime}(y)$ respectively and that $\eta_{\bb,a}$ is continuously differentiable in $(\bb,a)$.

Recall that under a log--Gamma model, $y_i=\eta_0(\bbe_0\trasp \bx_i)+ \epsilon_i$ with $\epsilon_i\sim \log\left(\Gamma(\gamma_0,1)\right)$ independent of $\bx_i$, so  $d(y,a)=\exp(y-a)-(y-a)-1$ and $\phi(y,a,c)=\rho\left(\sqrt{d(y,a)}/c\right)$. If we define $d^{\ast}(u)=\exp(u)-u-1$, we have that $d(y,\eta_{0}(\bbe_0\trasp \bx))=d^{\ast}(y-\eta_{0}(\bbe_0\trasp \bx))$.   Let
\begin{eqnarray}
\psi(y,a,c)&=& \frac{\partial}{\partial a} \phi(y,a,c) 
\label{phi1}\\
\chi(y,a,c)&=& \frac{\partial}{\partial a} \psi(y,a,c) \,.
\label{xi1}
\end{eqnarray}
Hence, $\psi(y,\eta_{0}(\bbe_0\trasp \bx),c)=\psi^{\ast}(y-\eta_{0}(\bbe_0\trasp \bx),c)$ and $\chi(y,\eta_{0}(\bbe_0\trasp \bx),c)= \chi^{\ast}(y-\eta_{0}(\bbe_0\trasp \bx),c)$, where
\begin{eqnarray*}
\psi^{\ast}(u,c)&=& \frac{1}{2c} \; \Psi\left( \frac{\sqrt{d^{\ast}(u)}}{c}\right)  \frac{1-  \exp(u)}{\sqrt{ d^{\ast}(u)}}\\
 \chi^{\ast}(u,c)&=&  \frac{1}{4c^2} \; \Psi^{\prime}\left(\frac{\sqrt{d^{\ast}(u)}}{c}\right) \frac{(1- \exp(u))^2}{d^{\ast}(u)}+ \;  \frac{1}{4c}  \; \Psi\left(\frac{\sqrt{d^{\ast}(u)}}{c}\right) 
\left[ \frac {2 \exp(u)}{\sqrt{d^{\ast}(u)}}- \frac{(1-\exp(u))^2}{ d^{\ast}(u)^{3/2}}\right] \,.
\end{eqnarray*}
Define  $\bB =\esp_0\left[\chi\left(y_1,\eta_{0}(\bbe_0\trasp \bx_1),c\right) \tau(\bx) \; \bnu_1(\bbe_0,\bbe_0\trasp\bx_1)\bnu_1(\bbe_0,\bbe_0\trasp\bx_1)\trasp\right]$, 
where
\begin{eqnarray}
\bnu_i(\bb,t)={ \frac{\partial}{\partial\bbech}
\eta_{\bbech}(s)|_{(\bbech,s)=(\bb,t)}+
\frac{\partial}{\partial s}
\eta_{\bbech}(s)|_{(\bbech,s)=(\bb,t)}\;\bx_i} \; . \label{nues}
\end{eqnarray}
Due to the independence between the errors and the covariates, $\bB$ can be written as
\begin{equation} 
\bB 
=\esp\left(\chi^{\ast}\left(\epsilon_1,c\right)\right)\, \wtbB\; ,
\label{matrizB}
\end{equation} 
where $\wtbB=\esp\left[ \tau(\bx_1) \bnu_1(\bbe_0,\bbe_0\trasp\bx_1) \bnu_1(\bbe_0,\bbe_0\trasp\bx_1)\trasp\right]$.
Furthermore, consider the matrix
\begin{eqnarray}
\bSi&=& 4 \, \esp_0\left\{\psi^2\left(y_1,\eta_{0}(\bbe_0\trasp \bx_1),c\right) \tau^2(\bx_1)\bnu_1(\bbe_0,\bbe_0\trasp\bx_1)\bnu_1(\bbe_0,\bbe_0\trasp\bx_1)\trasp\right\} =    \esp\left\{\psi^{\ast\, 2}\left(\epsilon_1,c\right)\right\}\,  \wtbSi \,,\label{matrizSigma}
\end{eqnarray}
with $ \wtbSi= 4 \esp\left\{  \tau^2(\bx_1)\bnu_1(\bbe_0,\bbe_0\trasp\bx_1)\bnu_1(\bbe_0,\bbe_0\trasp\bx_1)\trasp\right\}$.
Let $\bB_1$, $\wtbB_1$, $\bSi_1$ and $ \wtbSi_1$ be the left superior matrices  of dimension
$(q-1)\times (q-1)$ of  $\bB$, $\wtbB$, $\bSi$ and $ \wtbSi$, respectively. Assume that $\bB$ is  non--singular, $\wbbe\convprob \bbe_0$,  $\bx_i$ are random vectors with distribution with compact support $\itX$ and the bandwidth $h=h_n$ satisfies $nh^4\to 0$ and ${nh^2}/{\log{(1/h)}}\to\infty$.  Then, using analogous arguments to those considered in Rodriguez (2007) for the case of fixed nuisance parameter and taking into account that  $\wc_n\convprob c$, we obtain that
\begin{eqnarray}
 \sqrt n (\wbeta_q-\beta_{0q})&\convprob & 0 \label{convbetaq}\\
 \sqrt{n} ({\wbbe^{(q-1)}-\bbe^{(q-1)}_0}) &\convdist &
N(0,\bB_1^{-1}\bSi_1(\bB_1^{-1})\trasp)\; ,
\label{convnor}
\end{eqnarray}
where for any $\bb\in \real^{q}$, $\bb^{(q-1)}=(b_1,\dots,b_{q-1})\trasp$. 
  
Hence, using (\ref{matrizB}) and (\ref{matrizSigma}), we get that
$$
\sqrt
n ({\wbbe^{(q-1)}-\bbe^{(q-1)}_0})\convdist
N\left(0,\frac{ \esp\psi^{\ast\, 2}\left(\epsilon_1,c\right)}{\left(\esp\chi^{\ast}\left(\epsilon_1,c\right)\right)^2}\;\wtbB_1^{-1}\wtbSi_1(\wtbB_1^{-1})\trasp\right)\; .$$
Since the classical estimator of the single index parameter corresponds to the choice $\rho(s)=s^2$,  its  asymptotic covariance matrix is of the form 
$({1}/{\gamma_0})\;\wtbB_1^{-1}\wtbSi_1(\wtbB_1^{-1})\trasp $,
therefore, the asymptotic efficiency with respect to the classical estimator is given by
$$e=\frac{1}{\gamma_0}\; \frac{\left(\esp\chi^{\ast}\left(\epsilon_1,c\right)\right)^2}{ \esp_0\psi^{\ast\, 2}\left(\epsilon_1,c\right)} \, ,$$
which equals the efficiency of the $MM-$regression estimator described in Bianco \textsl{et al.} (2005). 

Let us consider the parametrization given in (\ref{modelotheta}) and let $\bthe=(\bthe^\star,\theta_q)$ with $\bthe^\star=(\theta_1,\dots,\theta_{q-1}) \in \real^{q-1}$ and $\theta_q=1$. Using that the parameter $\bthe$ equals $\bbe/\beta_q$ ($\beta_q>0$), we have that $\bthe^\star=\bbe^{(q-1)}/\beta_q$. Hence, the relation  between both parameters suggests to estimate $\bthe^\star$  by means of $\wbthe^\star=\wbbe^{(q-1)}/\wbeta_q$. Thus, from
$$
\sqrt{n}
 (  \wbthe^\star-   \bthe^\star_0) = \sqrt{n} \left( \frac{\wbbe^{(q-1)}}{\wbeta_q}-  \frac{\bbe^{(q-1)}_0}{\beta_{0q}} \right)= \frac{1}{\wbeta_q}\sqrt{n} \left( \wbbe^{(q-1)} -  \bbe^{(q-1)}_0\right) +
  \sqrt{n} \frac{(\beta_{0q}-\wbeta_q)}{\wbeta_q \beta_{0q}} \, \bbe^{(q-1)}_0 \;,
$$
it is easy to see that $\sqrt{n}
 (\wbthe^\star- \bthe^\star_0) \convdist
N\left(0,({1}/{\beta^2_{0q}})\, \bB_1^{-1}\bSi_1(\bB_1^{-1})\trasp\right)$, 
since  (\ref{convbetaq}) and  (\ref{convnor}) entail  that
$\sqrt{n}
 (  \wbthe^\star-   \bthe^\star_0) =  \sqrt{n} ( \wbbe^{(q-1)} -  \bbe^{(q-1)}_0)/{\beta_{0q}} + o_{\prob}(1) \;.$

\section{Empirical Influence Curve}{\label{eif}}

In this section, we derive the empirical influence function of the single index parameter estimator under a log--Gamma model. 
The empirical influence function ($\EIF$), introduced by Tukey (1977), measures the robustness of an estimator with respect to a single outlier. Essentially, it assesses the impact on an estimator of adding an arbitrary observation to the sample.  Diagnostic measures with the purpose of outlier identification  can be defined from the empirical influence functions.  
Mallows (1974) defines a finite version of the influence function, introduced by Hampel (1974), that is computed at the sample empirical distribution.  The $\EIF$ has been widely used in parametric statistics, but has retrieved less attention in nonparametric literature. Foremost, Manchester (1996) introduces a simple graphical procedure to display the sensitivity of a scatter plot smoother to perturbations in the data. Tamine (2002)  defines a
smoothed influence function in the context of nonparametric regression with a fixed bandwidth that is based on A\"it Sahalia (1995) smoothed functional approach to nonparametric kernel estimators. 

Following  Boente and Rodriguez (2010), we consider an empirical influence function that is close to Manchester (1996) approach and at the same time, retains the spirit of the $\EIF$ definition introduced by Mallows (1974).

To be more precise, denote  $\wbbe$ the single index parameter estimator based on the original data set ${(y_i, \bx_i)}, 1 \le i \le n$. If $P_n$ is the empirical measure that gives weight $1/n$ to each datum in the sample, we have that $\wbbe = \wbbe(P_n)$. Let $P_{n,\varepsilon}$ be the empirical measure that gives mass $(1 -\varepsilon)/n$ to each $(y_i, \bx_i), 1 \le i \le n$ and mass $\varepsilon$ to the arbitrary observation $(y_0, \bx_0)$. In other words, we have a new sample with the original data set accounting an $1-\varepsilon$ proportion and the new observation an $\varepsilon$ proportion. Now, denote $\wbbe_{\varepsilon}= \wbbe(P_{n,\varepsilon})$ the single index parameter estimator for the new sample. We compute the empirical influence function of $\wbbe$ at a given point $(y_0, \bx_0)$  as 
$$\EIF(\wbbe, (y_0, \bx_0)) = \lim_{\varepsilon \to 0} \frac{\wbbe_{\varepsilon}-\wbbe}{\varepsilon} \, .$$
It is easy to see that  the single index estimator is equivariant under orthogonal transformations. Hence, without loss of generality,  we can assume  that $\wbbe=\be_q$, the $q-$th canonical vector  of $\real^q$.  To obtain the empirical influence function, we will assume that the matrix $\bB_1$ given in (\ref{matrizB}) is non--singular, as required when deriving the asymptotic distribution of $\wbbe$ in Section \ref{asymp}. 
Furthermore, for simplicity, we will assume that the tuning parameter $c$ is fixed. 

To avoid burden notation,   denote $\EIF(\wbbe)=\EIF(\wbbe, (y_0, \bx_0))$,  $\EIF(\weta_{\bbech}(u))=\EIF(\weta_{\bbech}(u), (y_0, \bx_0))$, \linebreak $\EIF({\partial \, \weta_{\bbech}(u)}/{\partial\bbe} )=\EIF({\partial \, \weta_{\bbech}(u)}/{\partial\bbe}, (y_0, \bx_0))$ and $\EIF({\partial \, \weta_{\bbech}(u)}/{\partial u})=\EIF({\partial \, \weta_{\bbech}(u)}/{\partial u}, (y_0, \bx_0))$. 
Moreover, from now on, given $\bb \in \real^q$, $\bb^{(q-1)}=(b_1,\dots,b_{q-1})\trasp$ stands for the vector of its first $q-1$ elements. Besides, given the kernel $K$ and its first  derivative $K'$, we define    $K'_h(u)=(1/h)\,K'(u/h)$.

Let us assume that $\rho$ is three times continuously differentiable. As in (\ref{phi1}) and (\ref{xi1}), $\psi(y,a,c)$ and $\chi(y,a,c)$ stand for the derivatives with respect to $a$ of  $ \phi(y,a,c)$ and $\psi(y,a,c)$, respectively, and we further define
$$
\chi_1(y,a,c)= \frac{\partial}{\partial a} \chi(y,a,c) \,.
$$
Then, if $\Psi^{\prime \prime}$ stands for the second order derivative of $\Psi$, we have that $\chi_1(y,a,c)=\chi_1^{\star}(y-a,c)$ with
\begin{eqnarray*}
\chi_1^{\star}(u,c)&=&-\left\{   \frac{1}{8\,c^3} \; \Psi^{\prime \prime}\left(\frac{\sqrt{d^{\ast}(u)}}{c}\right) \frac{(1- \exp(u))^3}{d^{\ast}(u)^{3/2}} + \;  \frac{1}{4\,c}  \; \Psi\left(\frac{\sqrt{d^{\ast}(u)}}{c}\right) 
\left[ \frac{2 \,\exp(u)}{\sqrt{d^{\ast}(u)}}- \frac {3\,(1- \exp(u))^3}{2\,d^{\ast}(u)^{5/2}}\right]\right.\\
&&+\left. \;  \frac{1}{4\,c^2}  \; \Psi^{\prime}\left(\frac{\sqrt{d^{\ast}(u)}}{c}\right) 
\left[  \frac {(1- \exp(u))^3}{2\,d^{\ast}(u)^2}- \frac{\exp(u)(1-\exp(u))}{ d^{\ast}(u)}\right]\right\} \,.
\end{eqnarray*}
It is worth noticing that  if $\rho$ is three times continuously differentiable and the kernel $K$ is continuously differentiable, we have that $\weta_{\bbech}(u)$ defined in \textbf{Step LG1} is continuous and has continuous partial derivatives with respect to $\bbe$ and $u$.
\vskip0.2in

\noi \textbf{Proposition  1.} \textsl{Assume that $\rho$ is three times continuously differentiable, the kernel $K$ is continuously differentiable and that $ \bB_1$, the left superior matrix of dimension $(q-1)\times (q-1)$ of the matrix $\bB$ given in (\ref{matrizB}) is non--singular. Denote as}
\begin{eqnarray}
\Bell_n&=& \frac 1n \sum_{i=1}^n \chi \left(y_i,\weta_{\bbech}({\wbbe\trasp} \bx_i),c\right) \; \tau(\bx_i) \; {\left.\EIF\left(\weta_{\bbech}(u)\right)\right|_{(\bbech,u)=\widehat{\bese}_i}}
\; \wbnu_i({\wbbe},{\wbbe\trasp} \bx_i)   \nonumber\\
&+&  \frac 1n \sum_{i=1}^n 
\psi \left(y_i,\weta_{\bbech}({\wbbe\trasp} \bx_i),c\right) \;  \tau(\bx_i)  \left\{
 \left.\EIF\left(\frac{\partial}{\partial\bbe} \weta_{\bbech}(u)\right)\right|_{(\bbech,u)=\widehat{\bese}_i} + \left.\EIF\left(\frac{\partial}{\partial u} \weta_{\bbech}(u)\right)\right|_{(\bbech,u)=\widehat{\bese}_i} \bx_i\; \right\} \nonumber\\
&+& \psi \left(y_0,\weta_{\bbech}({\wbbe\trasp} \bx_0),c\right) 
\wbnu_0({\wbbe},{\wbbe\trasp} \bx_0) \tau(\bx_0)  \label{Bell}\\
 \bM_n&=& \left[\frac 1n \sum_{i=1}^n \chi \left(y_i,\weta_{\bbech}({\wbbe\trasp} \bx_i),c\right)   
\; \tau(\bx_i) \; \wbnu_i({\wbbe},{\wbbe\trasp} \bx_i) \;\wbnu_i({\wbbe},{\wbbe\trasp} \bx_i)\trasp   \right. \nonumber\\
&+&\left. \frac 1n \sum_{i=1}^n  \psi \left(y_i,\weta_{\bbech}({\wbbe\trasp} \bx_i),c\right) 
\bV(\widehat{\bese}_i) \tau(\bx_i)  \right] \label{bMn}\,,
\end{eqnarray} 
\textsl{where $\widehat{\bese}_i=(\wbbe,\wbbe\trasp \bx_i)$ and $\wbnu_i(\bb,t)$ are estimates of the quantities given in (\ref{nues}), that is, 
\begin{eqnarray*}
\wbnu_i(\bb,t)&=& { \frac{\partial}{\partial\bbech}
\weta_{\bbech}(s)|_{(\bbech,s)=(\bb,t)}+
\frac{\partial}{\partial s}
\weta_{\bbech}(s)|_{(\bbech,s)=(\bb,t)}\;\bx_i}\\
\wbnu_0(\bb,t)&=& { \frac{\partial}{\partial\bbech}
\weta_{\bbech}(s)|_{(\bbech,s)=(\bb,t)}+
\frac{\partial}{\partial s}
\weta_{\bbech}(s)|_{(\bbech,s)=(\bb,t)}\;\bx_0}\, 
\end{eqnarray*}
and
$$\bV(\widehat{\bese}_i)= \left[\left.\frac{\partial^2 \weta_{\bbech}(u) }{\partial^2\bbech}  \right|_{(\bbech,u)=\widehat{\bese}_i} +
\left.\frac{\partial^2 \weta_{\bbech}(u) }{\partial^2 u} \right|_{(\bbech,u)=\widehat{\bese}_i}
\bx_i \bx_i\trasp 
+\left.\frac{\partial^2\weta_{\bbech}(u)  }{\partial
u \partial \bbech} \right|_{(\bbech,u)=\widehat{\bese}_i} \bx_i\trasp
+ \left.\frac{\partial^2 \weta_{\bbech}(u) }{\partial \bbech \partial u}  \right|_{(\bbech,u)=\widehat{\bese}_i}
\bx_i\trasp \right] \; .
$$
Then, if $\bM_{n,1}$, the left upper $(q-1) \times (q-1) $ submatrix of $ \bM_n$, is invertible, we have that} 
\begin{itemize}
\item[a)] \textsl{$\EIF(\wbbe)_q=0$ and $\EIF(\wbbe^{(q-1)}) = - \bM_{n,1}^{-1} \Bell_n^{(q-1)}$.}
\item[b)] \textsl{the empirical influence functions at $(y_0,\bx_0)$, $ \EIF(\weta_{\bbech}(u))$, $ \EIF({\partial \, \weta_{\bbech}(u)}/{\partial\bbe})$ and $\EIF({\partial \, \weta_{\bbech}(u)}/{\partial u})$ are given by
\begin{eqnarray*}
\EIF(\weta_{\bbech}(u)) &=&
-\frac{K_h(\bbe\trasp \bx_0-u) \psi \left(y_0,\weta_{\bbech}(u),c\right)}{\dst D_n} \\
\EIF\left(\frac{\partial \weta_{\bbech}(u)}{\partial\bbech}\right) &=& 
-\frac{\dst\frac{1}{h} K_h^{\prime}(\bbe\trasp \bx_0-u) \psi \left(y_0,\weta_{\bbech}(u),c\right) \bx_0+ K_h (\bbe\trasp \bx_0-u) \chi \left(y_0,\weta_{\bbech}(u),c\right)\dst\frac{\partial}{\partial \bbech} \weta_{\bbech}(u)}{D_n}\\
&&+ \frac{K_h(\bbe\trasp \bx_0-u) \psi \left(y_0,\weta_{\bbech}(u),c\right)}{D_n^2}\,  \left[ \frac 1h \bg_n + F_n\dst\frac{\partial}{\partial \bbech} \weta_{\bbech}(u)\right] 
\\
\EIF\left(\frac{\partial \weta_{\bbech}(u)}{\partial u}\right) &=& \frac{1}{D_n}\left\{\dst\frac{1}{h} K_h^{\prime}(\bbe\trasp \bx_0-u) \psi \left(y_0,\weta_{\bbech}(u),c\right) - K_h (\bbe\trasp \bx_0-u) \chi \left(y_0,\weta_{\bbech}(u),c\right)\dst\frac{\partial}{\partial u} \weta_{\bbech}(u)\right\}\\
&& \,-\, \frac{K_h(\bbe\trasp \bx_0-u) \psi \left(y_0,\weta_{\bbech}(u),c\right)}{\dst D_n^2} \left( F_n\dst\frac{\partial}{\partial u} \weta_{\bbech}(u)- \frac{1}{h} E_n\right) \, ,
\end{eqnarray*}
where  
$$
\frac{\partial}{\partial u} \weta_{\bbech}(u) = \frac 1h\, F_n^{-1}  \, E_n\,, \qquad  
\frac{\partial}{\partial \bbech} \weta_{\bbech}(u) = -\frac 1h F_n^{-1}  \, \bg_n
$$
and
\begin{eqnarray*}
D_n= \frac 1n\sum_{i=1}^n K_h(\bbe\trasp \bx_i-u) \psi \left(y_i,\weta_{\bbech}(u),c\right)\,,&\qquad &
E_n = \frac{1}{n}\sum_{i=1}^n K_h^{\prime}\left( \bbe\trasp \bx_i-u \right) \psi\left(y_i,\weta_{\bbech}(u),\alpha\right)\,, \\
F_n = \frac 1n \sum_{i=1}^n K_h\left( \bbe\trasp \bx_i-u \right) \chi\left(y_i,\weta_{\bbech}(u),\alpha\right)\,,&\qquad &
\bg_n = \frac{1}{n}\sum_{i=1}^n K_h^{\prime}\left(\frac{\bbe\trasp \bx_i-u}{h}\right) \psi\left(y_i,\weta_{\bbech}(u),\alpha\right) \bx_i  \,.
\end{eqnarray*} }
\end{itemize} 

\vskip0.2in
\noi \textbf{Remark 3.} In Proposition 1,  the left upper submatrix $\bM_{n,1}$ is assumed to be non--singular. Using the conditional Fisher--consistency, that is,
$\esp_0\left(\psi\left(y,\eta_{\bbech_0}(\bbe_0\trasp\bx),c_0\right)| \bx\right)=0$,  we have that 
$$\frac 1n \sum_{i=1}^n  \psi \left(y_i,\weta_{\bbech}({\wbbe\trasp} \bx_i),c\right) 
\bV(\widehat{\bese}_i) \tau(\bx_i) \convprob 0 \; ,$$
under mild conditions, while
$$\frac 1n \sum_{i=1}^n \chi \left(y_i,\weta_{\bbech}({\wbbe\trasp} \bx_i),c\right) \;  
\; \tau(\bx_i) \; \wbnu_i({\wbbe},{\wbbe\trasp} \bx_i) \wbnu_i({\wbbe},{\wbbe\trasp} \bx_i)\trasp  
\convprob \bB \; .$$
Therefore,  $\bM_n\convprob \bB$ which implies that $\bM_{n,1}\convprob \bB_1$. Hence, taking into account that we have assumed that $\bB_1$ is invertible, we get that with probability converging to 1 $\bM_{n,1}$ is non--singular. 

It is worth noticing that even when considering a bounded loss function, such as the Tukey's bisquare function,  the empirical influence function may not be bounded in directions orthogonal to $\wbbe$, since the term involving $\bx_0$ in $\wbnu_0({\wbbe},{\wbbe\trasp} \bx_0) $ may not be bounded, unless the function $\tau(\bx_0)$ controls large values of the covariates. This behaviour is similar to that arising with projection--pursuit estimators when estimating the principal directions (see Croux and Ruiz--Gazen, 2005).

\section{Selection of the smoothing parameters}{\label{KCV}}
The estimation of the nonparametric component of the model involves a smoothing parameter both in the first and third steps. Each step may require a different degree of smoothness and for this reason, the bandwidths may be chosen different.  The effect of the bandwidth is crucial on the performance of the nonparametric estimator; the smoothing parameter must warranty a balance between bias and variance. The problem of bandwidth selection has been widely studied in nonparametric and semiparametric models and leave-one-out cross validation procedures have been extensively used for this purpose. $K$--fold--cross validation criteria are also a reasonable choice with a computationally cheaper cost.

However, it is well known that classical cross--validation criteria are very sensitive to outliers. It is worth noticing that robust criteria for the selection of the smoothing parameter are needed even when robust estimators are considered. 
Leung \textsl{et al.} (1993), Wang and Scott (1994), Boente \textsl{et al.} (1997), Cantoni and Ronchetti (2001) and Leung (2005) discuss these ideas in the fully nonparametric framework, while  Bianco and Boente (2007) and Boente and Rodriguez (2008) consider robust cross--validation in semiparametric models.

For the initial and final smoothing steps performed in \textbf{Steps 1} and \textbf{3} of the proposed method,  we consider a  robust version of the classical $K$--fold cross--validation criterion   based on the deviance to select the bandwidths.
More precisely, let us first randomly split the data set into $K$ subsets  of similar size, disjoint and exhaustive,  with indexes $\itI_j$, $1 \le  j \le K$, such that  $\cup_{j=1}^K \itI_j=\{ 1,\dots,n\}$.
Let ${\cal H}^1_n \subset \real$ be the set of bandwidths to be considered in the first step of the proposed procedure. Denote $\wtbbe_h^{(-j)}$ the robust regression estimator computed in \textbf{Step 2} without the observations with indexes in the set $\itI_j$ and using as smoothing
parameter $h \in {\cal H}^1_n$ in the previous step and let $\weta^{(-j)}_{\wtbbech,h}(u)$ be the corresponding nonparametric robust estimator computed in \textbf{Step 1}. 

Taking into account that for each $i$, $1 \le i \le n$, there exists $j$,  $1 \le  j \le K$, such that $i \in  {\cal I}_j$,
we define the prediction of observation $y_i$ as $\wy_i=\weta^{(-j)}_{\wtbbech,h}(\bx_i\trasp\wtbbe_h^{(-j)})$. Noticing that in the actual setting, the deviance residuals are a suitable measure of the discrepancy between an observation and its predictor, the 
robust $K-$th fold cross--validation smoothing parameter is defined as $\wh_1=\argmin_{h \in {\cal H}^1_n} RCV(h)$, where
\begin{equation}
RCV(h) \, = \,\sum_{i=1}^n \rho\left(\frac{\sqrt{d\left(y_i,\wy_i\right)}}{c}\right) \, , 
\label{RKCV}
\end{equation}
for a given tuning constant $c$.
Denote $\wbbe$ the robust estimator based on the whole sample when the smoothing parameter is the optimal $h=\wh_1$.

It is worth noticing that the robust $K-$th fold cross--validation $RCV(h)$ given in (\ref{RKCV}) is a robustified version of its classical counterpart that seeks  for the smoothing parameter   minimizing
\begin{equation}
CCV(h) \, = \,\sum_{i=1}^n {d\left(y_i,\wy_i\right)} \, , 
\label{CKCV}
\end{equation}
where $\wy_i$ are based on the classical estimators. 

In order to select the second bandwidth to be used in the local linear nonparametric estimator described in \textbf{Step 3}, we consider a similar procedure. That is, we take  ${\cal H}^2_n \subset \real$ the set of bandwidths to be considered in the third step and denote $\weta^{(-j)}_{\wbbech,h}(u)$ the robust nonparametric estimator without the observations with indexes in the set ${\cal I}_j$ and using as smoothing parameter $h \in {\cal H}^2_n$ and $\wbbe$. 
Again, reasoning as above, for each $1 \le i \le n$, we define the predictor of observation $y_i$ as $\wy_i=\weta^{(-j)}_{\wbbech,h}(\bx_i\trasp\wbbe)$ and so, the 
robust $K-$th fold cross--validation linear smoothing parameter is defined as $\wh_2=\argmin_{h \in {\cal H}^2_n} RCV(h)$. Once the data driven--bandwidth $\wh_2$ is obtained, the final nonparametric  estimator denoted $\weta_{\wbbech,\wh_2}$ can be computed as in \textbf{Step 3} from the whole sample using this bandwidth.

\section{Numerical results}{\label{monte}}


In this Section, we summarize the results of a simulation study designed to  compare the performance of the proposed estimators with the classical ones under a log--Gamma model. 

We have performed $N=1000$ replications with samples of  size $n=100$.
For the clean samples the covariates $\bx_i$ are generated as $\bx_i\sim  \itU((0,1)\times(0,1))$, while the response variables follow  the log--Gamma single--index model  $y_i=\eta_{0}(\bbe_0\trasp\bx_i)+\epsilon$, with  $ \eta_{0} (u) = \sin(2 \pi u)$, $\bbe_0=(1/\sqrt{2},1/\sqrt{2})\trasp$ and $\epsilon\sim \log(\Gamma(3,1))$.

In all Tables, the results for the uncontaminated samples  are denoted as $C_0$. Furthermore, the robust estimators introduced in this paper are subindicated with \textsc{r}, while  their classical counterparts based on the deviance are subindicated with \textsc{cl}.  To be more precise, the robust estimators correspond to those controlling large values of the deviance.  In this case, the robust estimators were computed using the Tukey's bisquare loss function with adaptive tuning constants computed as in Bianco \textsl{et al.} (2005). 
On the other hand, the classical estimators correspond to choose the loss function equal to the deviance.  
With respect to the weight or trimming function, in order to make a fair comparison between the classical and robust estimators, we choose  $\tau(\bx)= \|\bx-\bc\| \indica_{[0,b_n]}$, with $\bc=(0.5,0.5)$ and $b_n=0.4 \, \sqrt{\log{(\log{(n)})}}$ for both estimators. The value $b_n$ is selected  as in Sherman (1994) to avoid the density of $\bbe_0\trasp \bx$ to be too small.

The smoothing parameters were selected as described in Section \ref{KCV} using a $5$-fold cross--validation procedure. For the classical estimators we use the criterion (\ref{CKCV}) in each step, while for  the robust estimates,  we used the robust $5-$fold method (\ref{RKCV}) with $c=1.6394$ that under the central model corresponds to an asymptotic efficiency of 0.90. In all these cases,  the set ${\cal H}^1_n$ of candidates for the initial bandwidth $h$ was taken as an equidistant grid of length 13 between 0.05 and 0.35, while for the local linear smoothing parameter we choose ${\cal H}^2_n$ as an equidistant grid of length 25 between 0.05 and 0.35.  To simplify the notation, henceforth we denote $\wbbe_{\rob}$ and $\weta_{\rob}$ the robust estimators computed with the two robust cross--validation bandwidths, while $\wbbe_{\cl}$ and $\weta_{\cl}$ stand for the classical estimators computed with   the  bandwidths obtained minimizing (\ref{CKCV}).

To evaluate the performance of each estimator we compute different measures. For the parametric component, given an estimator $\wbbe$ of the true single index parameter $\bbe_0$, we consider $\mbox{MSE}_{\wbbech}$ as the mean values over replications of $\|\wbbe-\bbe_0\|^2$. For the nonparametric component,   we compute $\mbox{MSE}_{\wetach}$ as the mean over replications of  $(1/n)\sum_{i=1}^n (\eta_0(\bx_i\trasp\bbe_0)-\weta(\bx_i\trasp\wbbe))^2$ and also $\mbox{MedSE}_{\wetach}$ as the median over replications of $\mbox{median}_{i=1:n} (\eta_0(\bx_i\trasp\bbe_0)-\weta(\bx_i\trasp\wbbe))^2$, where $\weta$ is a given estimator of the function $\eta_0$.

In order to assess the behaviour of the estimators under contamination, we have considered two types of contaminations and samples $(y_{i,c}, \bx_{i,c})$ generated from them. The first set of contaminations introduces moderate outlying points, while with the second one we expect a more dramatic effect on the classical estimators.

Three different models, labelled $M_1$, $M_2$ and $M_3$ in all Tables and Figures are considered in the moderate contamination scheme. To obtain the contaminated samples, we have first generated a sample   $u_{i} \sim \itU(0,1)$ for $1\leq i \leq n$ and then, we  introduce large values on the responses as
\begin{equation}
y_{i,c}=\left\{
\begin{tabular}{ll}
$y_{i}$& \mbox{if $u_{i}\leq0.90$}\\
$y_{i}^{\star}$ & \mbox{if $u_{i}>0.90$,}
\end{tabular}
\right.
\label{contam}
\end{equation}
where   $y_i^{\star}= \log(k)+ \wteta(\bx_i)+ \epsilon_i$, with $\epsilon_i\sim  \log(\Gamma(3,1))$,  $\wteta(\bx_i)= \eta(\bx_i\trasp\bbe_0^{\bot})$ where $\bbe_0^{\bot}$  is the unit vector orthogonal to the true single index parameter $\bbe_0$ and    $k= 3, 4$ and $5$ under $M_1$, $M_2$ and $M_3$, respectively.  

\begin{figure}[ht!]
\begin{center}
\small 
 $C_0$\\
\includegraphics[scale=0.3]{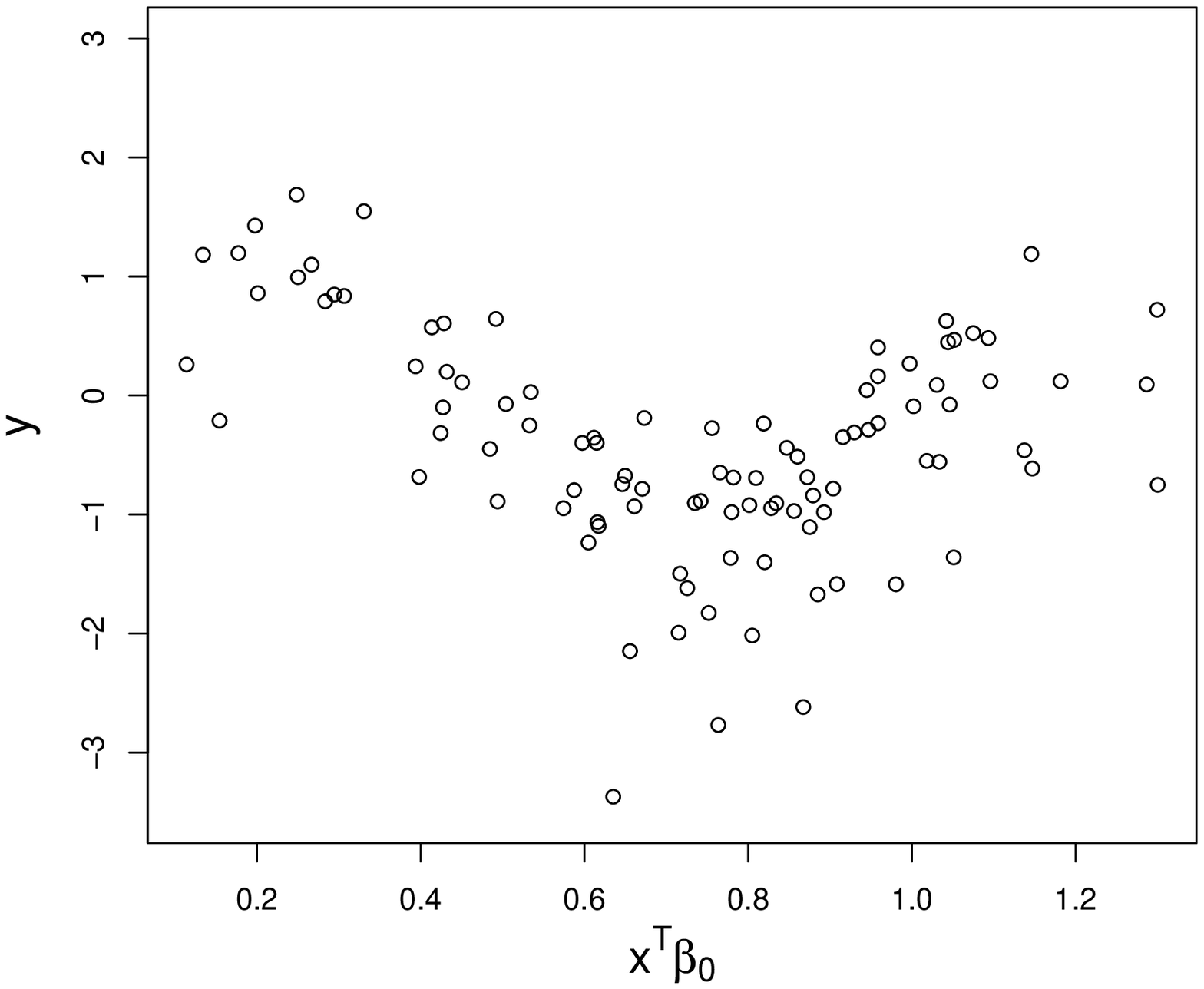}\\
$M_1$  \hspace{4.5cm} $M_2$ \hspace{4.5cm} $M_3$\\
\vspace{-0.3cm}\includegraphics[scale=0.3]{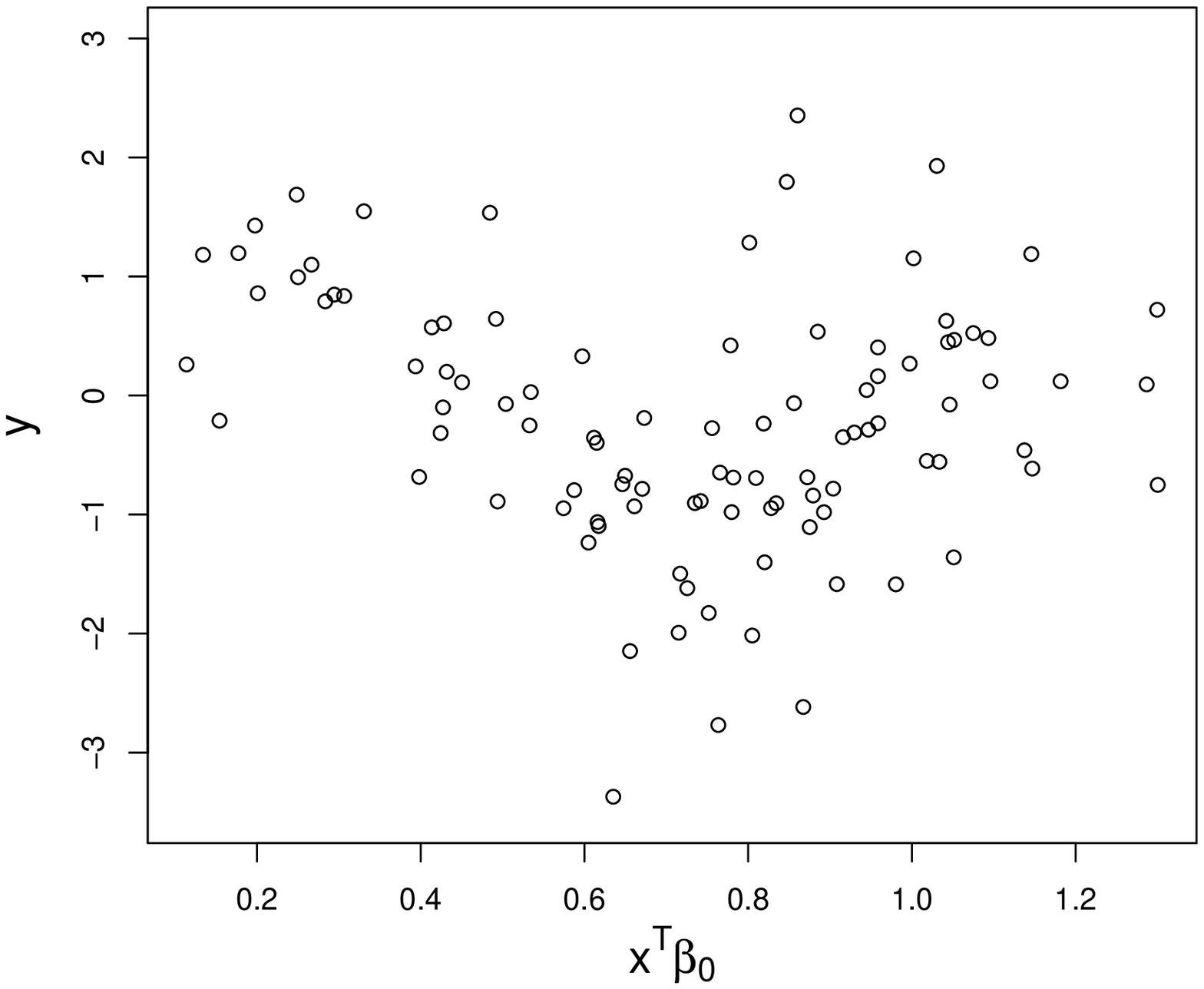} 
\includegraphics[scale=0.3]{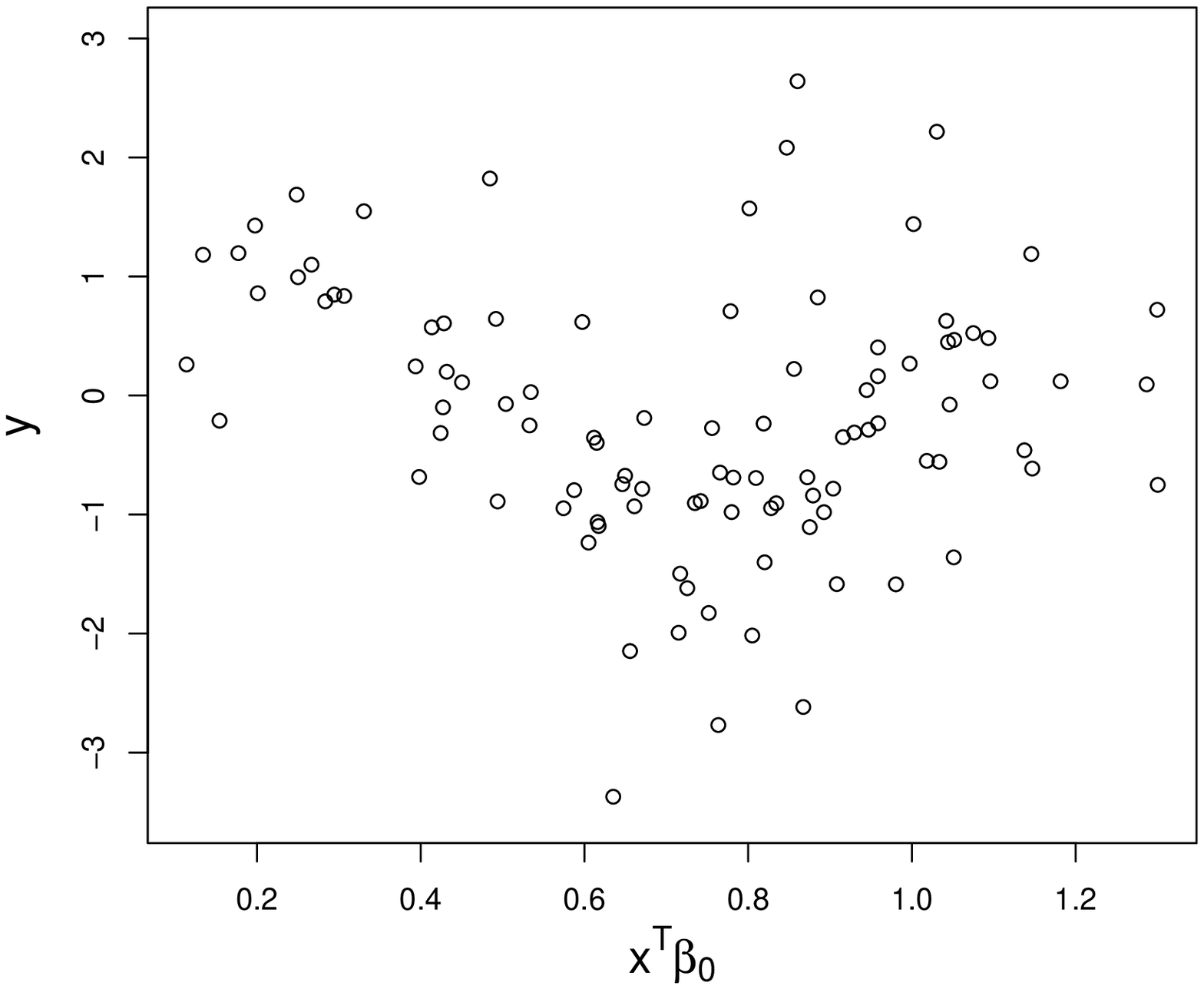} 
\includegraphics[scale=0.3]{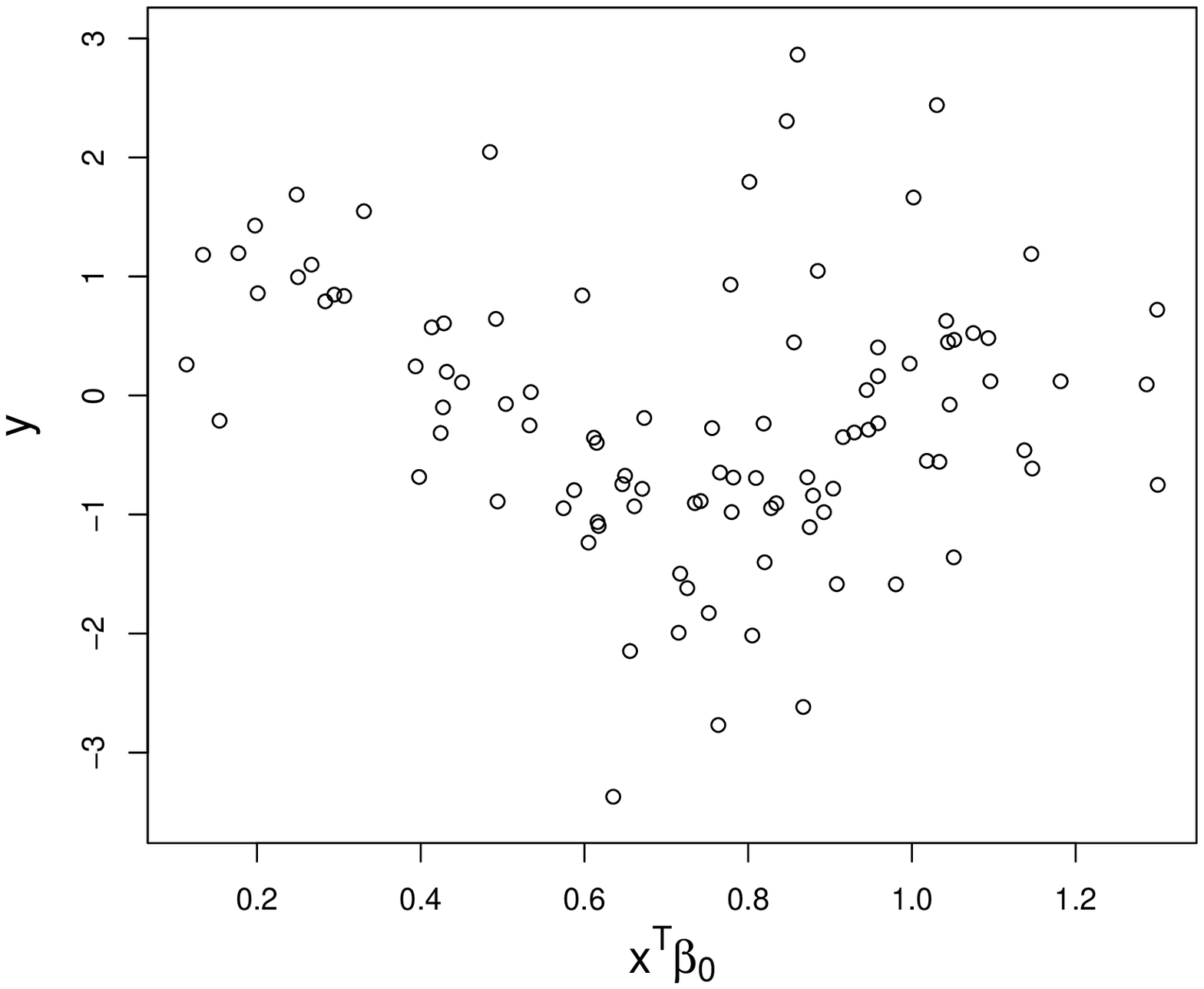} \\
$S_1$ \hspace{4.5cm}  $S_2$  \hspace{4.5cm} $S_3$\\
\vspace{-0.3cm}\includegraphics[scale=0.3]{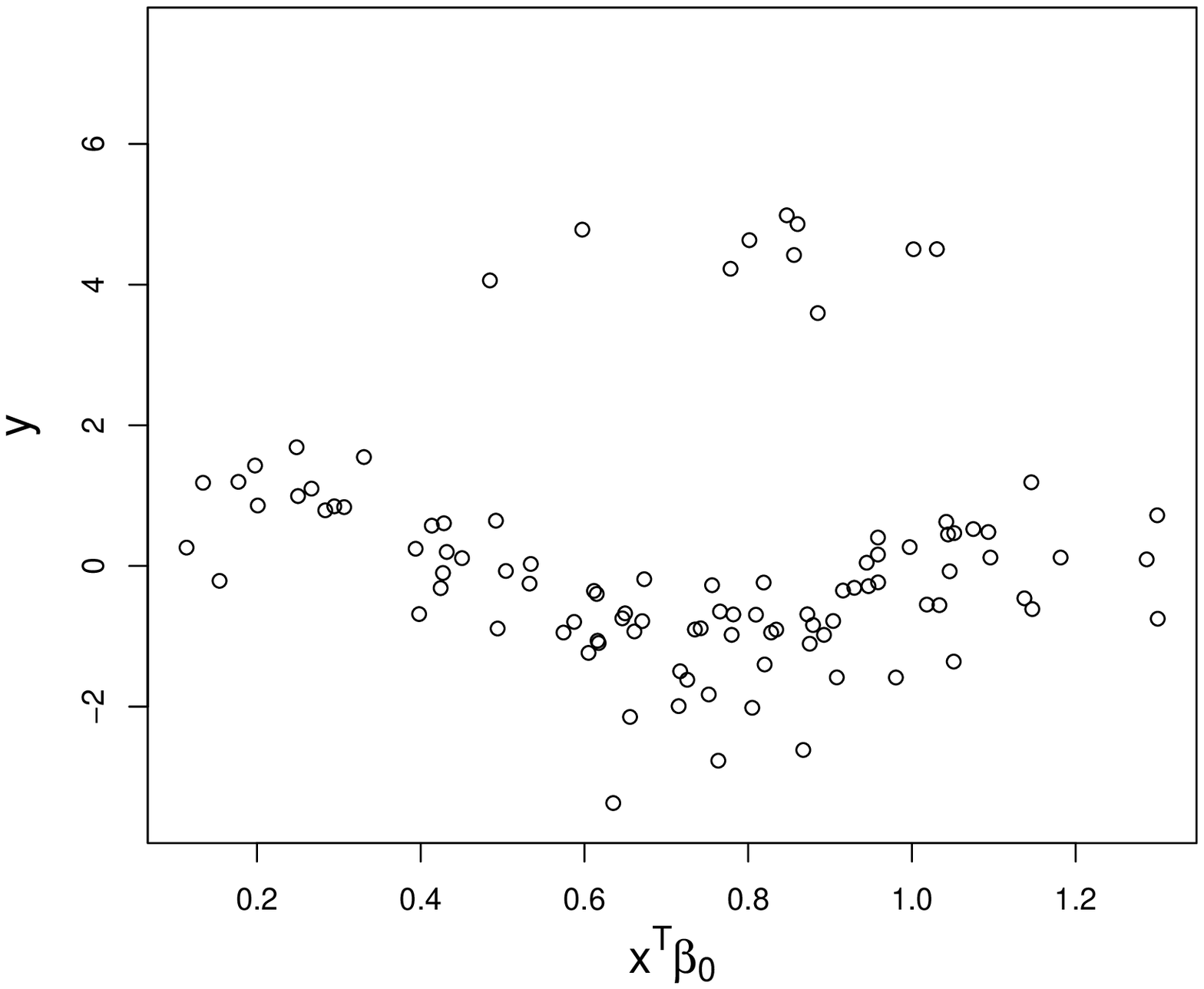} 
\includegraphics[scale=0.3]{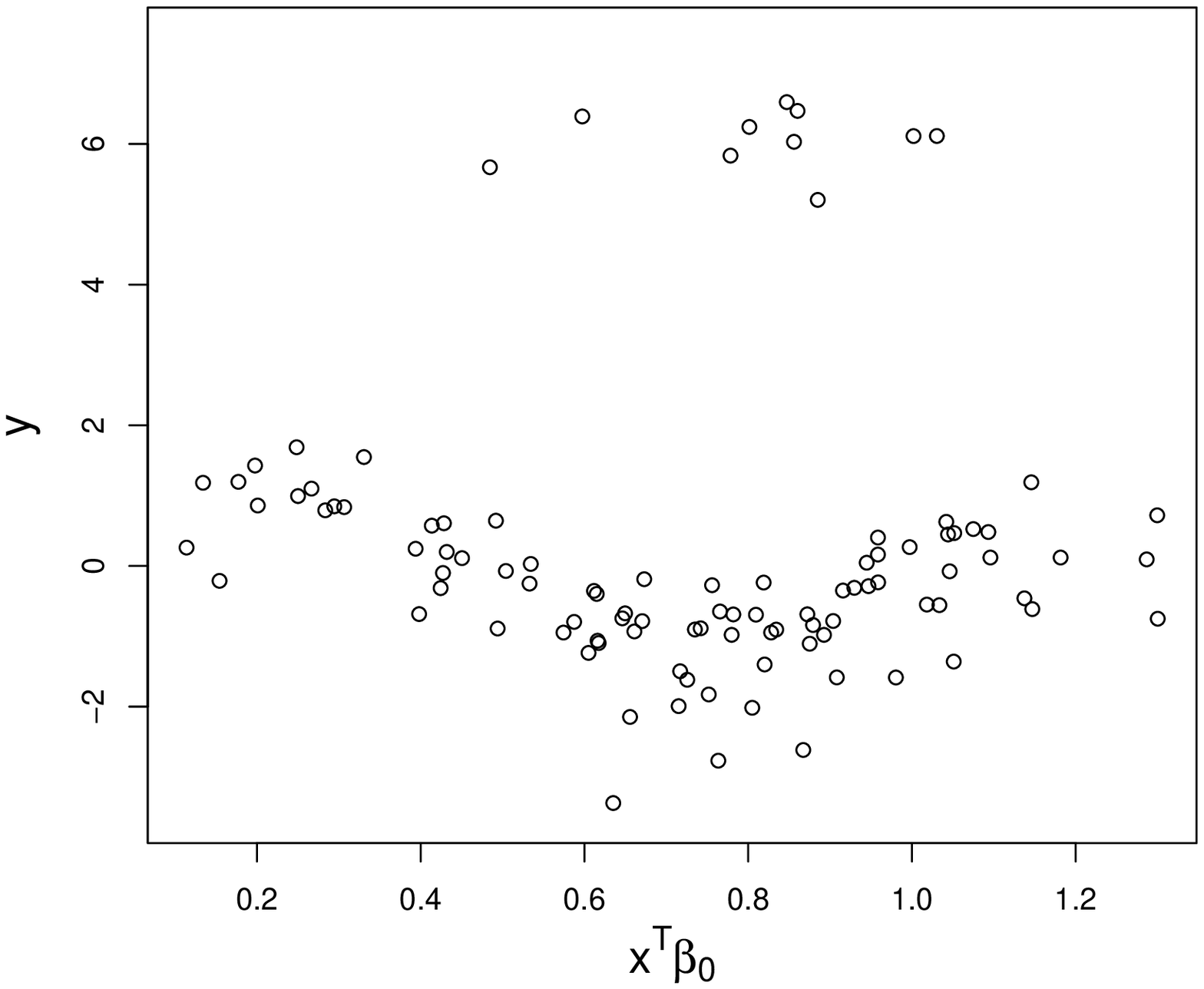} 
\includegraphics[scale=0.3]{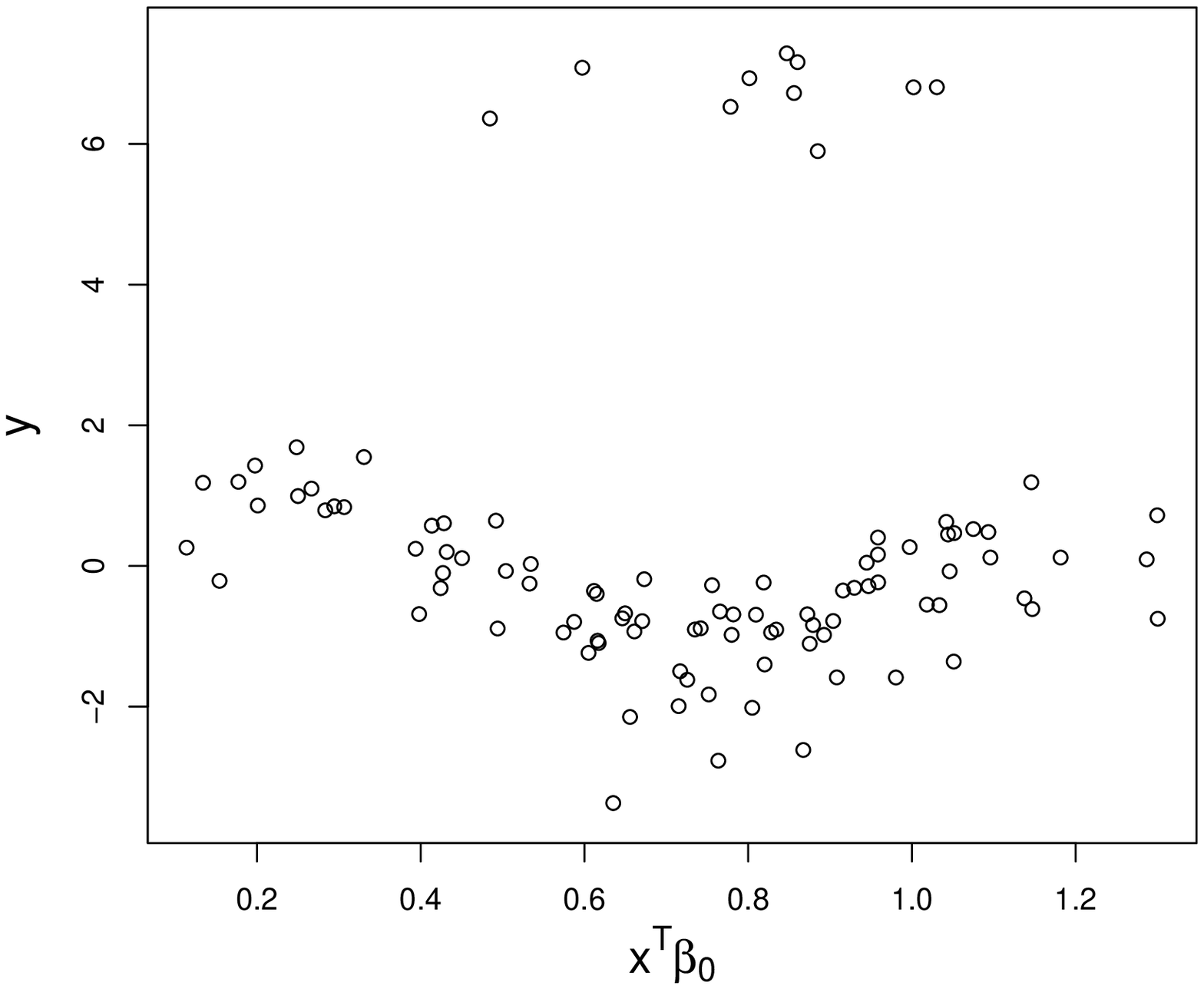} 
\caption{\label{fig:contaminaciones}\small  Generated sample when $\eta_0(u)=\sin(2 \pi u)$ and $\bbe_0=(1/\sqrt{2}, 1/\sqrt{2})\trasp$ under the central  and contaminated models.}
\end{center}
\end{figure}

The second  scheme accounts for more severe contaminations, labelled $S_1$, $S_2$ and $S_3$ in all Tables and Figures and we guess that its effect on the classical estimators would be more dramatic. To obtain the contaminated samples, the observations are generated as in (\ref{contam}) where now $y_i^{\star}=\log(k)+\epsilon_i$  where as above $\epsilon_i\sim \log(\Gamma(3,1))$ but $k=100,500$ or $1000$, respectively.
Figure \ref{fig:contaminaciones} illustrates the considered contaminations in a generated sample.

Table  \ref{tab:gamma}  summarizes the results  along the $N=1000$ replications. The reported results show the great stability of the robust procedure against  moderate and severe contaminations. As expected, when there is no contamination the classical estimators achieve the lowest square errors for both the parametric and nonparametric components. Nevertheless,  
the performance of the robust estimators is very satisfactory under $C_0$ since the loss of efficiency is very small.  
Focusing on the parametric component, under any of  the contaminated schemes, the performance of the classical estimator is very poor. Table \ref{tab:gamma} exhibits that the mean square error of the single index parameter increases more than forty times under the moderate contaminations and more than $200$ times under the severe ones, while the robust estimator remains very stable in all  considered scenarios.

\begin{table}[ht!]
\begin{center}
\renewcommand{\arraystretch}{1.1}
\begin{tabular}{c r r r r r r }
\hline
&\multicolumn{2}{c}{$\mbox{MSE}_{\wbbech}$}&\multicolumn{2}{c}{ $\mbox{MSE}_{\widehat{\eta}}$}&\multicolumn{2}{c}{$\mbox{MedSE}_{\widehat{\eta}}$}\\
& $\wbbe_{\cl}$ &  $\wbbe_{\rob}$ & $\weta_{\cl}$ &  $\weta_{\rob}$ & $\weta_{\cl}$ &  $\weta_{\rob}$\\
\hline
$C_0$ &  {0.005} &  {0.005} &  {0.041} &  {0.046} &  {0.020} &  {0.021}\\ 
\hline
$M_1$ & 0.209 & 0.008 & 0.294 &  0.061 & 0.164 &  0.031\\ 
$M_2$ & 0.357 & 0.007 & 0.408 &  0.060 & 0.226 &  0.030 \\ 
$M_3$ & 0.534 & 0.007 & 0.521 &  0.058 & 0.287 &  0.028 \\ 
\hline
$S_1$ & 1.064 & 0.013 & 5.393 &  0.059 & 4.510 &  0.024 \\ 
$S_2$ & 1.098 & 0.008 & 13.282&  0.057 & 13.282&  0.023 \\ 
$S_3$ & 1.106 & 0.006 & 18.057&  0.053 & 17.436&  0.022\\
\hline
\end{tabular}
\end{center}
\caption{\small\label{tab:gamma} Mean square errors for the estimators of $\bbe_0$, Mean over replications $MSE(\widehat{\eta})$ and Median over replications of $\mbox{median}_{i=1:n} (\eta_0(\bx_i\trasp\bbe_0)-\weta(\bx_i\trasp\wbbe))^2$.} 
\end{table}

\begin{figure}[H]
\begin{center}
\small 
Classical Method \hspace{4cm} Robust Method\\
\includegraphics[scale=0.45]{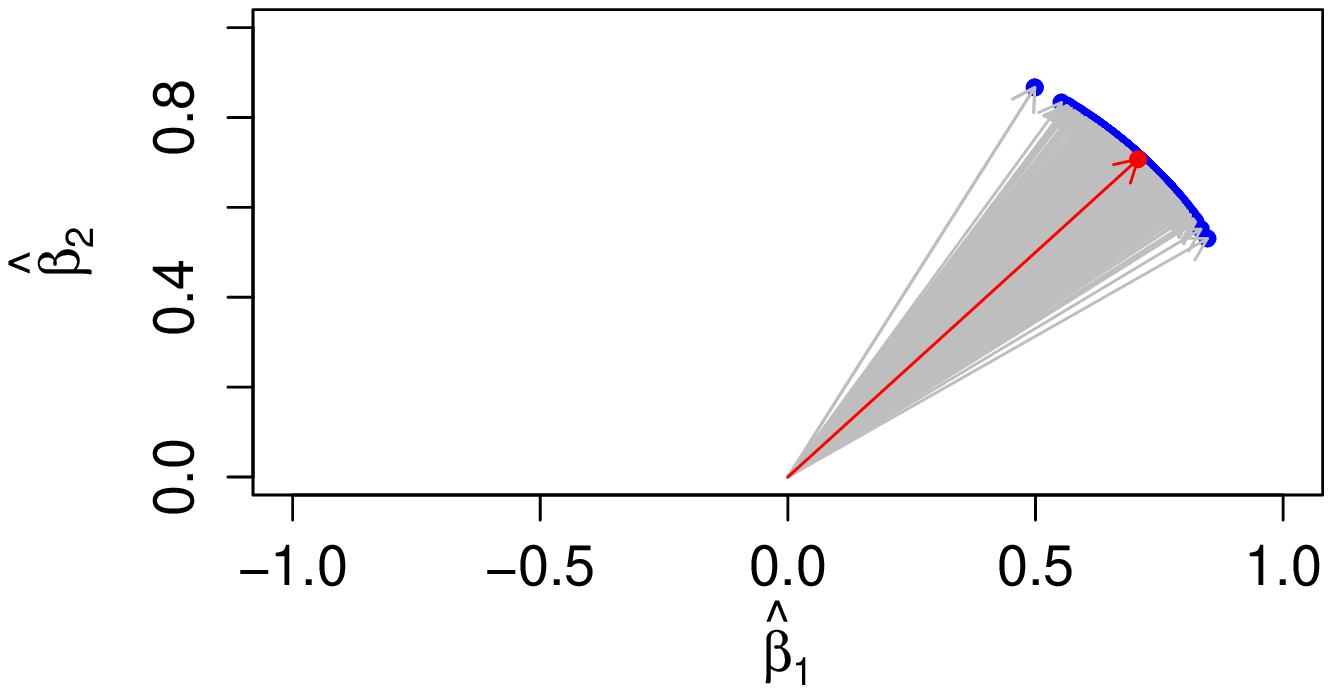} 
\includegraphics[scale=0.45]{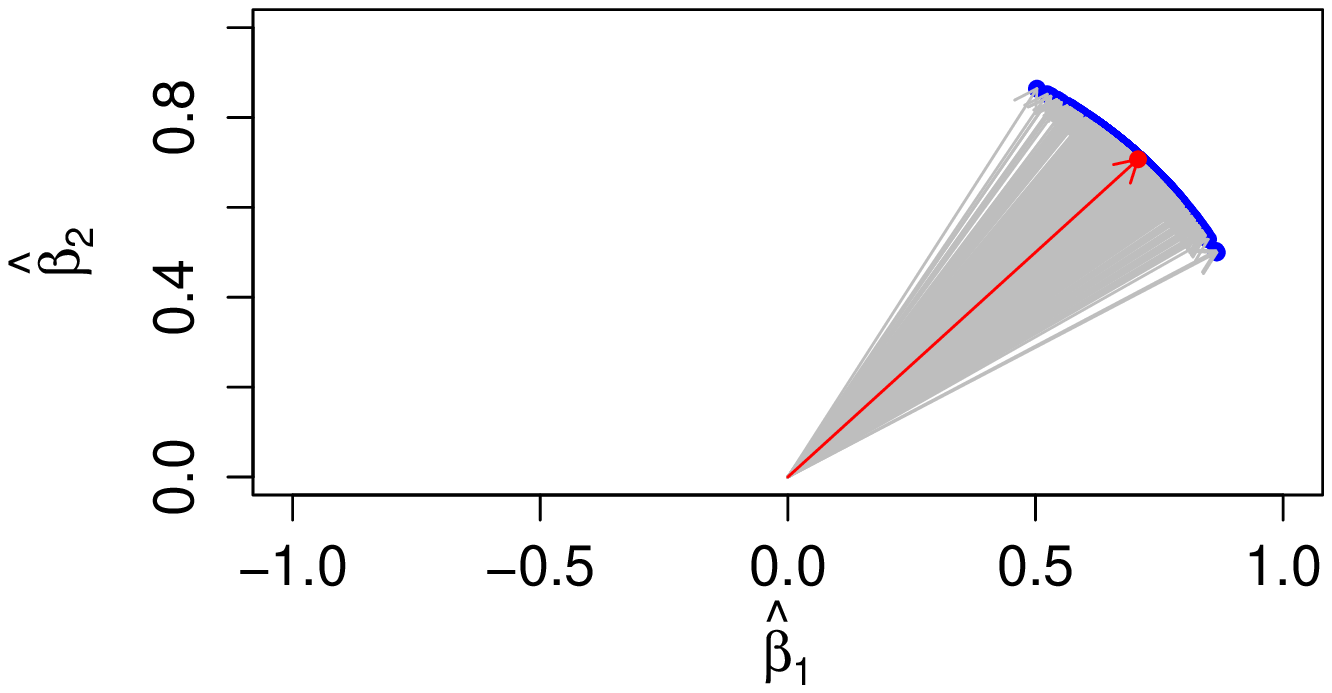} 
\vskip-0.1in
\caption{\small \label{fig:flecha_c0}\small{ Classical and robust estimators of the single index parameter under $C_0$. The red arrow  represents the true direction $\bbe_0=(1/\sqrt{2}, 1/\sqrt{2})\trasp$}, while the grey arrows are the estimates.}
\end{center}
\end{figure}

\begin{sidewaysfigure}
\centering
\small 
Classical Method\\
$\quad$\\
$M_1$ \hspace{6cm} $M_2$ \hspace{6cm} $M_3$\\
\hskip-0.2in\includegraphics[scale=0.53]{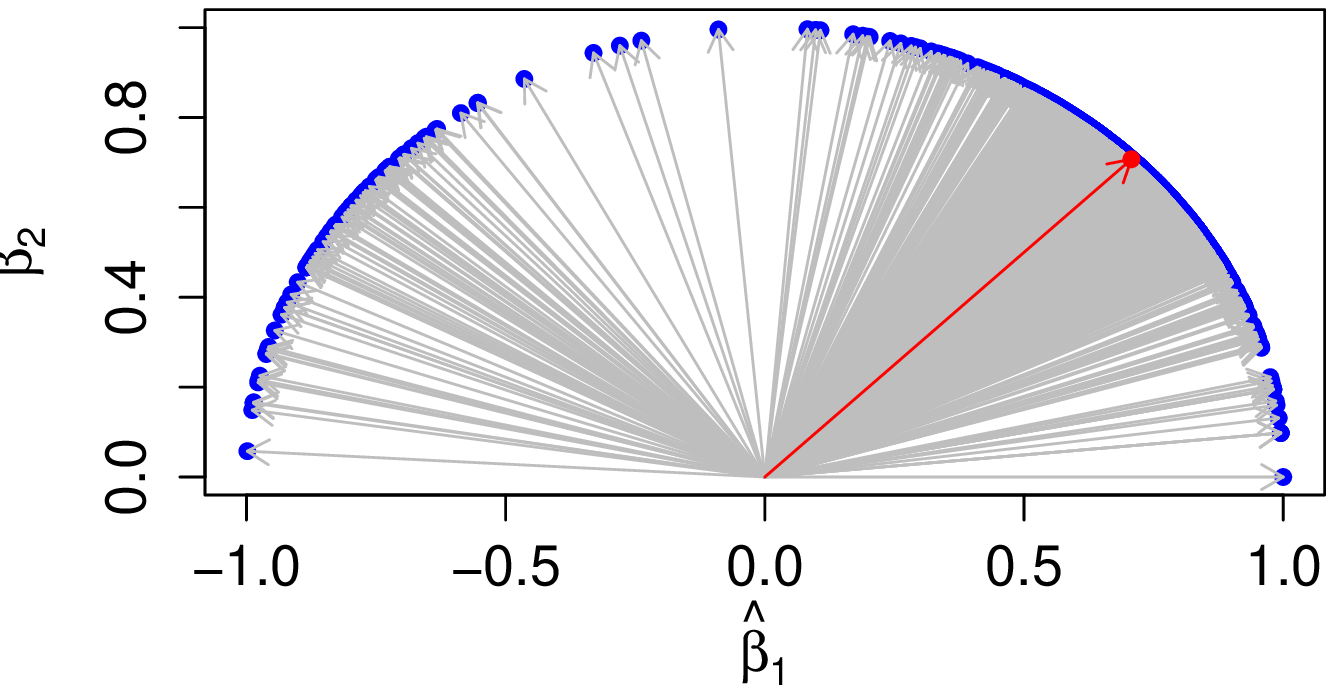} 
\hskip-0.2in\includegraphics[scale=0.53]{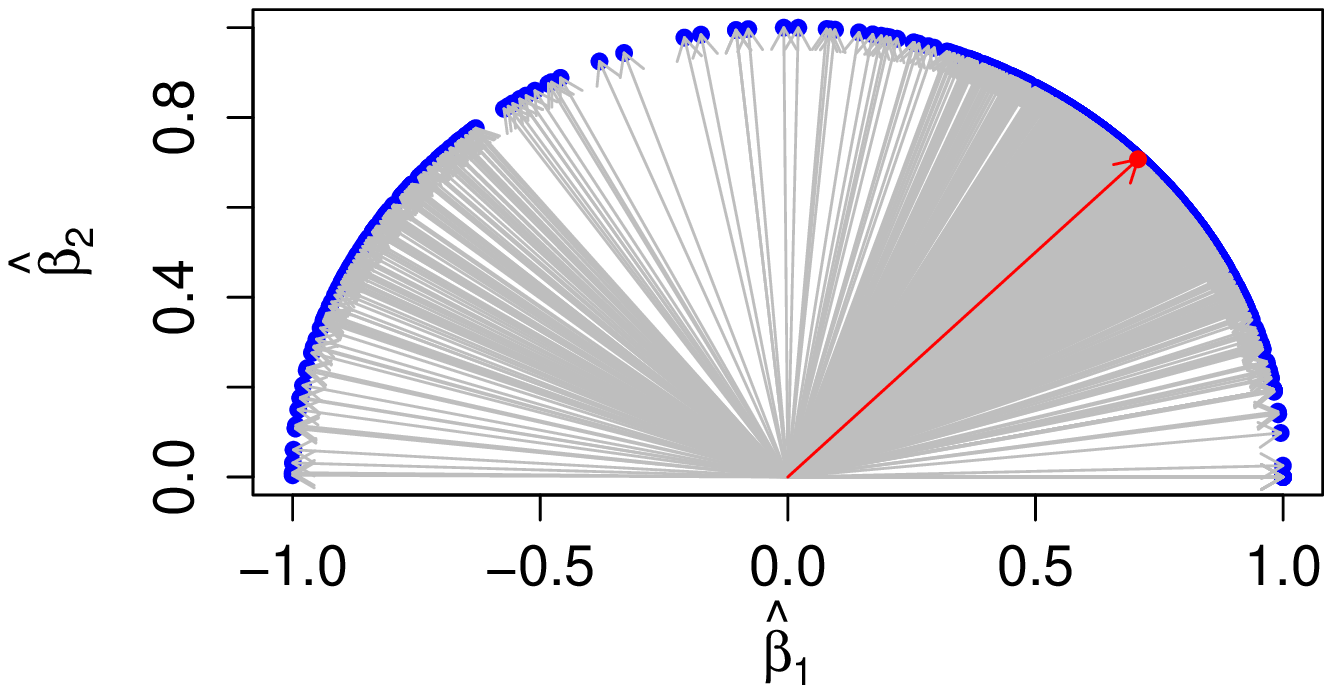}  
\hskip-0.2in\includegraphics[scale=0.53]{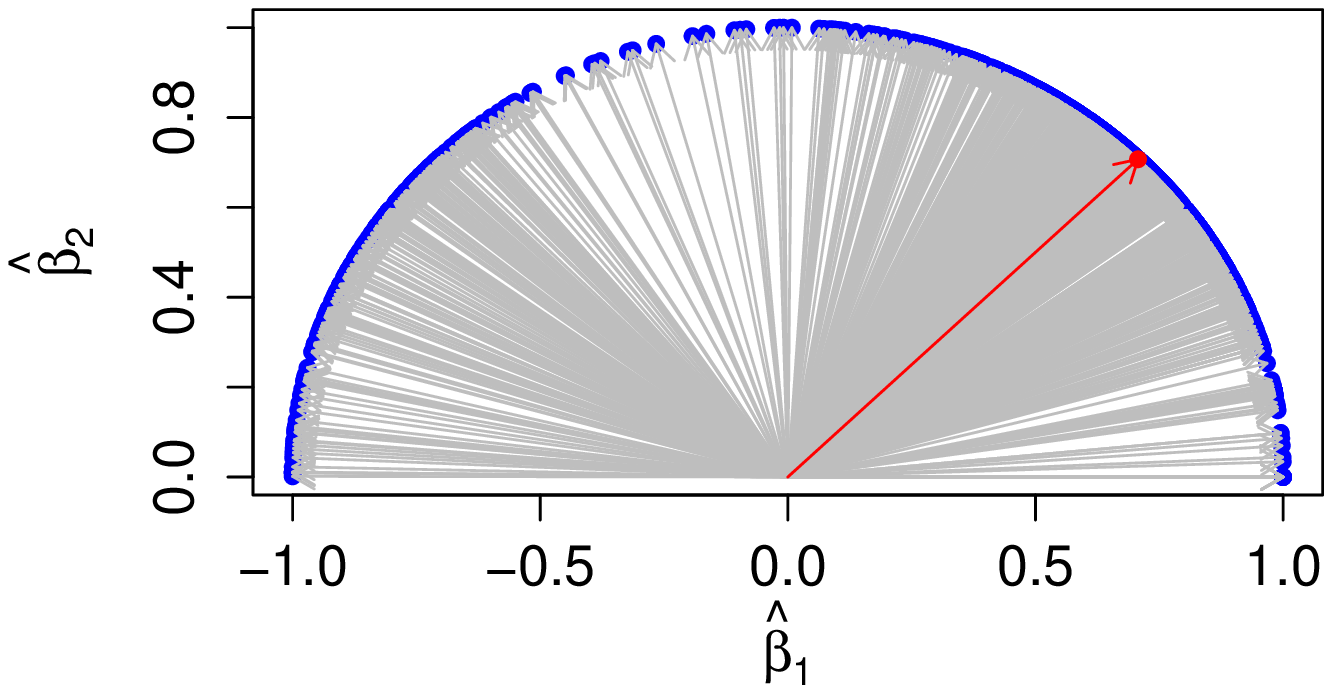}\\
Robust Method\\
\hskip-0.2in\includegraphics[scale=0.53]{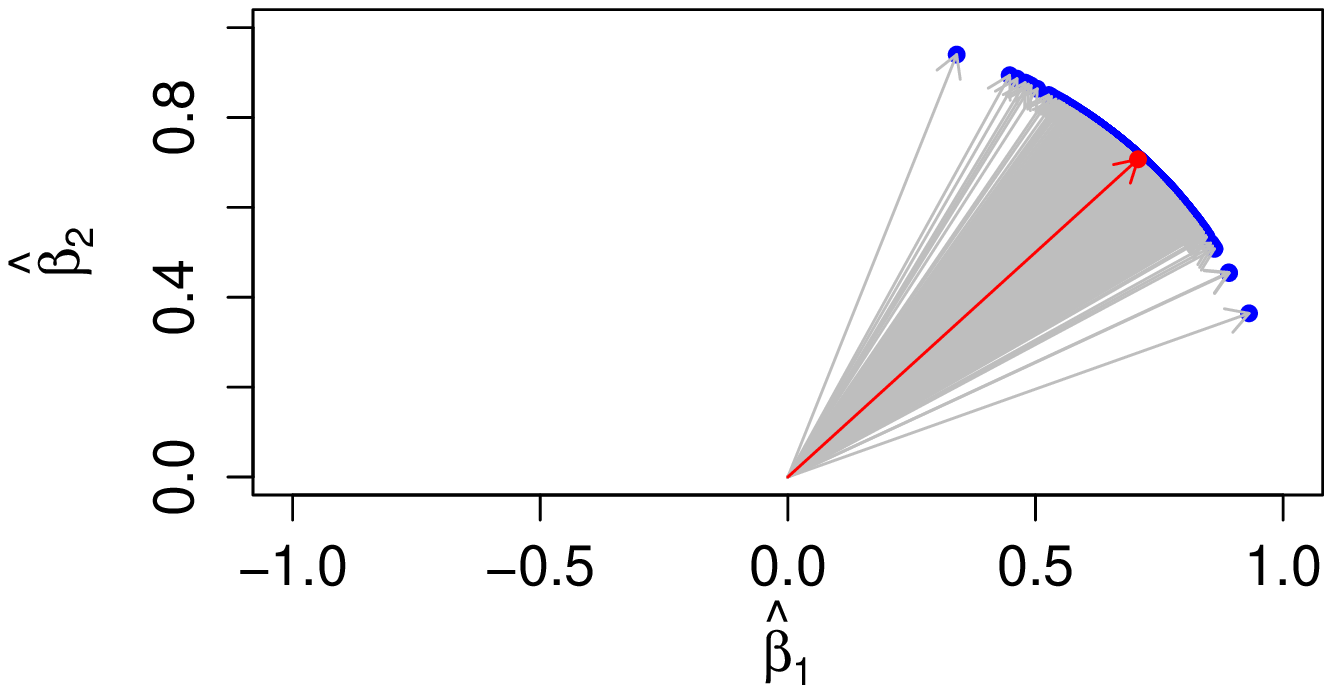} 
\hskip-0.2in\includegraphics[scale=0.53]{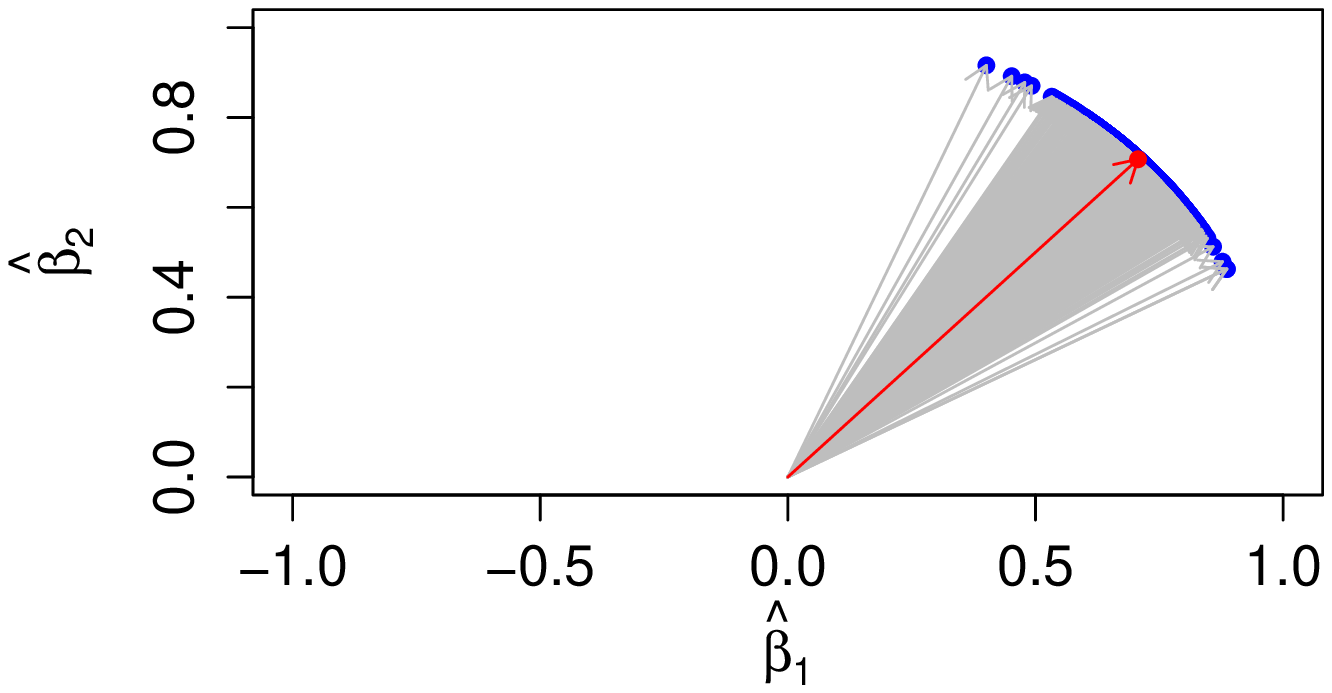}  
\hskip-0.2in\includegraphics[scale=0.53]{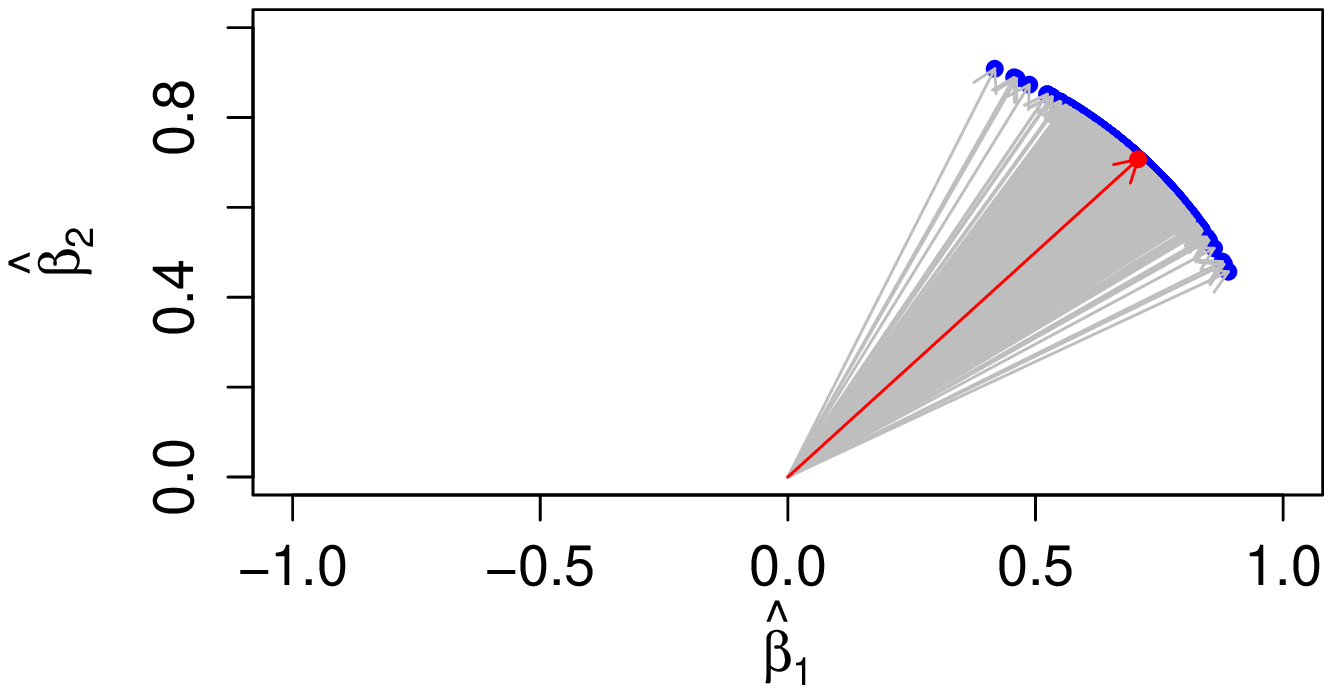}\\
\vskip-0.1in\caption{\small \label{fig:flecha_M}\small{ Classical and robust estimators of the single index parameter under $M_1$, $M_2$ and $M_3$. The red arrow  represents the true direction $\bbe_0=(1/\sqrt{2}, 1/\sqrt{2})\trasp$}, while the grey arrows are the estimates.}
\end{sidewaysfigure}

\begin{sidewaysfigure}
\centering
\small 
Classical Method\\
$\quad$\\
$S_1$ \hspace{6cm} $S_2$ \hspace{6cm} $S_3$\\
\hskip-0.2in\includegraphics[scale=0.53]{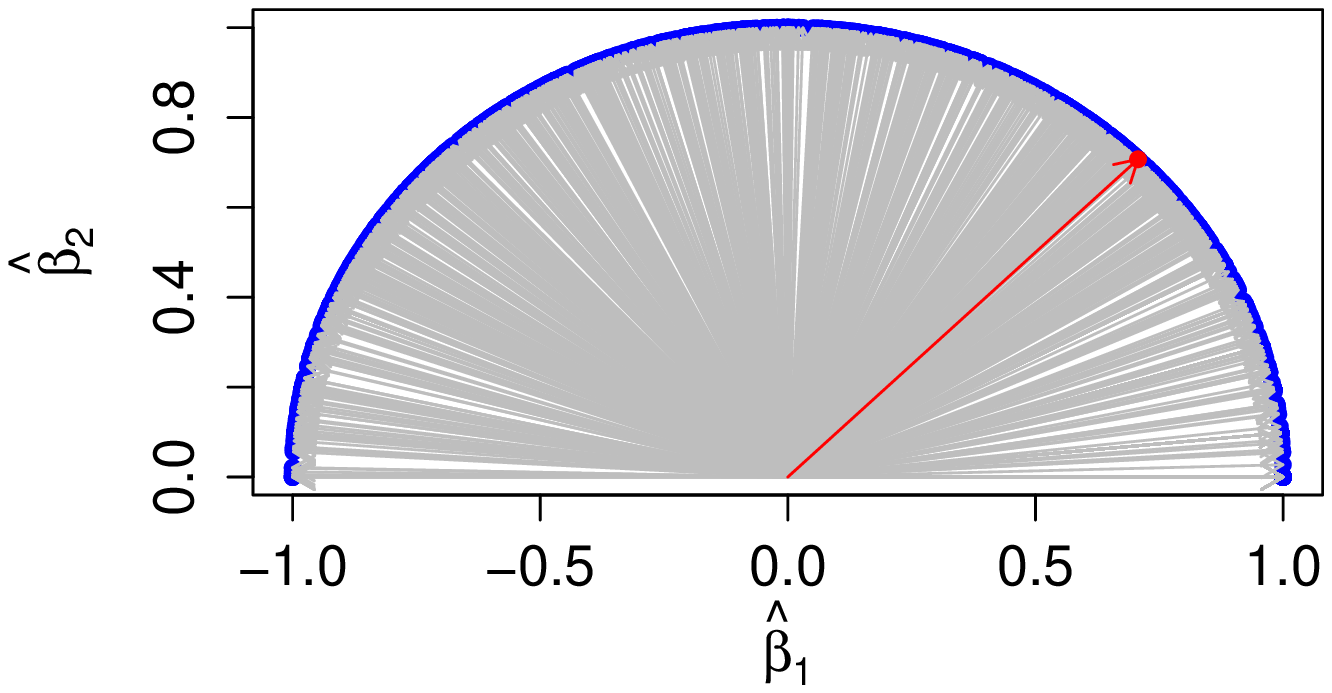} 
\hskip-0.2in\includegraphics[scale=0.53]{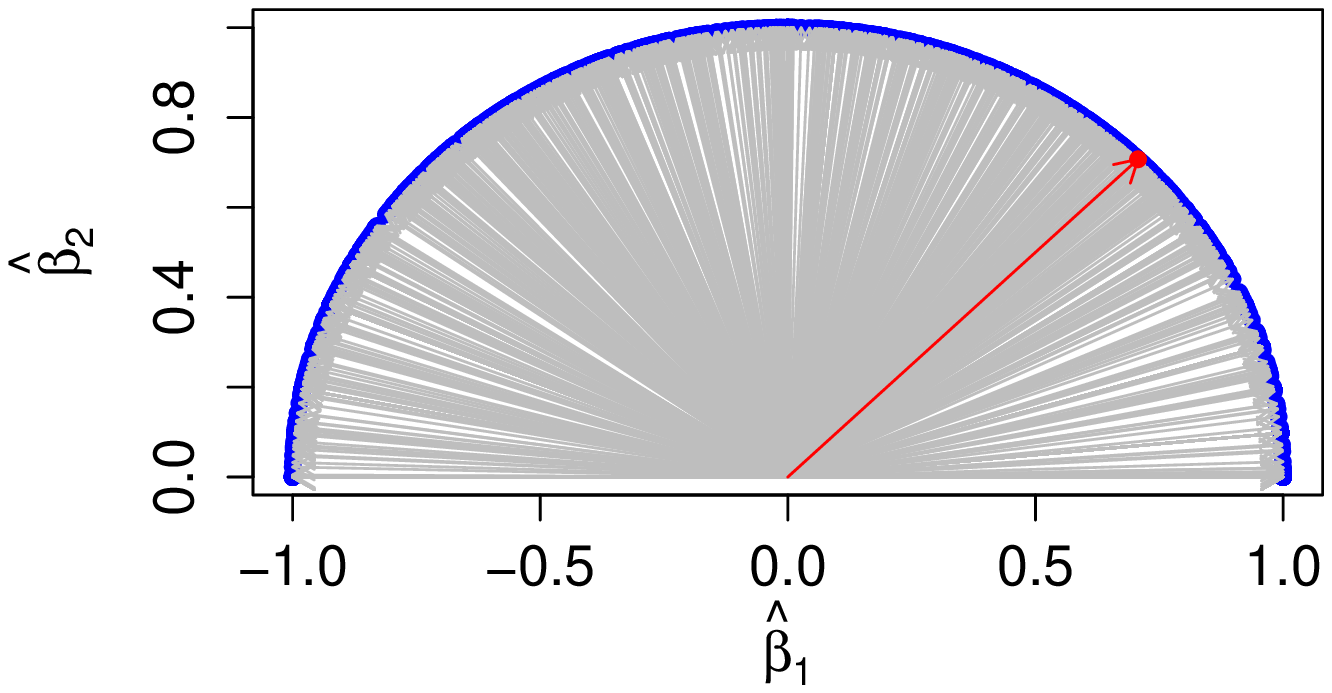}  
\hskip-0.2in\includegraphics[scale=0.53]{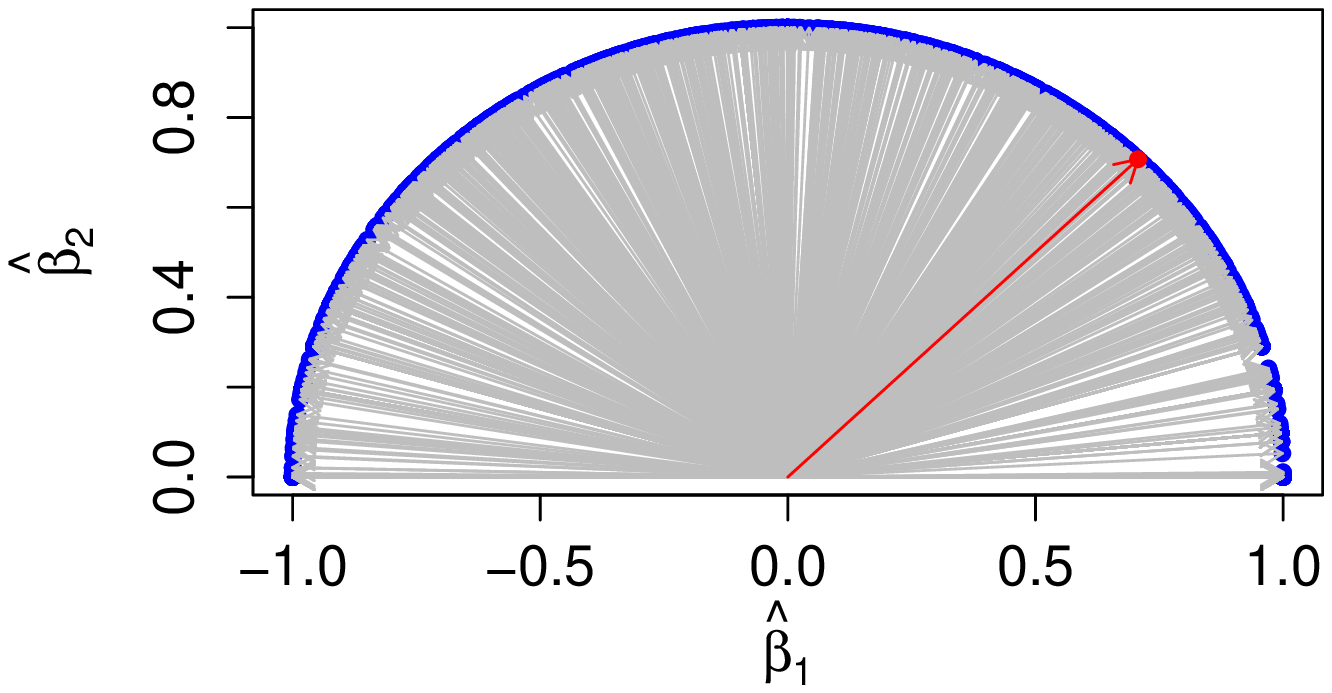}\\
Robust Method\\
\hskip-0.2in\includegraphics[scale=0.53]{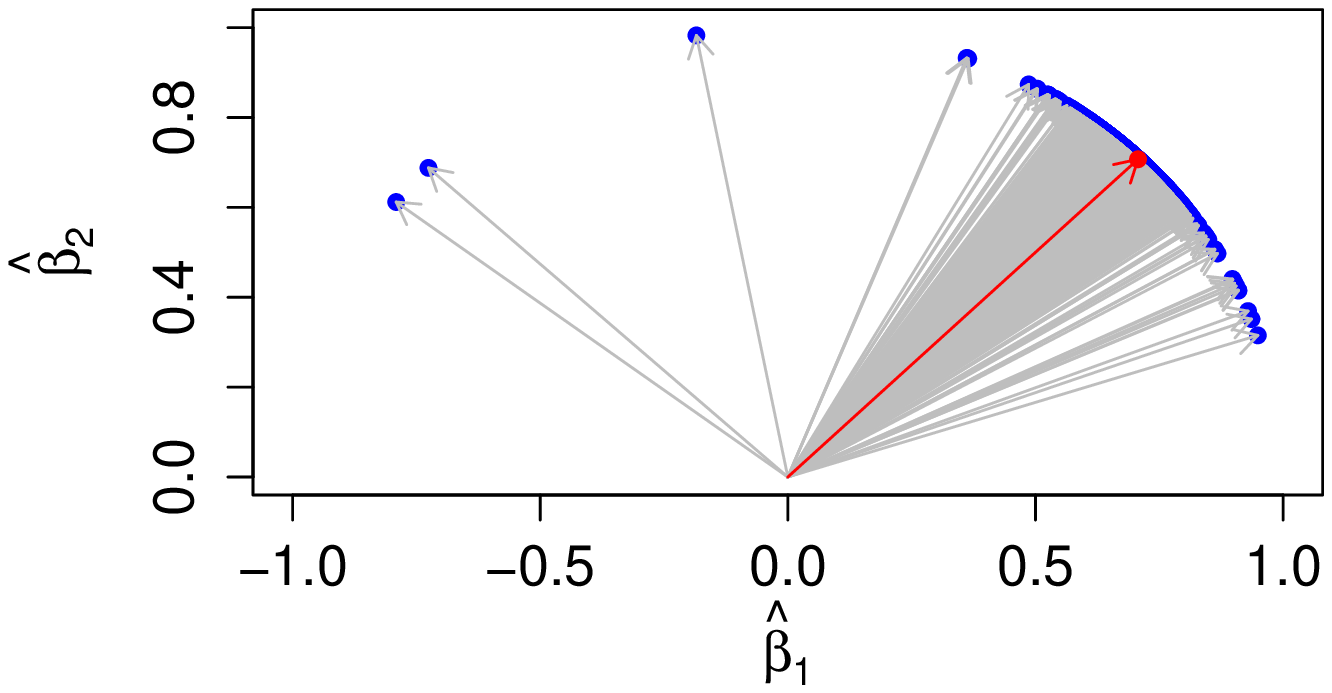}
\hskip-0.2in \includegraphics[scale=0.53]{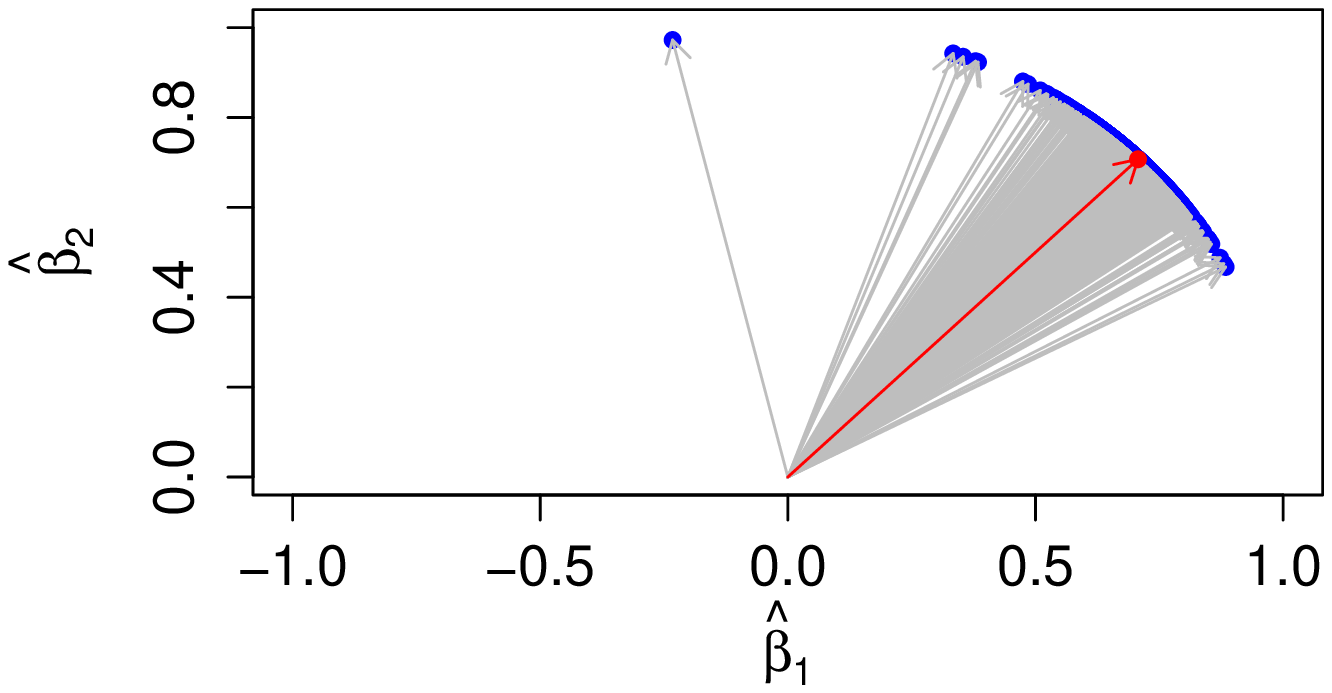}  
\hskip-0.2in \includegraphics[scale=0.53]{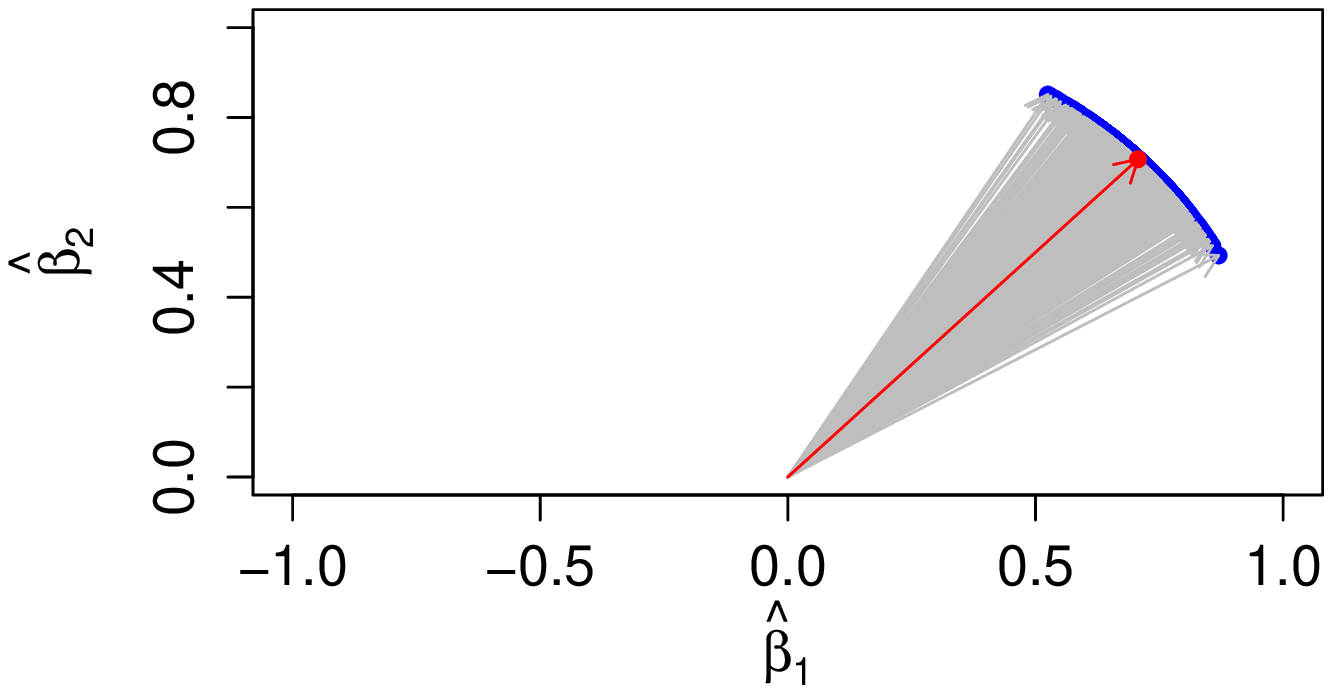}\\
\vskip-0.1in\caption{\small \label{fig:flecha_S}\small{ Classical and robust estimators of the single index parameter under $S_1$, $S_2$ and $S_3$. The red arrow  represents the true direction $\bbe_0=(1/\sqrt{2}, 1/\sqrt{2})\trasp$}, while the grey arrows are the estimates.}
\end{sidewaysfigure}

Since an important goal in this framework is to  capture the direction of the single index parameter $\bbe_0$, instead of presenting the traditional  boxplots of the estimates, in Figures \ref{fig:flecha_c0} to \ref{fig:flecha_S}  we present a two dimensional graph  that reflects the skill of the classical and robust estimators  to get the true  direction $\bbe_0$, for the clean and contaminated samples.
In these plots, the red arrow  represents the true direction $\bbe_0=(1/\sqrt{2}, 1/\sqrt{2})\trasp$, that corresponds to an angle $\theta_0=\pi/4$ and the grey ones to the estimated directions. 
These figures show that under $C_0$ the performance of the robust estimator of the parametric component is similar to that of the classical estimator since the robust estimates are more or less spread as the classical estimator around the target direction. It also becomes evident that in contaminated samples  the robust estimator of the parametric component is very stable  under all the contaminated scenarios,  while the classical estimator is completely spoiled. Indeed,  under $M_1$ to $M_3$, the classical estimates of the single index parameter tend to be concentrated not only on directions close  to the true value $\bbe_0$ but also   to its orthogonal direction $\bbe_0^{\bot}$, showing the impact of the contaminated points.  On the other hand, under the severe contaminations $S_1$ to $S_3$  the classical estimates cover almost all possible directions in the first and second quadrants, becoming completely unreliable. 

Regarding the estimation of the nonparametric component, Table \ref{tab:gamma} shows the large effect of the considered contaminations on the classical estimator of the nonparametric component, where the mean square error increases at least seven times under the moderate contaminations.  Under the severe contaminations $S_1$ to $S_3$, the  effect of the outliers on the classical estimator is devastating, while it is quite harmless for the robust estimator. It is worth noticing that under all the contamination schemes, the reported values of $\mbox{MedSE}_{\eta}$ for the classical estimator, which is a  more resistant measure based on the median, are very close to the corresponding values of $\mbox{MSE}_{\eta}$, making evident that in most replications the classical estimator of the nonparametric component is completely spoiled.

\begin{figure}[ht!]
\begin{center}
\small 
Classical Method \hspace{4.5cm} Robust Method\\
\vskip-0.1in
\includegraphics[scale=0.4]{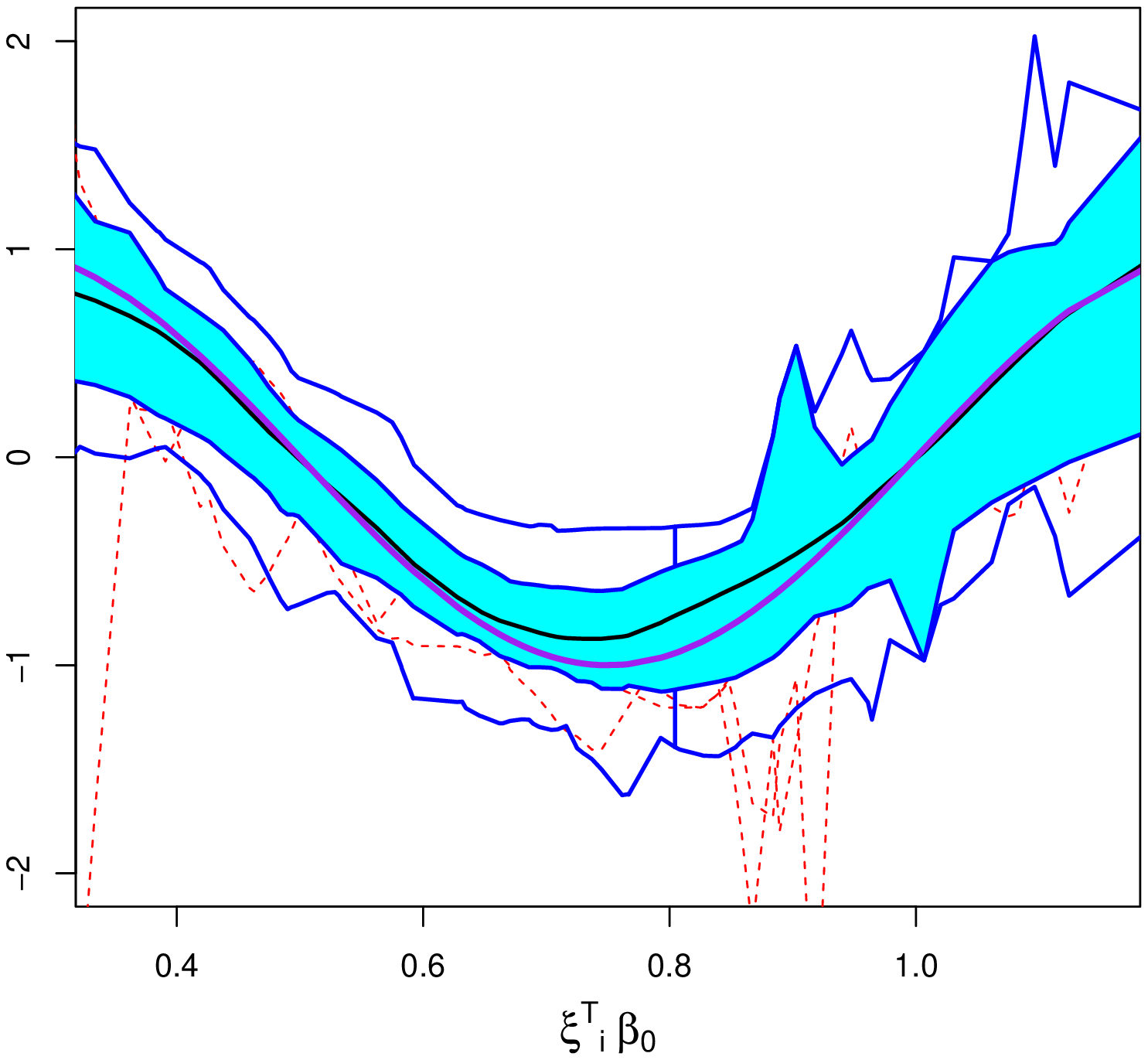} 
\includegraphics[scale=0.4]{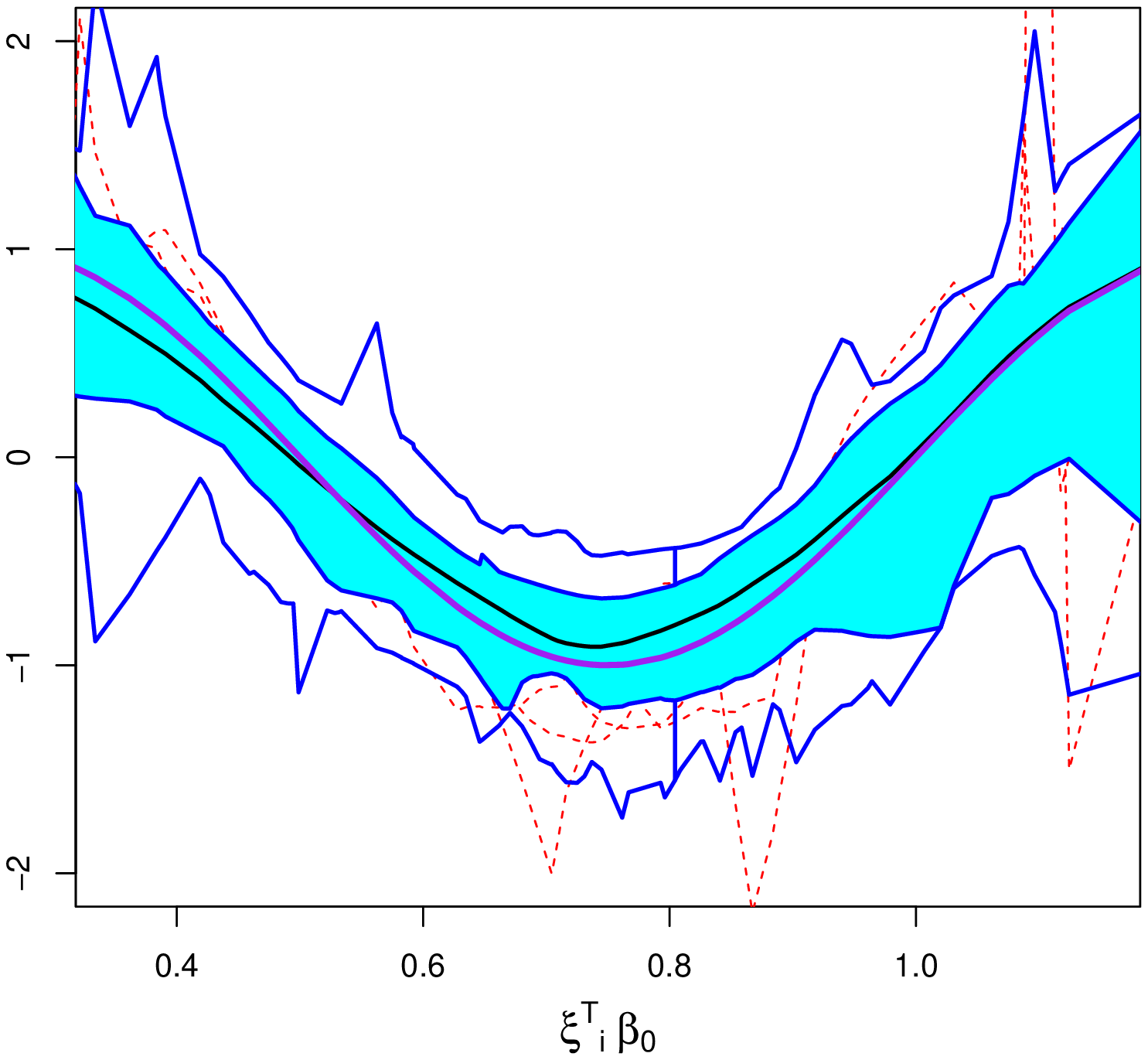} 
\vskip-0.1in
\caption{\small \label{fig:fda_c0}\small{Classical and robust estimators of $\eta_0$ under $C_0$.}}
\end{center}
\end{figure}

\begin{figure}[ht!]
\begin{center}
\small 
Classical Method\\
$\quad$\\
$M_1$ \hspace{4.5cm} $M_2$ \hspace{4.5cm} $M_3$\\
\vskip-0.1in
\hskip-0.2in\includegraphics[scale=0.3]{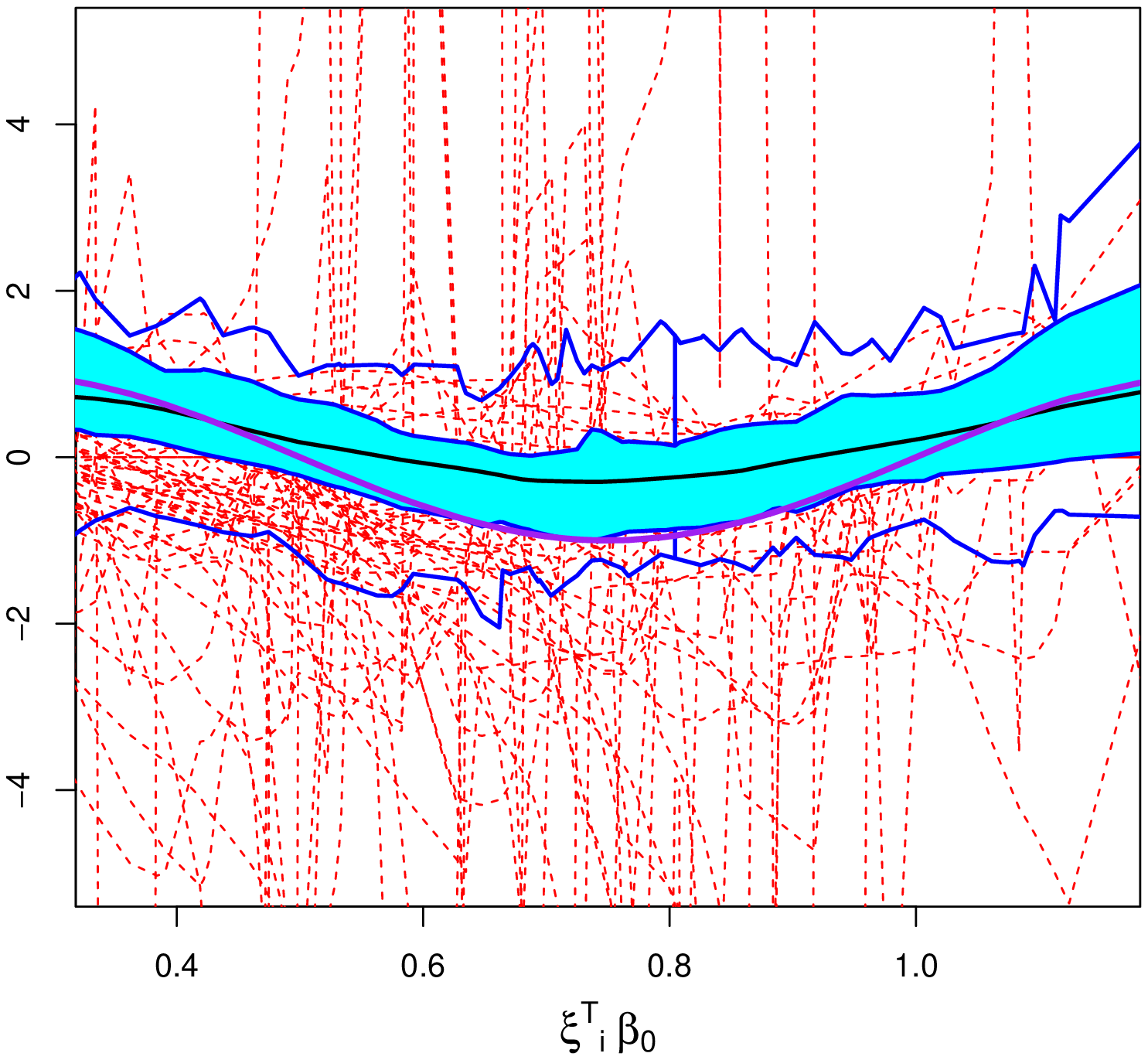} 
\hskip-0.2in\includegraphics[scale=0.3]{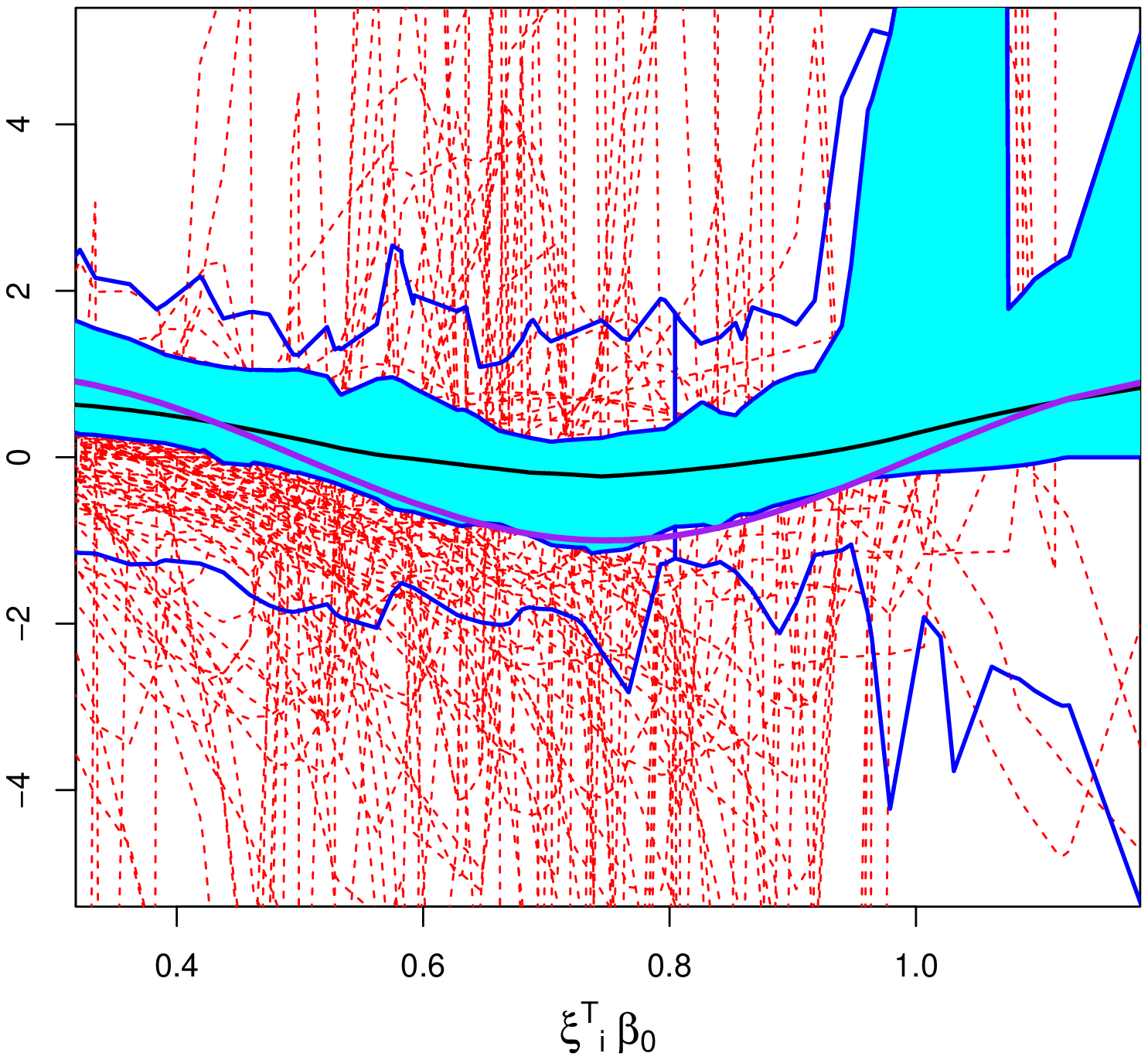}  
\hskip-0.2in\includegraphics[scale=0.3]{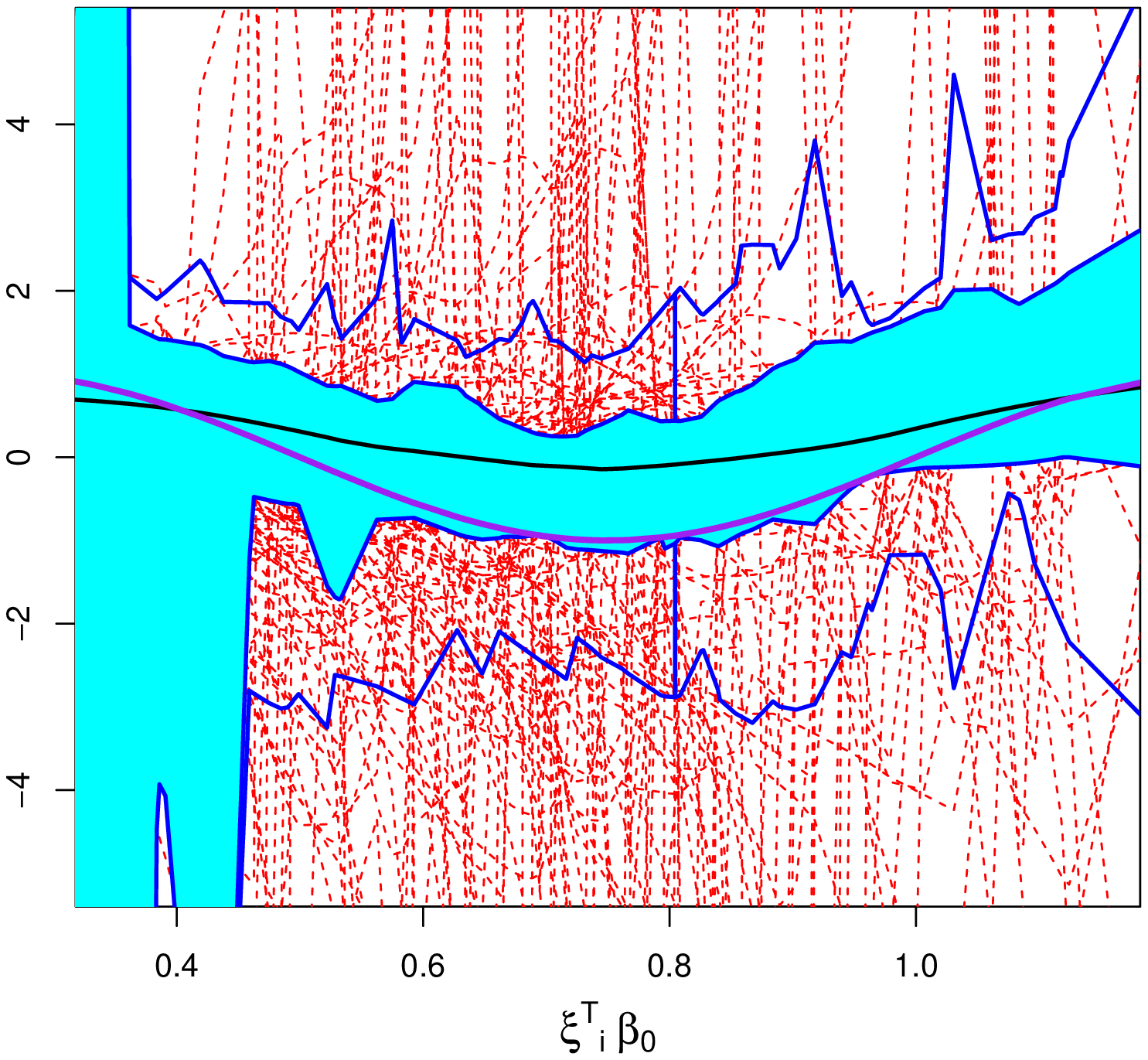}\\
Robust Method\\
\vskip-0.1in
\hskip-0.2in\includegraphics[scale=0.3]{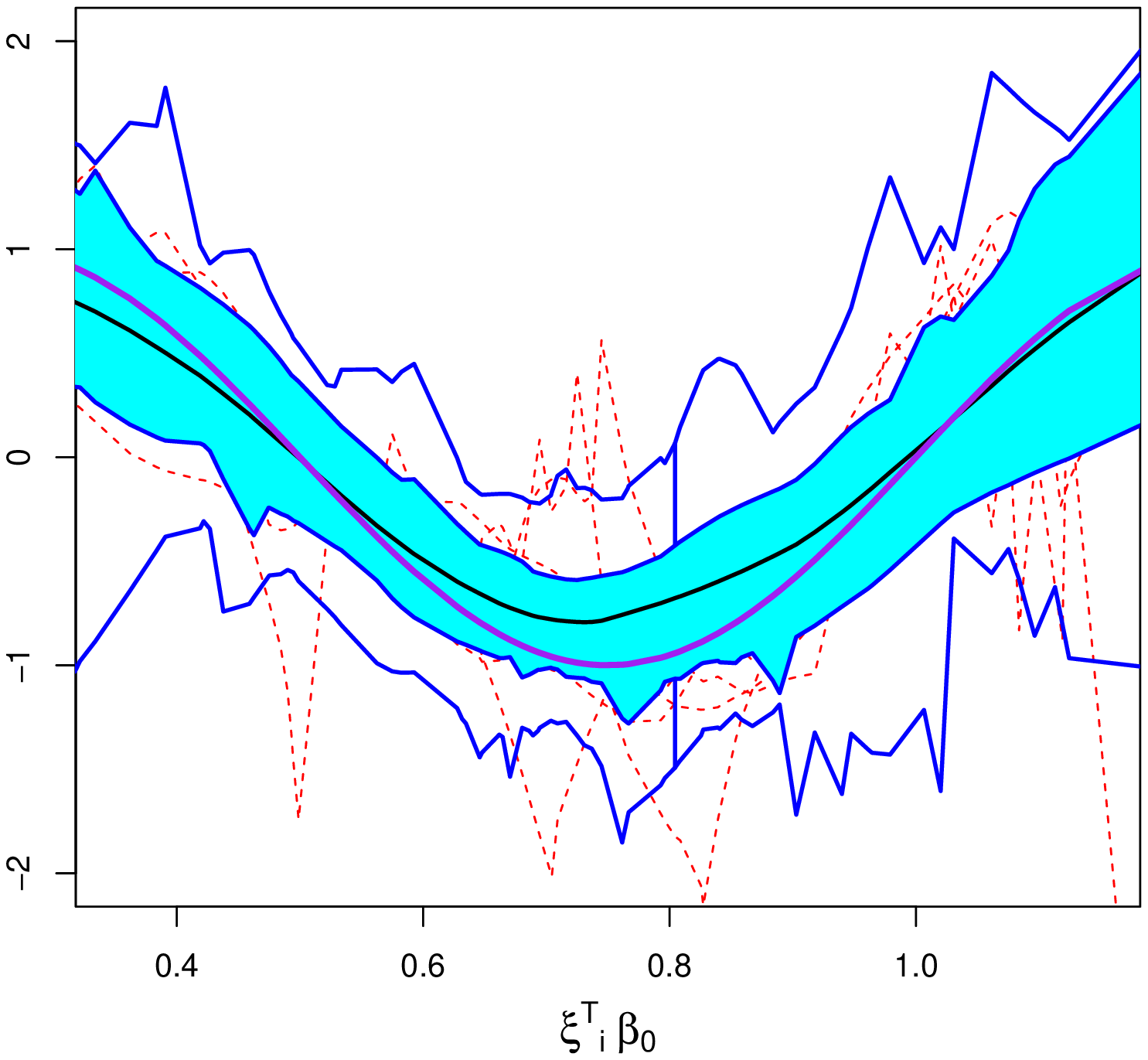} 
\hskip-0.2in\includegraphics[scale=0.3]{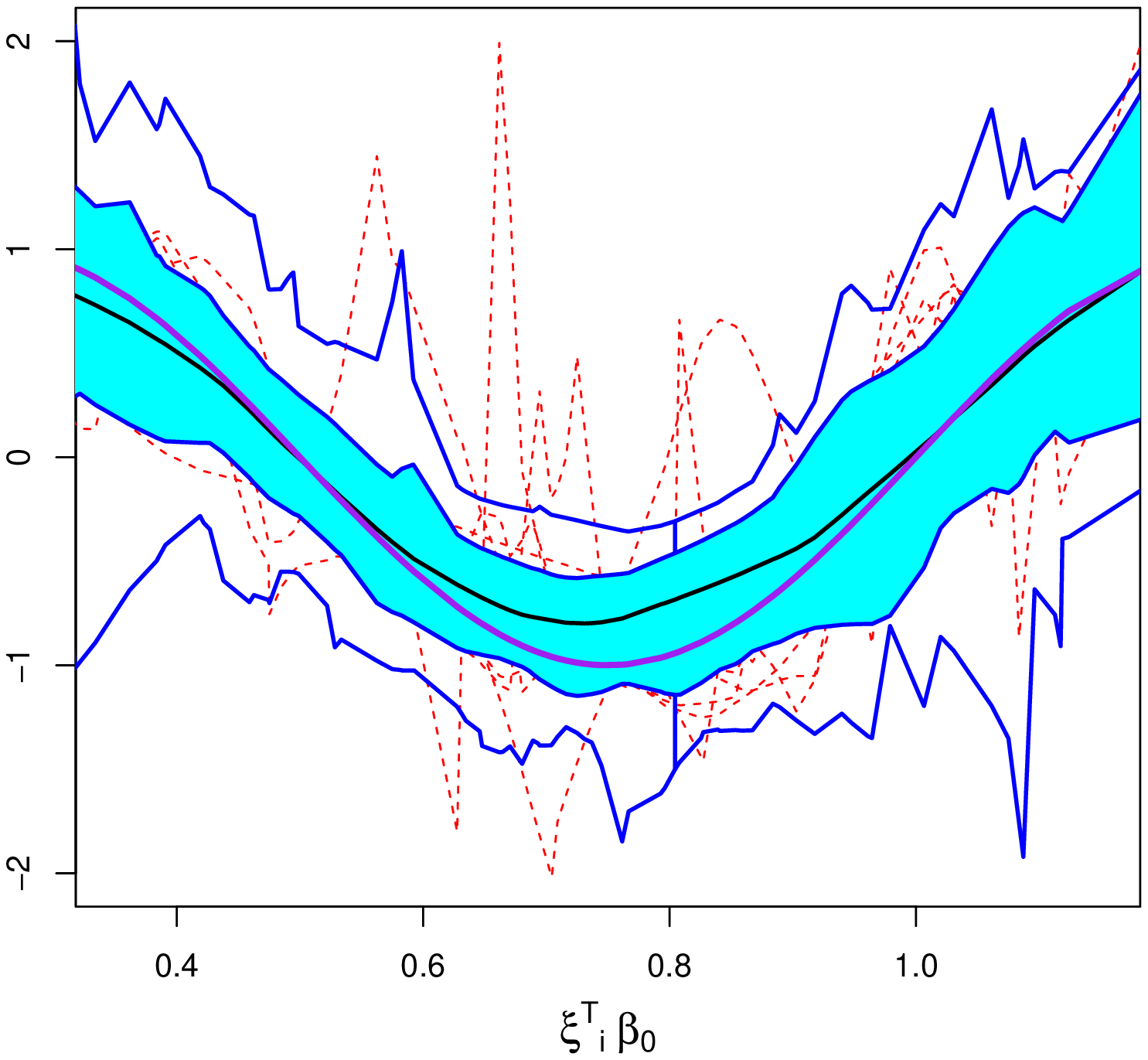}  
\hskip-0.2in\includegraphics[scale=0.3]{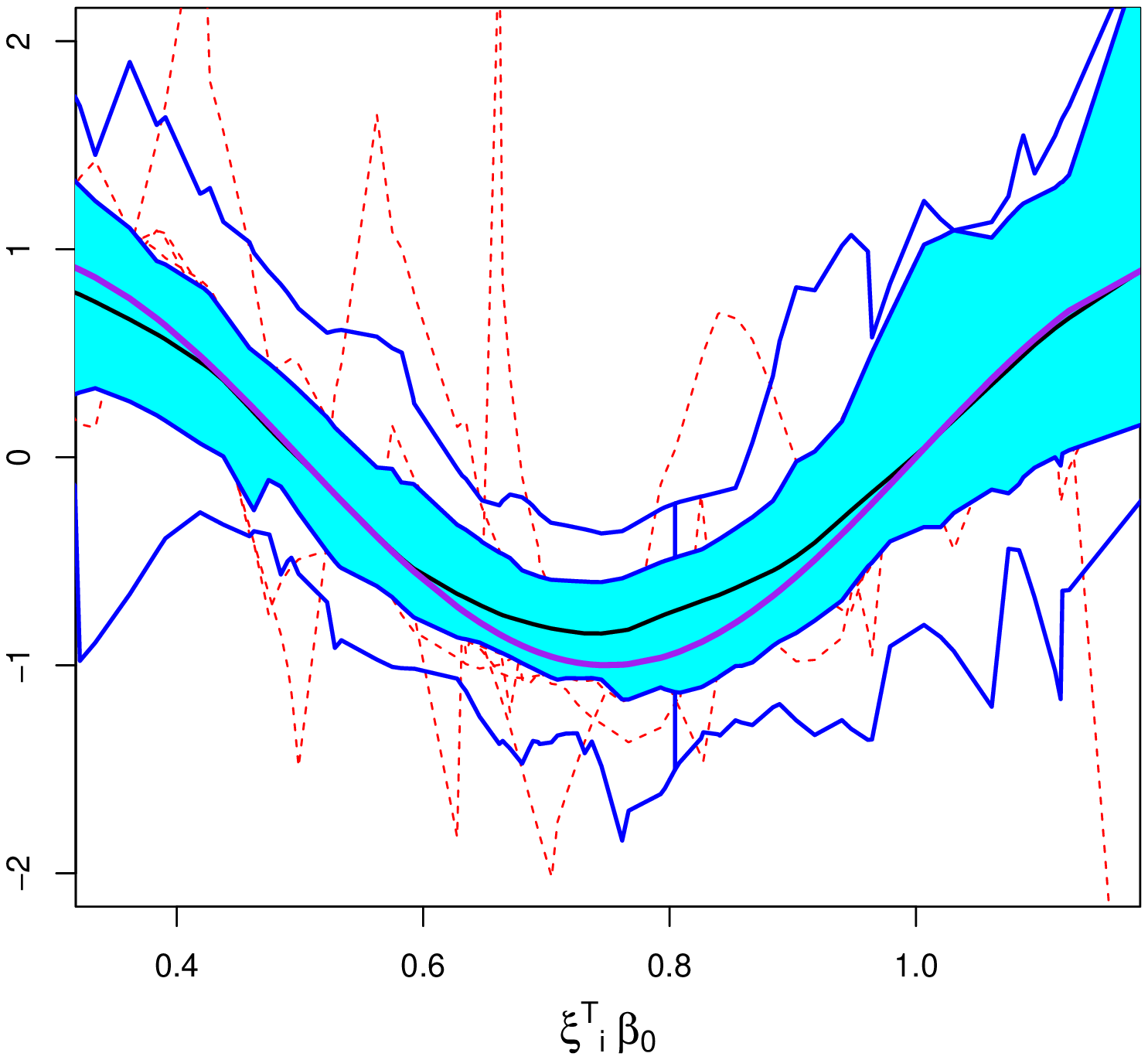}\\
\vskip-0.1in
\caption{\small\label{fig:fda_M}\small{ Classical and robust estimators of $\eta_0$ under  $M_1$, $M_2$ and $M_3$.}}
\end{center}
\end{figure}

\begin{figure}[ht!]
\begin{center}
\small 
Classical Method\\
$\quad$\\
$S_1$ \hspace{4.5cm} $S_2$ \hspace{4.5cm} $S_3$\\
\vskip-0.1in
\hskip-0.2in\includegraphics[scale=0.3]{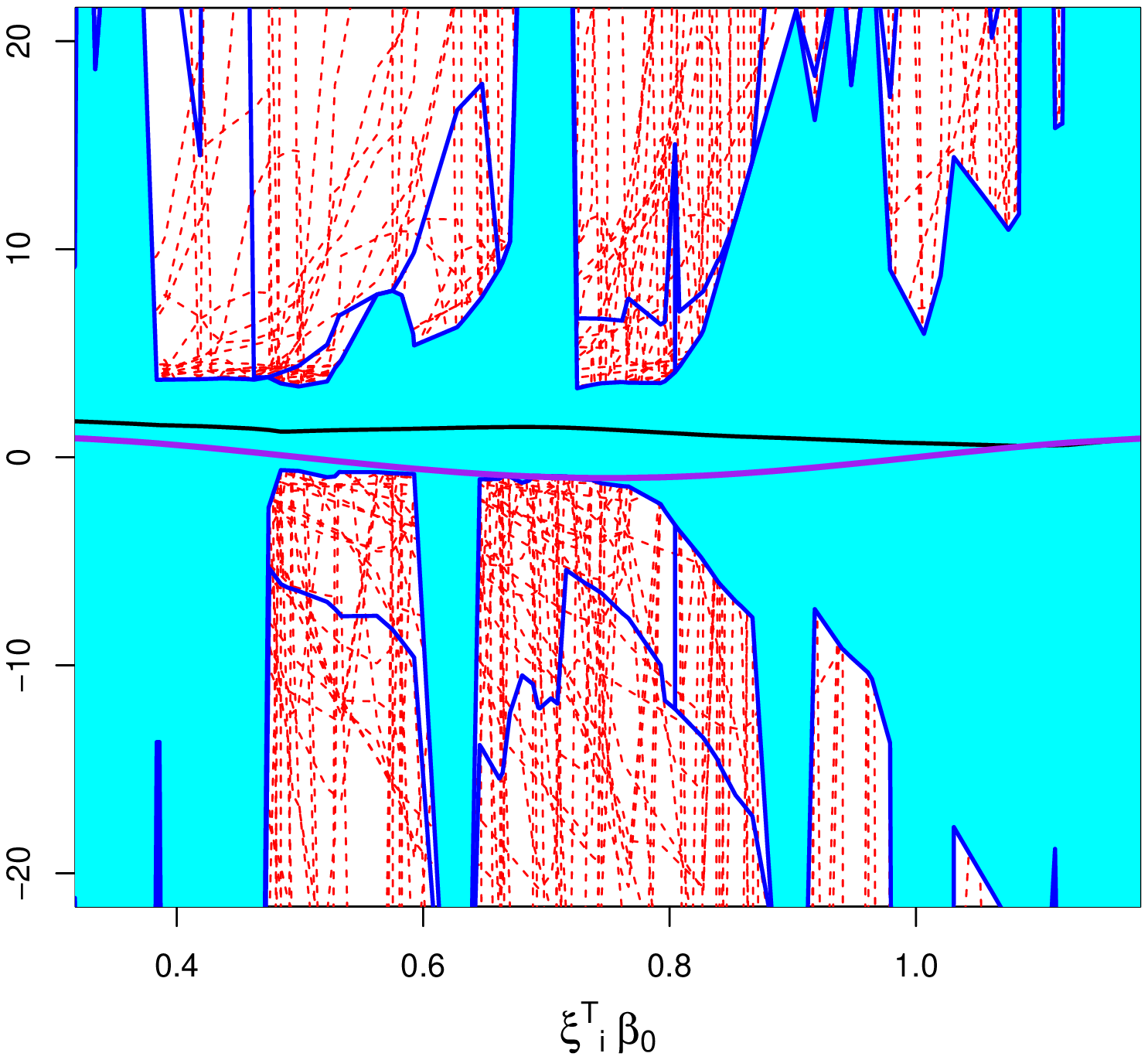} 
\hskip-0.2in\includegraphics[scale=0.3]{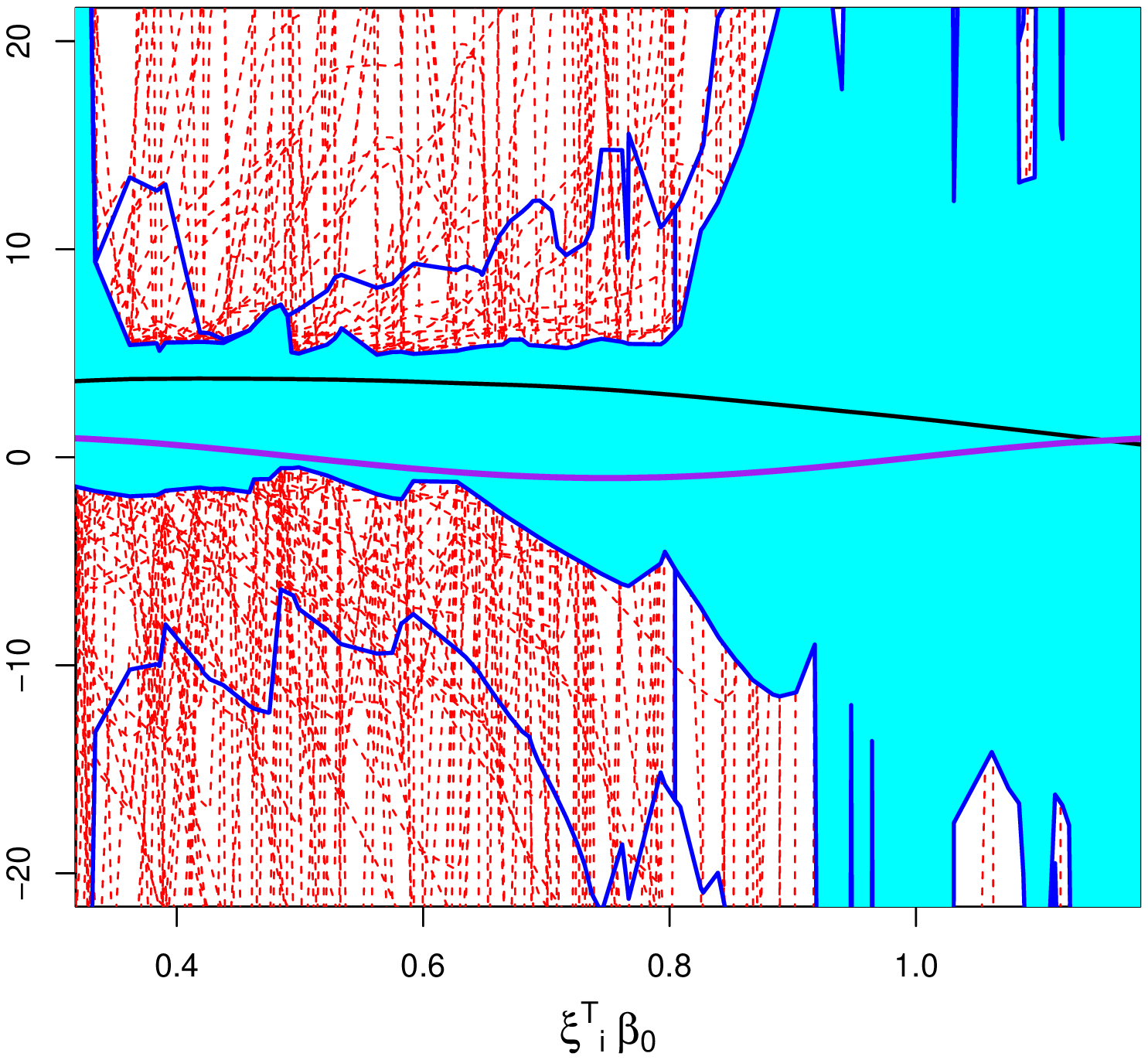}  
\hskip-0.2in\includegraphics[scale=0.3]{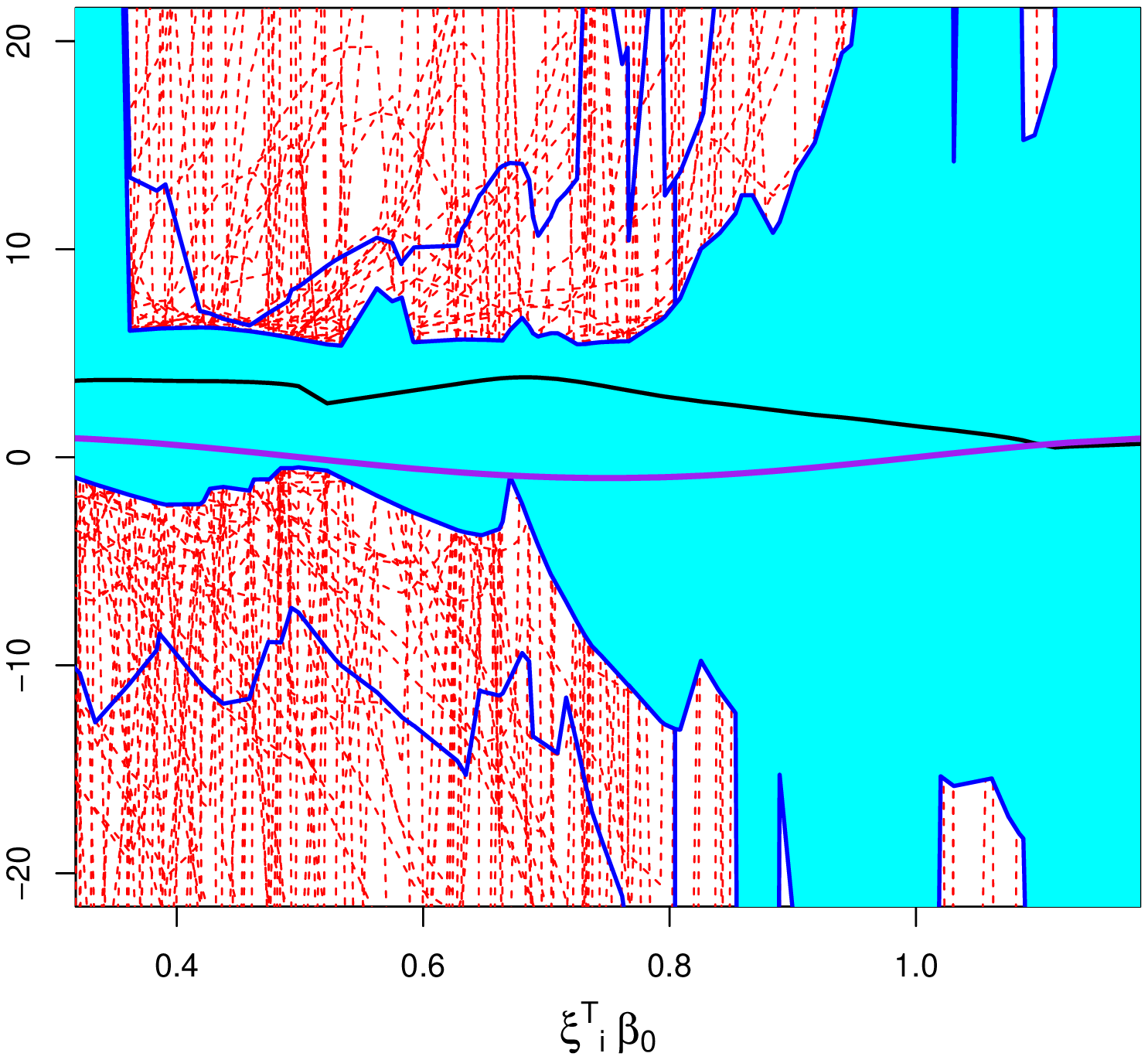}\\
Robust Method\\
\vskip-0.1in
\hskip-0.2in\includegraphics[scale=0.3]{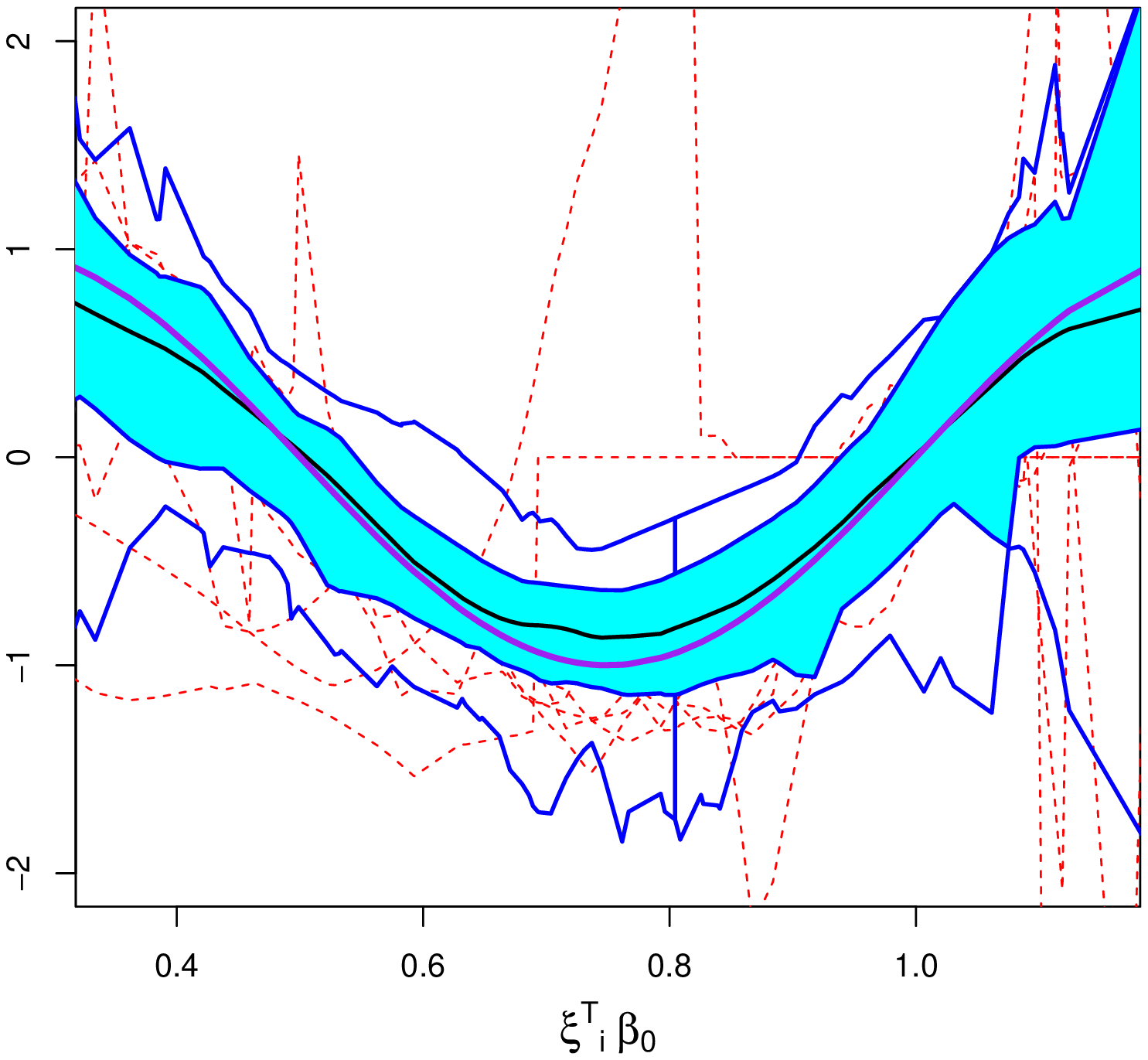} 
\hskip-0.2in\includegraphics[scale=0.3]{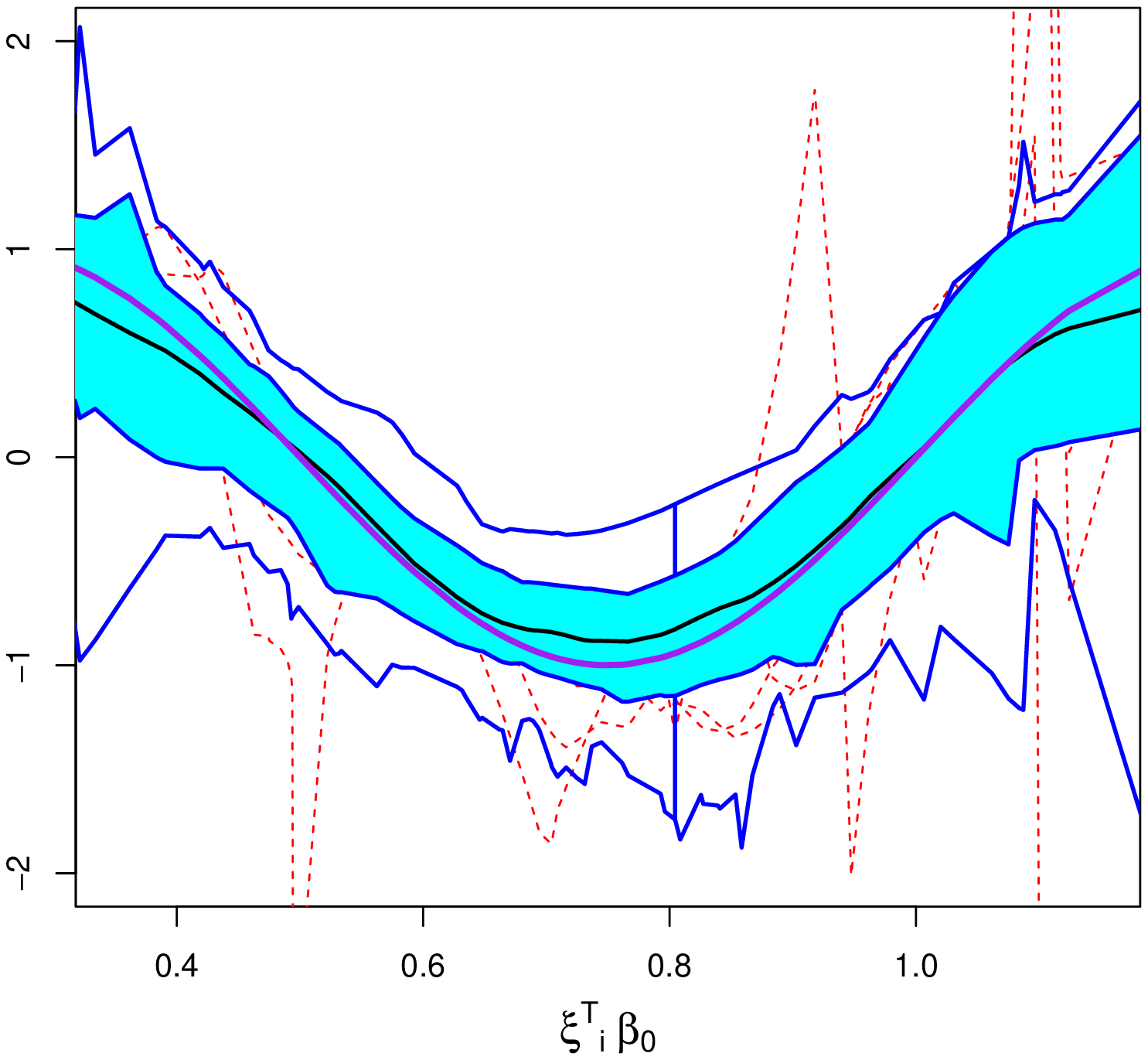}   
\hskip-0.2in\includegraphics[scale=0.3]{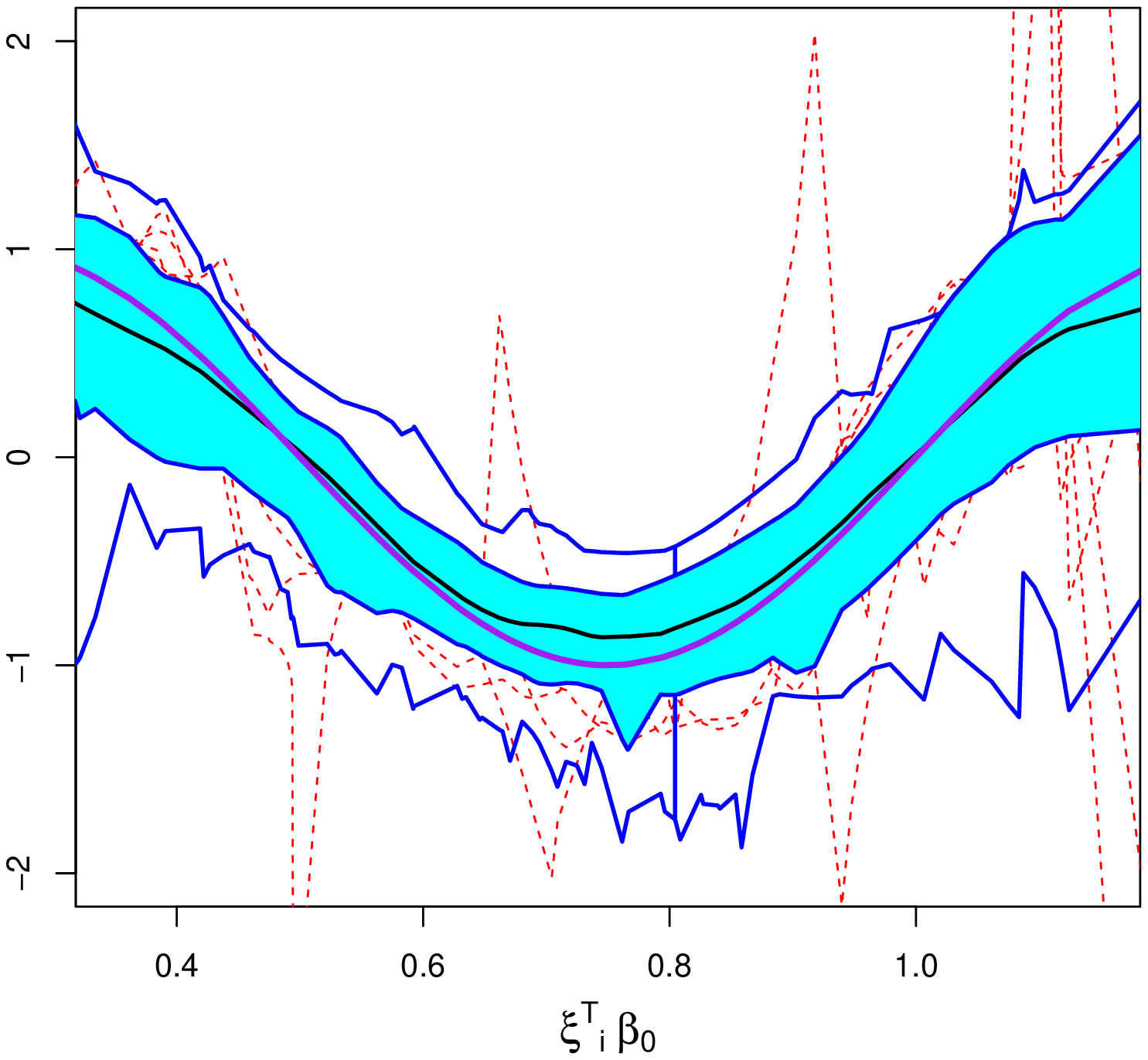}\\
\vskip-0.1in
\caption{\small \label{fig:fda_S}\small{Classical and robust estimators of $\eta_0$ under  $S_1$, $S_2$ and $S_3$.}}
\end{center}
\end{figure}
In order to give a full picture of the performance of both classical and robust estimators of $\eta_0$, Figures \ref{fig:fda_c0} to \ref{fig:fda_S} display their functional boxplots. Since the covariate $\bx$ is random, in order to obtain comparable estimations for $\eta_0$, we consider a fixed grid of  points $\bxi_j$, $j=1,\dots,100$ in  $[0,1] \times [0,1]$. Thus, for each replication, we estimate $\eta_0(\bxi_j\trasp \bbe_0)$ using the classical and robust procedures. In the functional boxplots, the area in light blue represents the central region, the dotted red lines correspond to outlying curves, the black line indicates the deepest curve, while the purple line is the true nonparametric function $\eta_0$. It is worth noticing that, for the contaminated settings, due to the effect of the outliers, some curves are out of range when using the classical procedure. For that reason, the functional boxplots of the classical estimators are plotted in a reduced range to allow a clear visualization of the central area.
Figure  \ref{fig:fda_c0} shows that the classical and robust nonparametric estimators of $\eta_0$ are quite similar under $C_0$, while Figures \ref{fig:fda_M} to \ref{fig:fda_S} exhibit the devastating effect of the contaminating points, even the moderate ones, on the classical estimator. The impact of the contaminations on the classical estimates is reflected either in the presence of a great number of outlying curves and also in the enlargement of the width of the bars of the functional boxplots. 
With respect to the robust estimates, despite the fact that a few outlying curves appear, the range of variation of the curves is almost the same than under $C_0$, the central region in light blue of all the boxplots  always contain the true function $\eta_0$ and most curves follow the pattern introduced by the sine function.
In general terms, the functional boxplots show the stability of the robust estimates of
$\eta_0$ which are reliable under the contaminated scenarios as well as the strong effect of the considered contaminations on the classical estimators of the nonparametric component.

\begin{figure}[ht!]
\begin{center}
\small 
Classical Method \hspace{4.5cm} Robust Method\\
\includegraphics[scale=0.4]{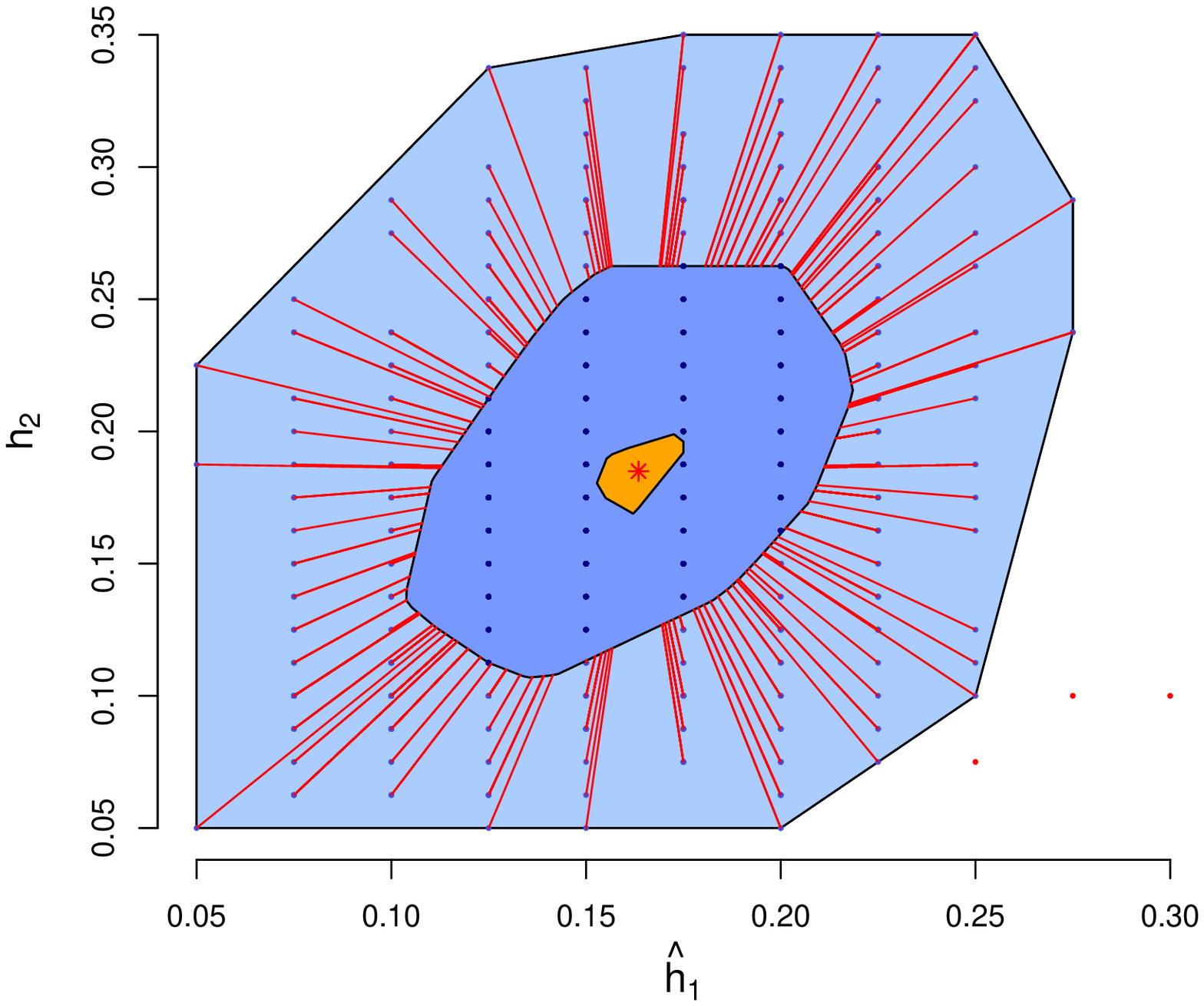} \includegraphics[scale=0.4]{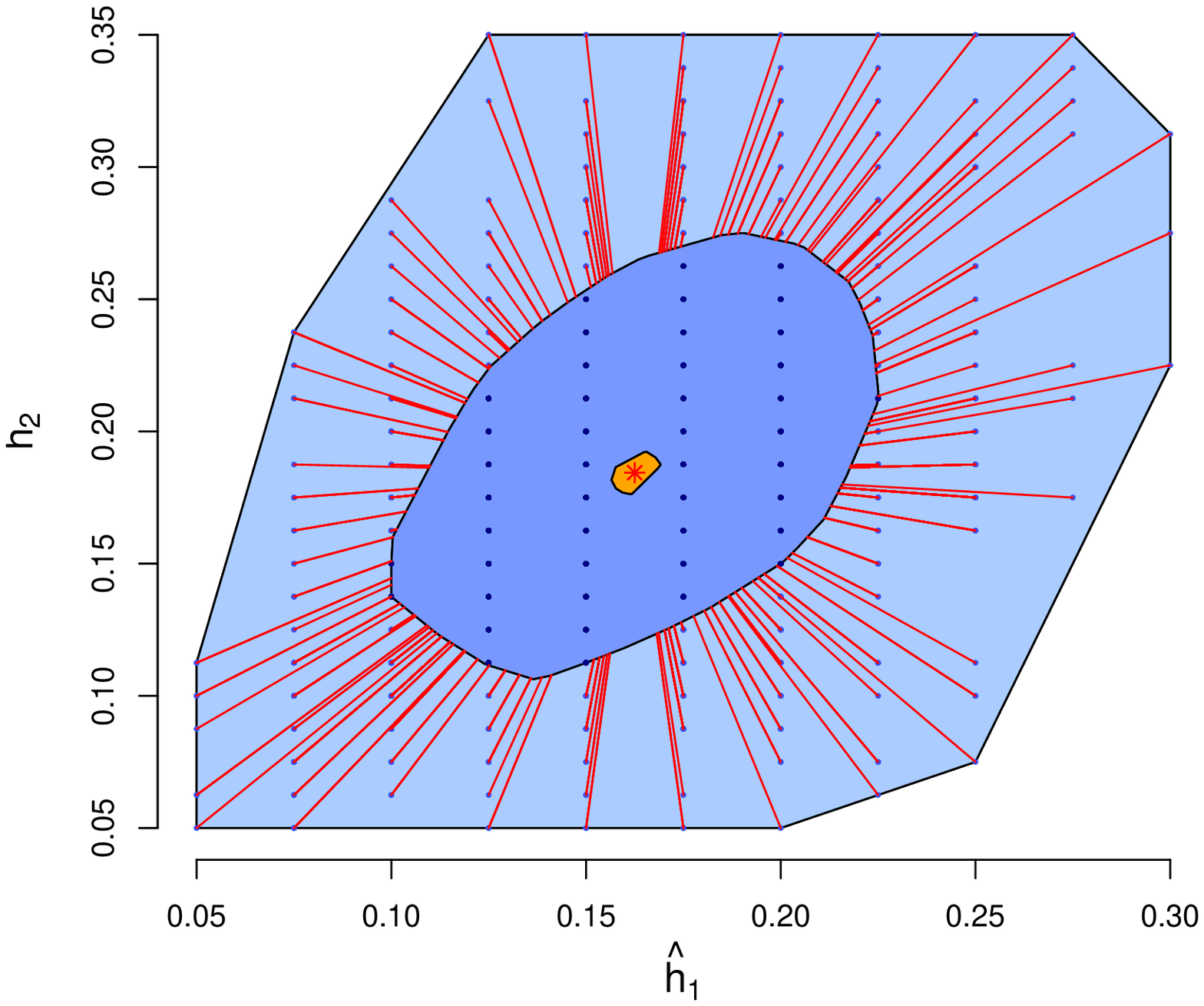} 
\caption{\small \label{fig:bag_c0}\small{Bagplots for $(\widehat{h}_1,\widehat{h}_2)$  chosen according the classical and robust cross--validation criteria under $C_0$.}}
\end{center}
\end{figure}

\begin{figure}[ht!]
\begin{center}
\small 
Classical Method\\
$\quad$\\
$M_1$ \hspace{4.5cm} $M_2$ \hspace{4.5cm} $M_3$\\
\vskip-0.1in
\hskip-0.2in\includegraphics[scale=0.3]{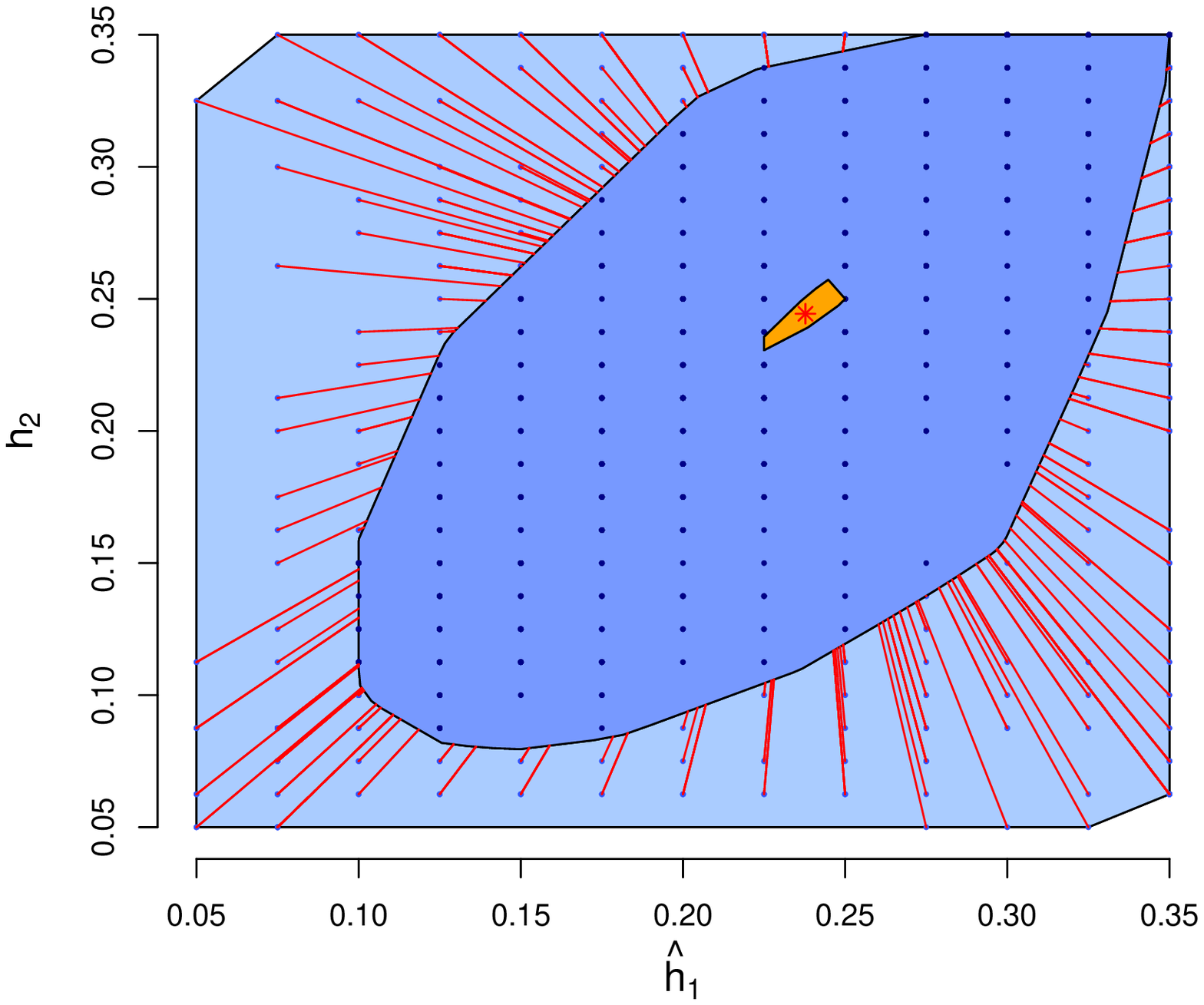} 
\hskip-0.2in\includegraphics[scale=0.3]{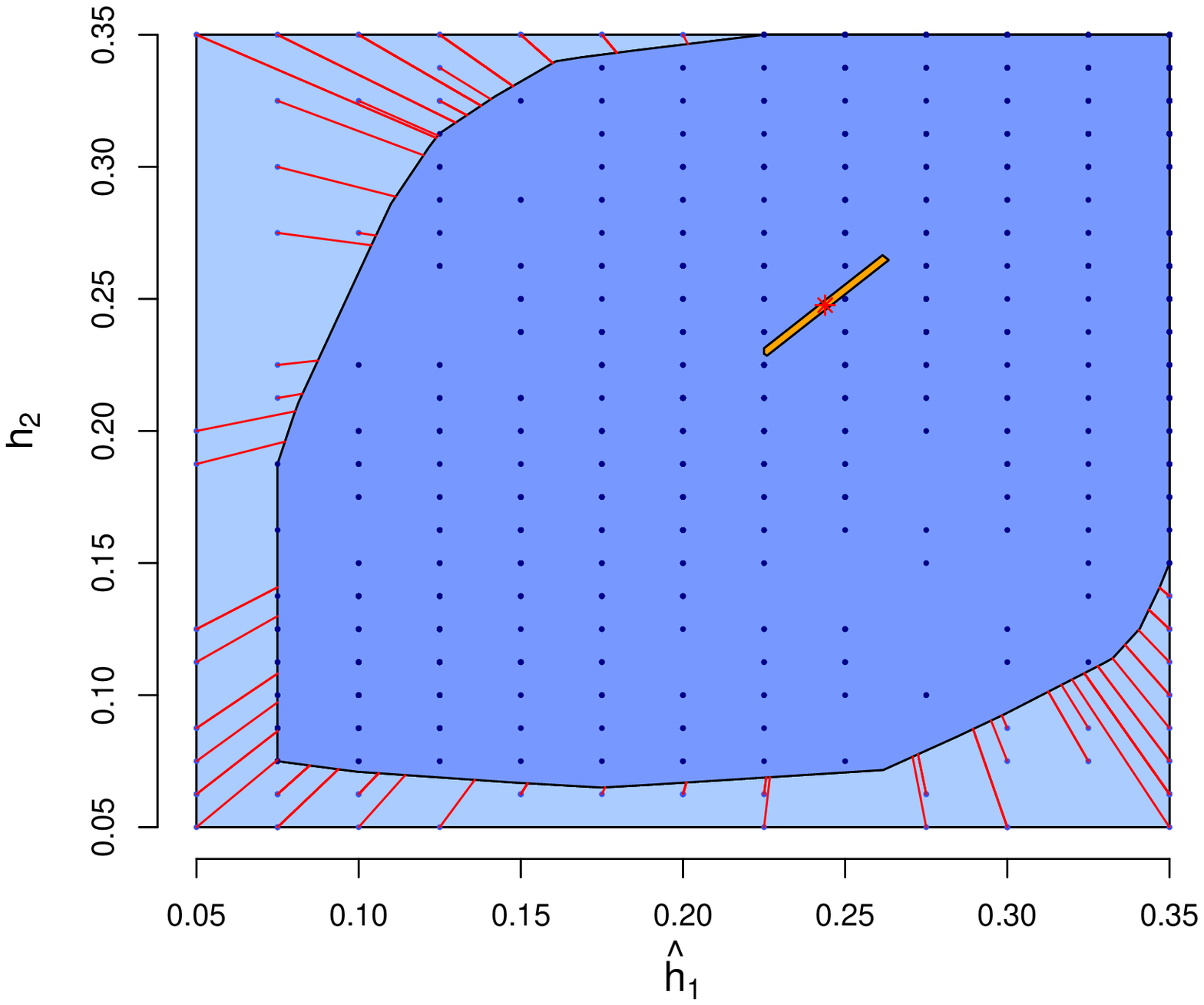}  
\hskip-0.2in\includegraphics[scale=0.3]{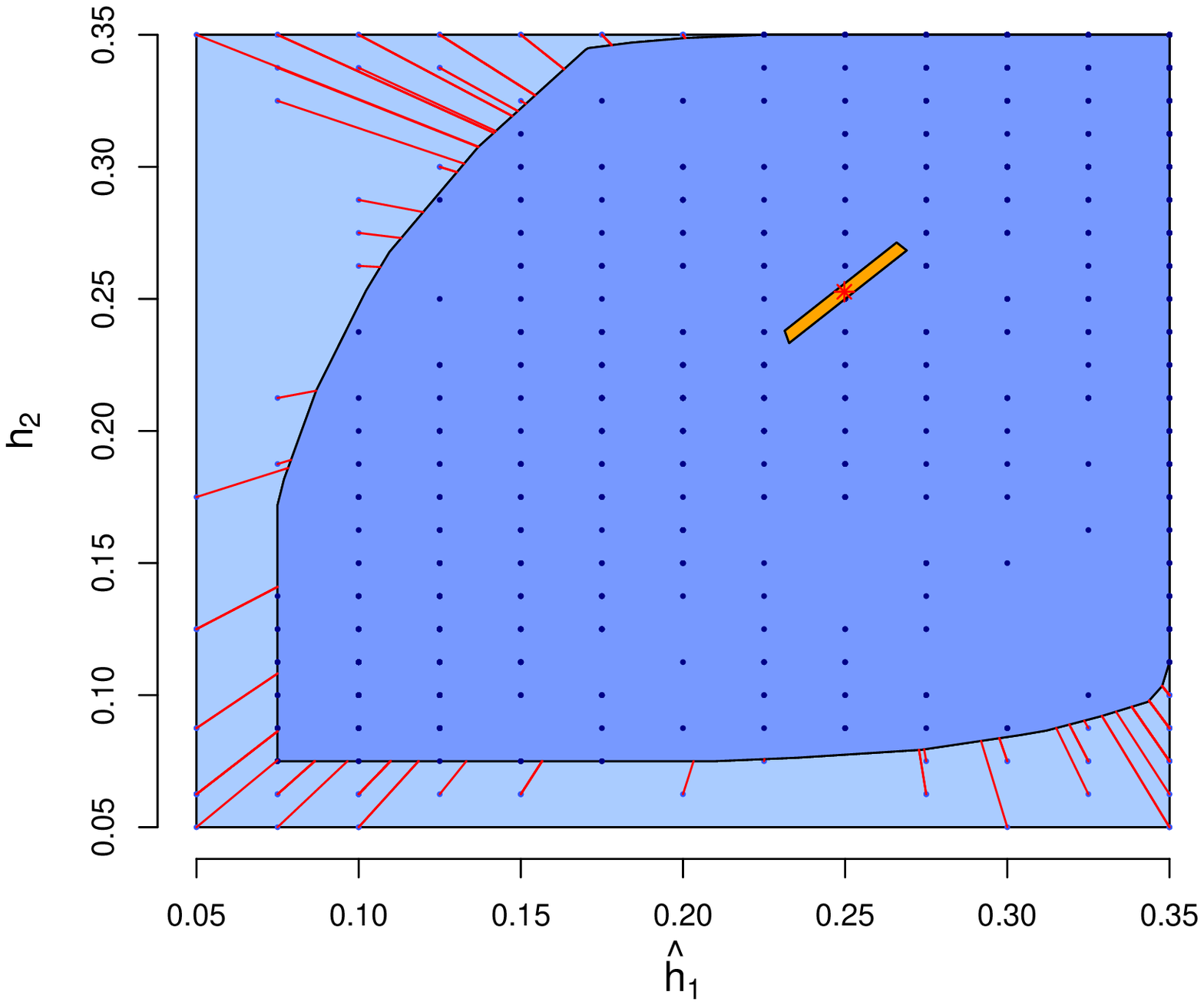}\\
Robust Method\\
\vskip-0.1in
\hskip-0.2in\includegraphics[scale=0.3]{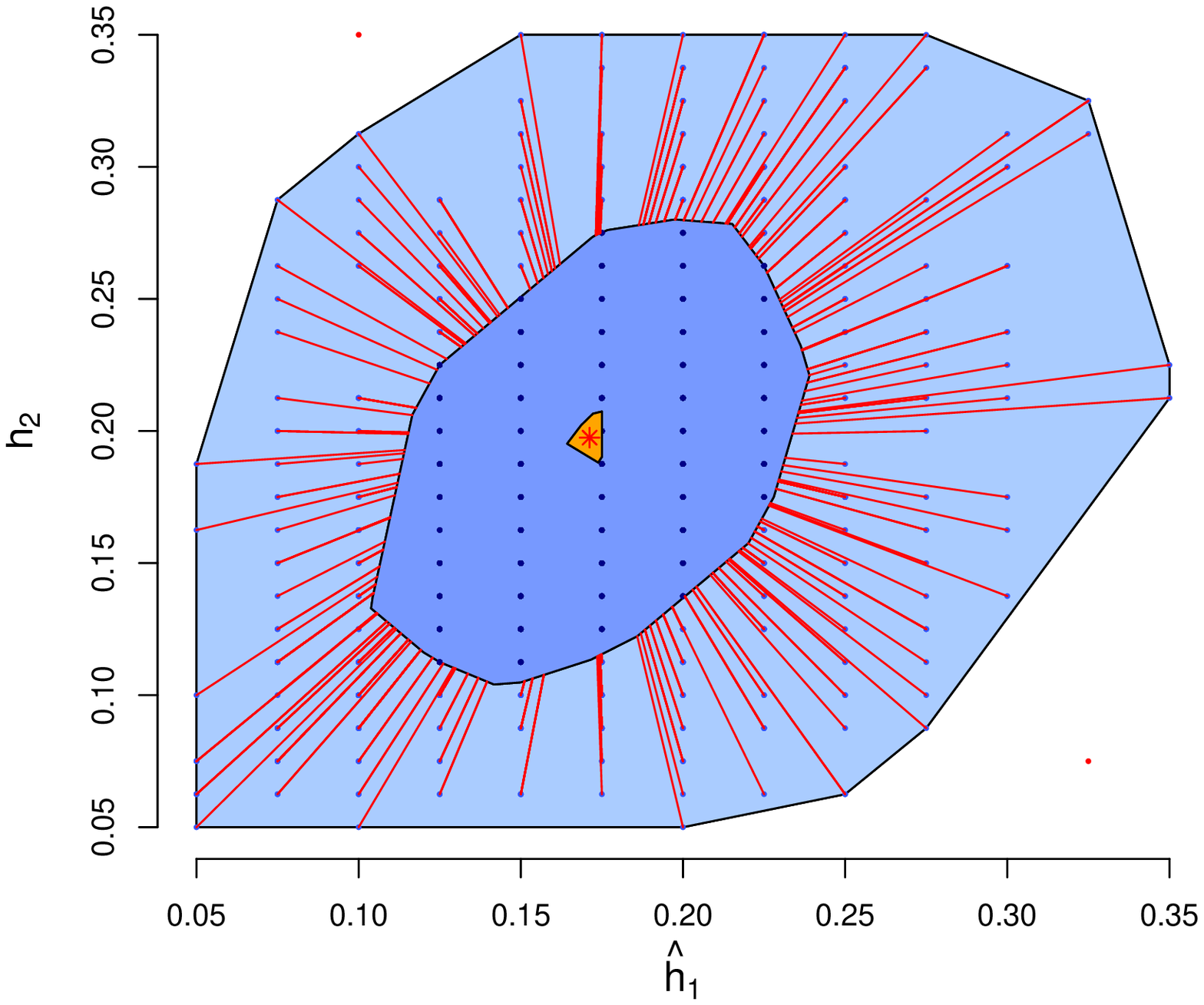} 
\hskip-0.2in\includegraphics[scale=0.3]{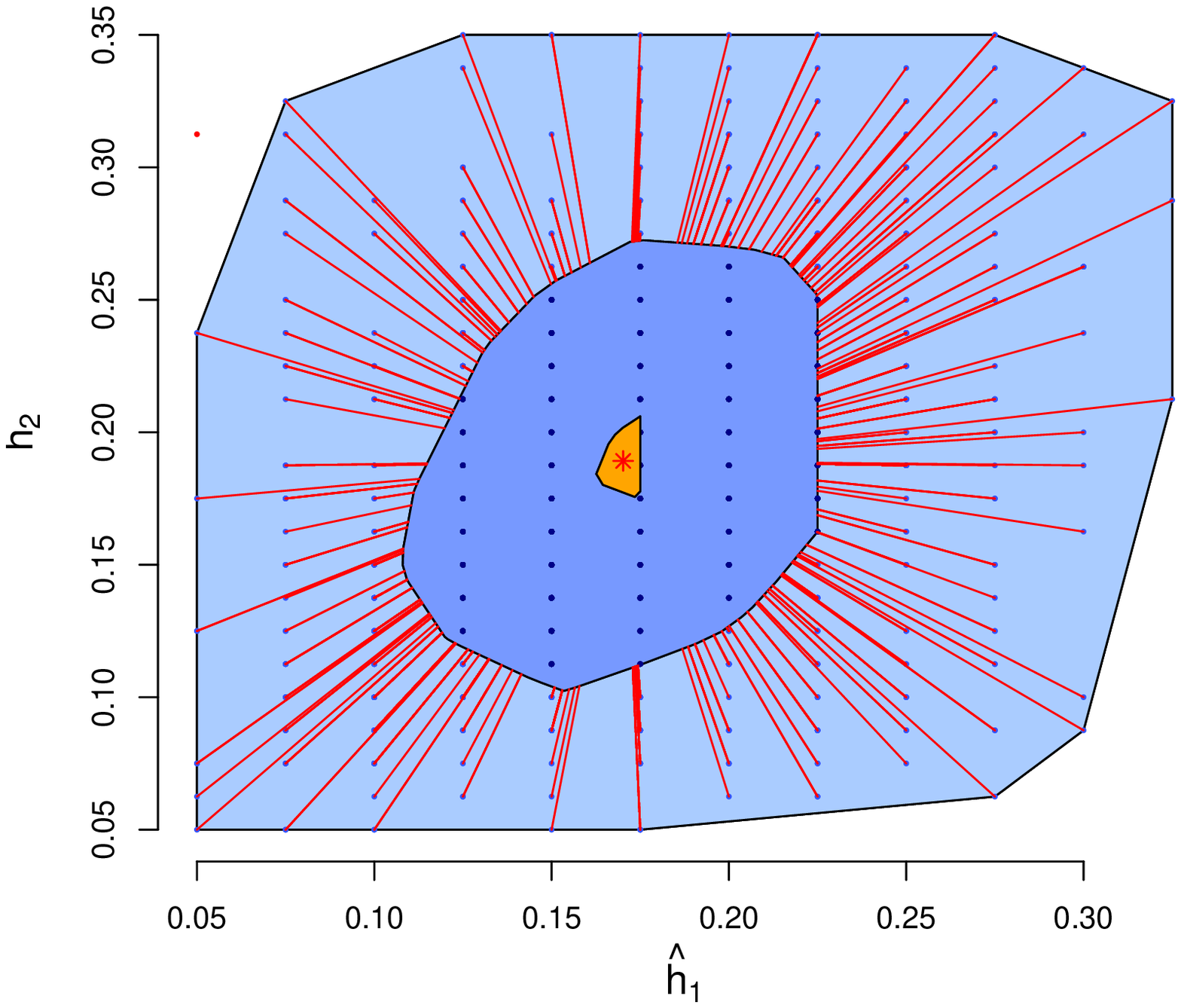}  
\hskip-0.2in\includegraphics[scale=0.3]{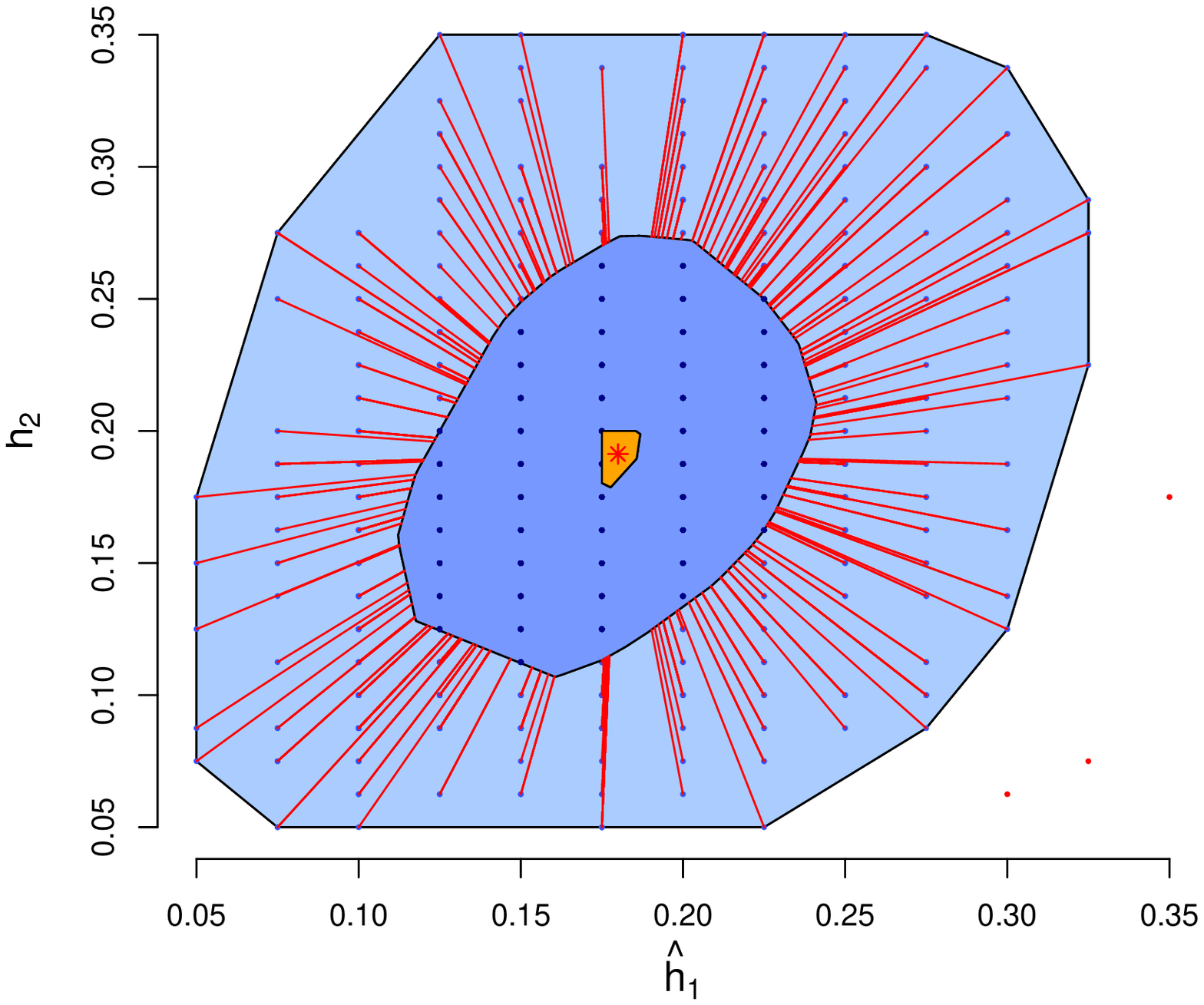}\\
\vskip-0.1in
\caption{\small\label{fig:bag_M}\small{Bagplots for $(\widehat{h}_1,\widehat{h}_2)$  chosen according the classical and robust cross--validation criteria under  $M_1$, $M_2$ and $M_3$.}}
\end{center}
\end{figure}

\begin{figure}[ht!]
\begin{center}
\small 
Classical Method\\
$\quad$\\
$S_1$ \hspace{4.5cm} $S_2$ \hspace{4.5cm} $S_3$\\
\vskip-0.1in
\hskip-0.2in\includegraphics[scale=0.3]{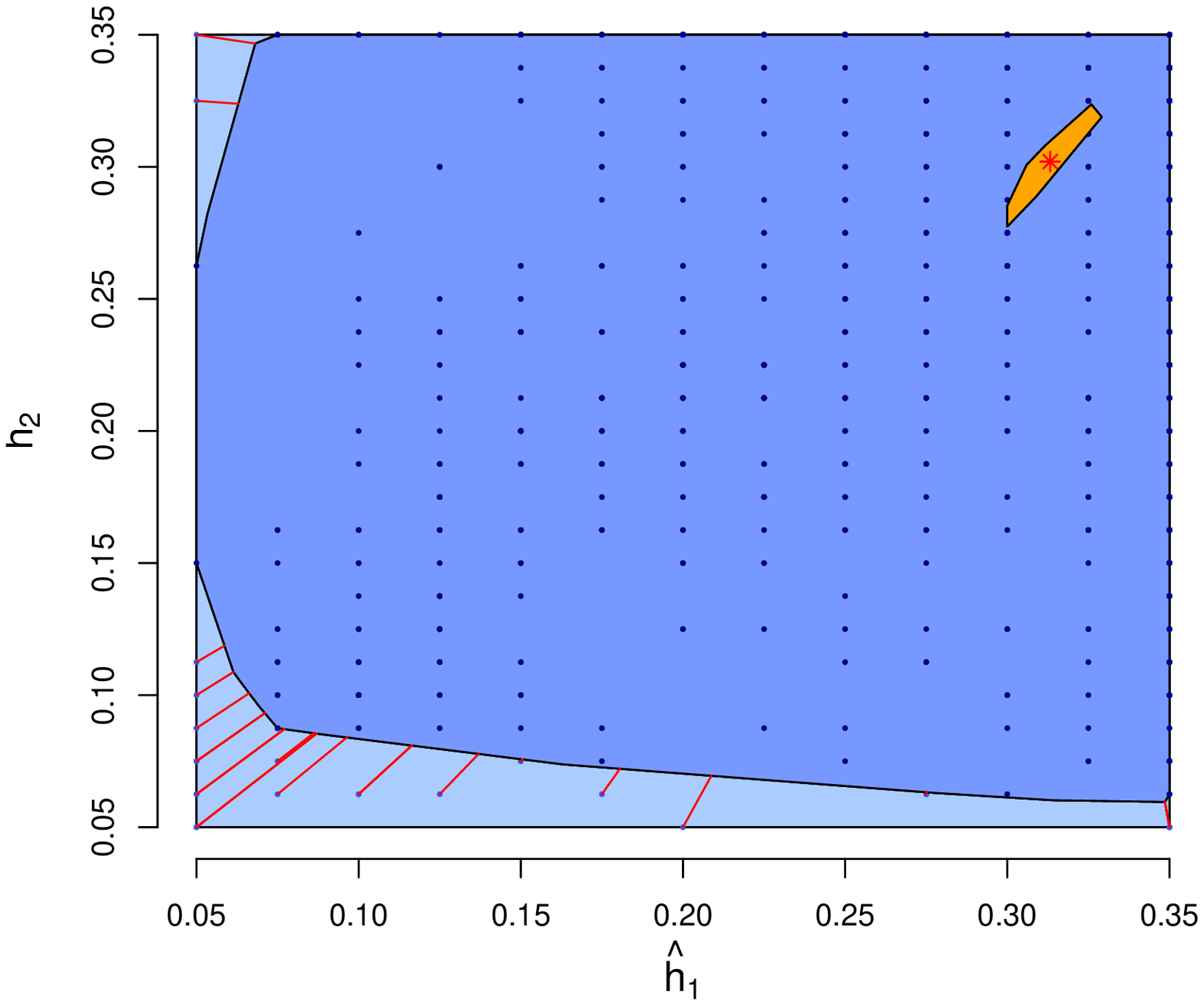} 
\hskip-0.2in\includegraphics[scale=0.3]{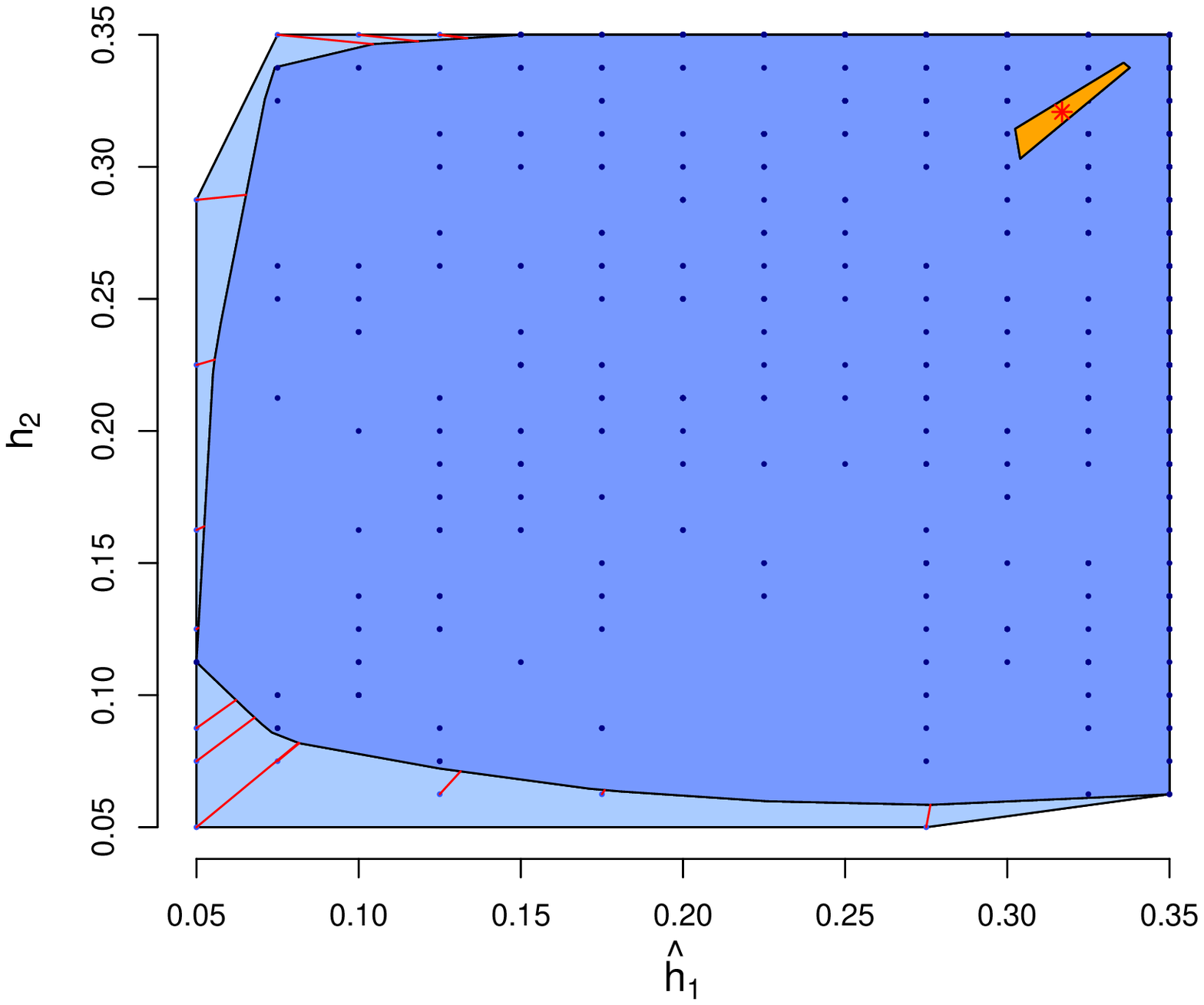}  
\hskip-0.2in\includegraphics[scale=0.3]{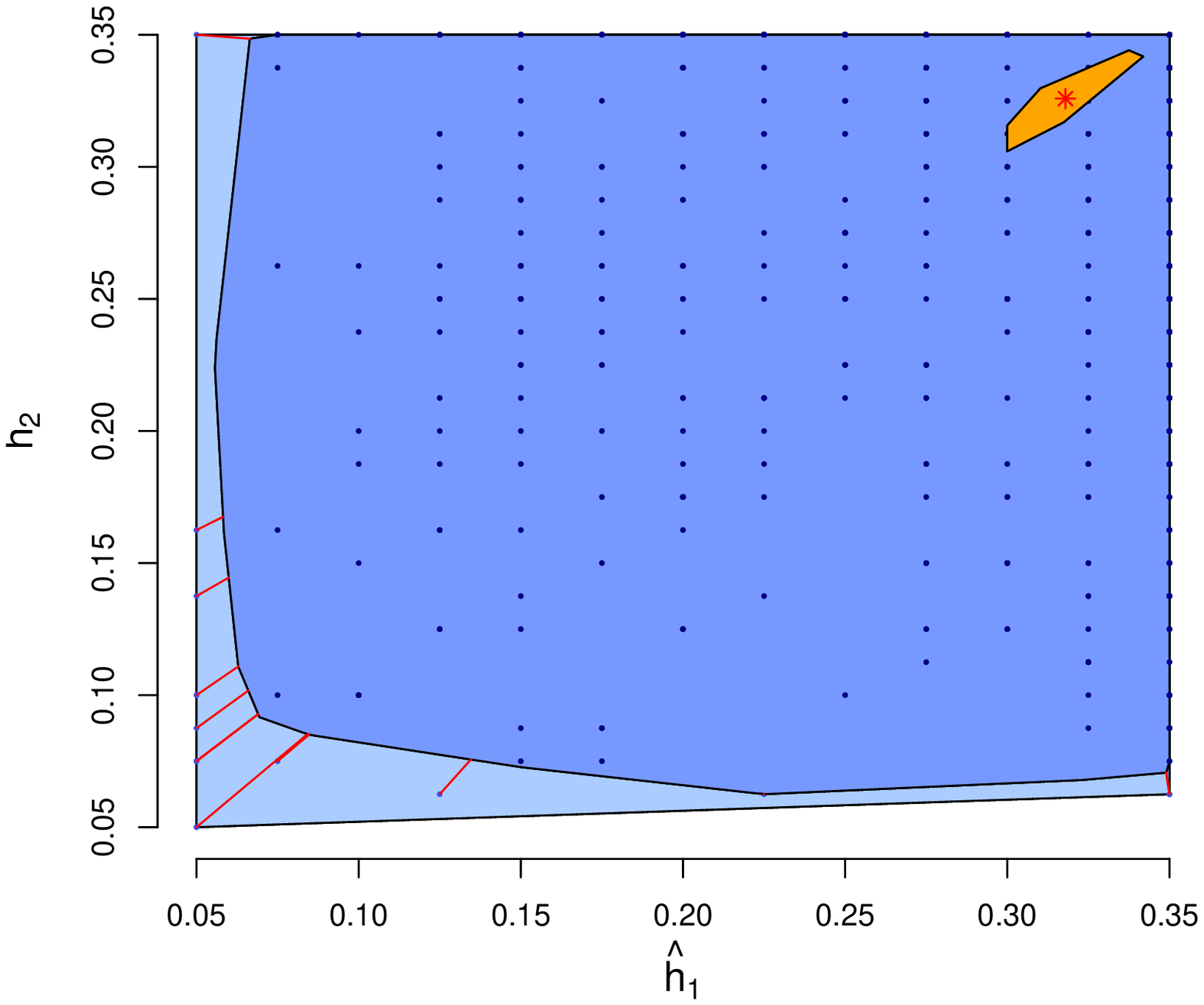}\\
Robust Method\\
\vskip-0.1in
\hskip-0.2in\includegraphics[scale=0.3]{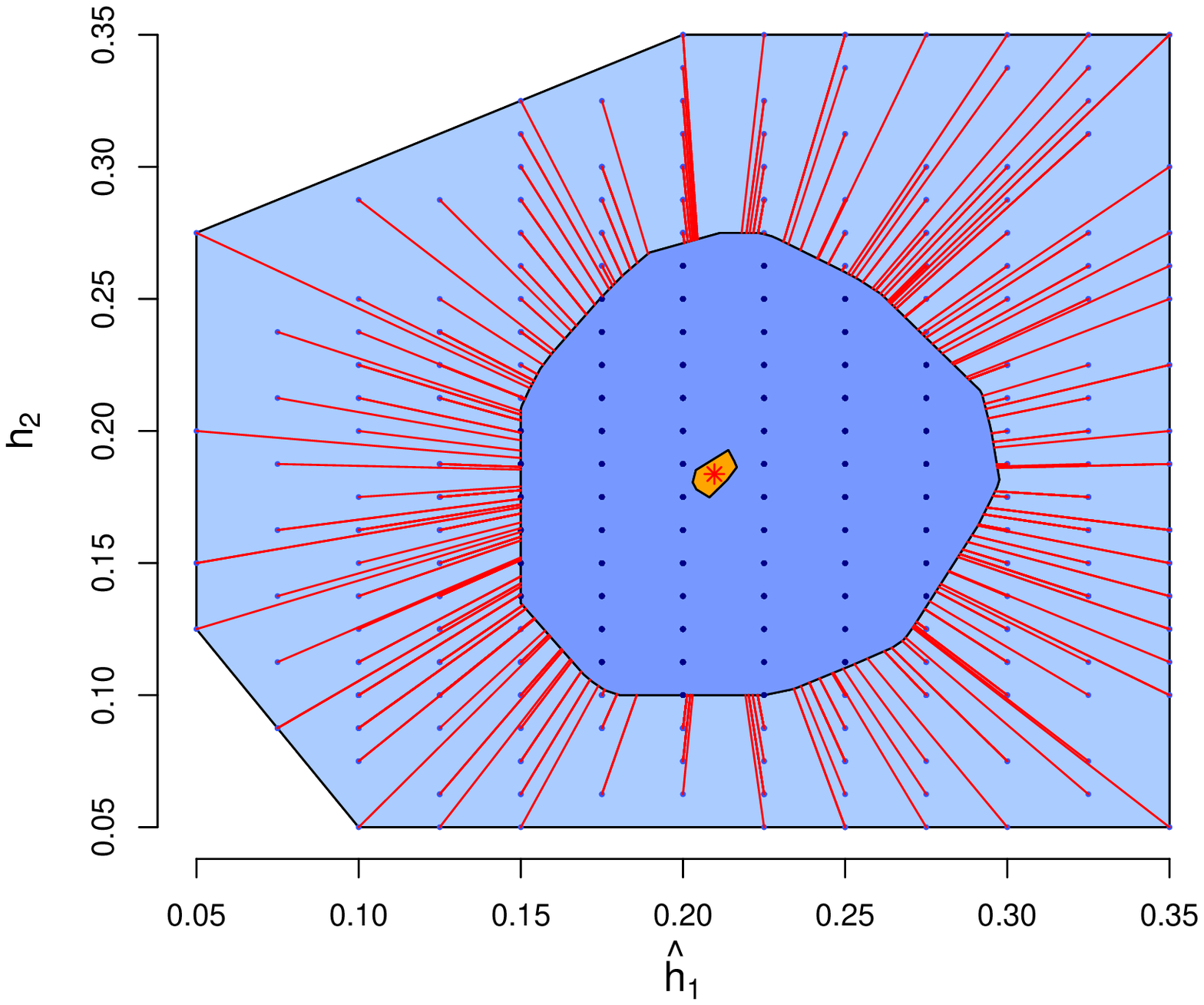} 
\hskip-0.2in\includegraphics[scale=0.3]{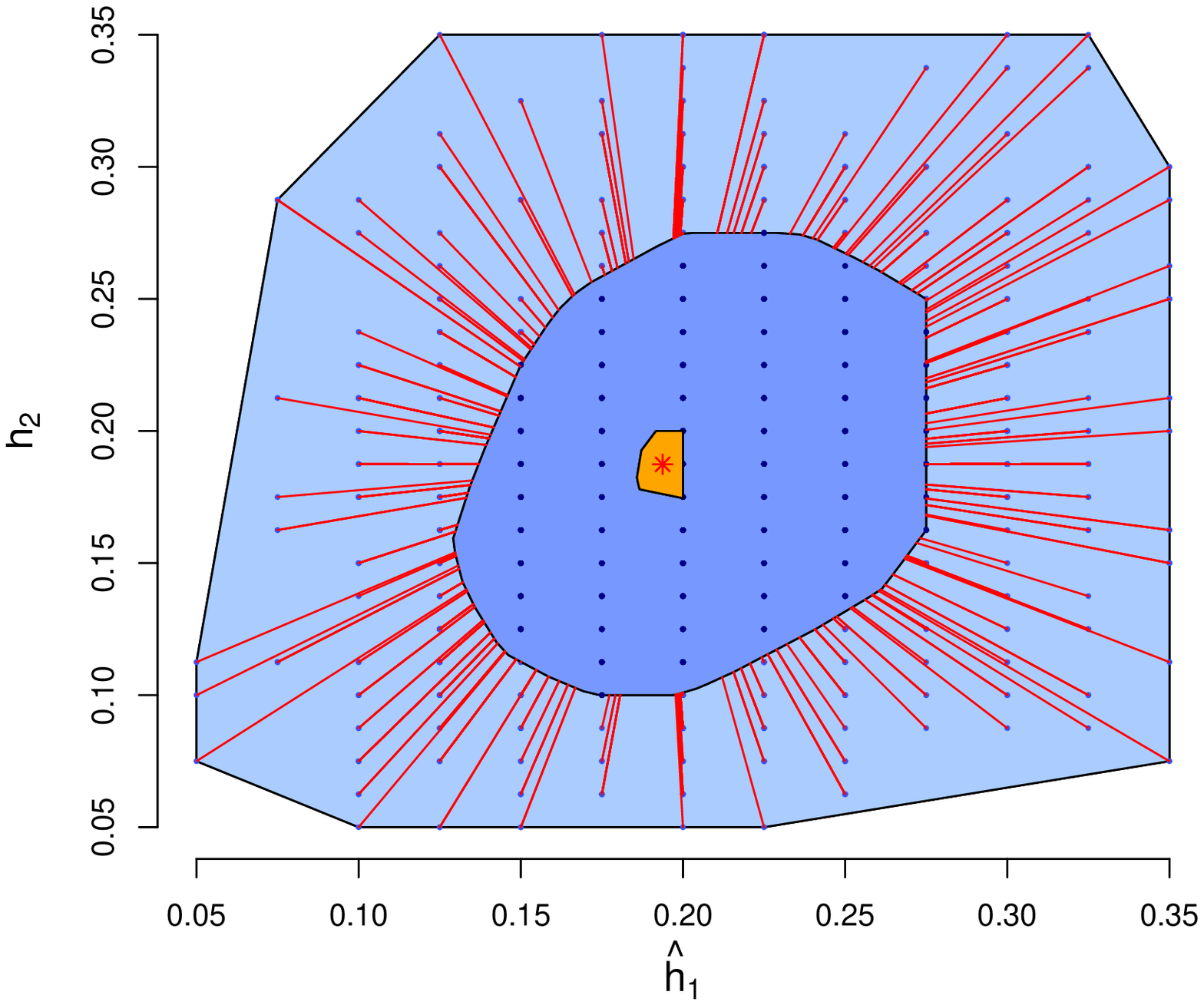}  
\hskip-0.2in\includegraphics[scale=0.3]{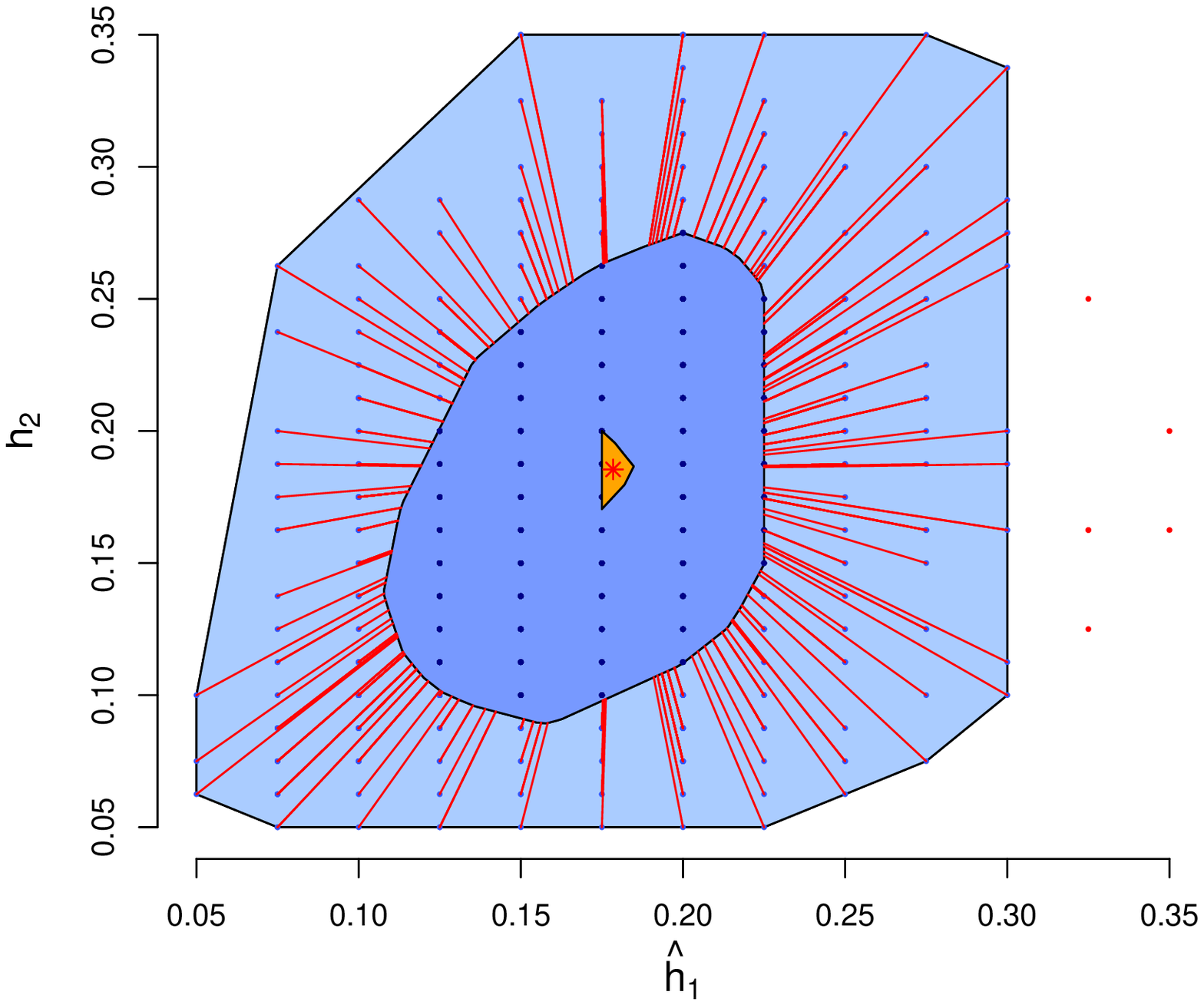}\\
\vskip-0.1in
\caption{\small\label{fig:bag_S}\small{Bagplots for $(\widehat{h}_1,\widehat{h}_2)$  chosen according the classical and robust cross--validation criteria under  $S_1$, $S_2$ and $S_3$.}}
\end{center}
\end{figure}

A careful study of the bandwidth behaviour is beyond the scope of the paper, however in order to have a  deeper insight of the performance of the selectors under $C_0$ and the considered contaminations,  we give a brief analysis of the data--driven parameters obtained in this numerical experiment.
Table \ref{tab:medianah_k} reports,   for both estimators,  the median over replications of the cross-validation data--driven bandwidths to be used in Steps 1 and 3 denoted $\wh_1$ and $\wh_2$ respectively. On the other hand, in Figures \ref{fig:bag_c0} to \ref{fig:bag_S} we present the   bagplots corresponding to $\wh_1$ and $\wh_2$  selected through the classical and robust cross--validation criteria. Under $C_0$ both criteria lead to similar data--driven smoothing parameters.
However, the lack of robustness of the classical cross--validation criterion under contaminations becomes evident from these plots. The classical cross--validation criterion under contaminations tends to choose greater bandwidths and this becomes evident, for instance, from  the behaviour of their medians reported in Table \ref{tab:medianah_k}. The poor behaviour of the classical data--driven bandwidths leads   towards over--smoothing which may explain the results reported in Table \ref{tab:gamma}. On the other hand, except for a few cases, the selected bandwidths obtained with the robust criterion remain stable in all circumstances.

\begin{table}[ht!]
\begin{center}
\renewcommand{\arraystretch}{1.1}
\begin{tabular}{c r r| r r r| r r r}
\hline
Method && $C_0$ & $M_1$ & $M_2$ & $M_3$& $S_1$ & $S_2$ & $S_3$ \\\hline
Classical&$\wh_1$ & 0.175 & 0.225 & 0.250 & 0.250 & 0.325 & 0.325 & 0.325\\
 &$\wh_2$ &  0.175 & 0.238 & 0.250 & 0.250  & 0.350 & 0.300 & 0.338\\\hline
Robust& $\wh_1$ & 0.175 & 0.175 & 0.175 & 0.175 & 0.175 & 0.200 & 0.200\\
 & $\wh_2$ & 0.188 & 0.188 & 0.188 & 0.188& 0.188 & 0.188 & 0.188\\
\hline
\end{tabular}
\end{center}
\caption{\small\label{tab:medianah_k} Median over replications of  cross-validation data--driven bandwidths to be used in Steps 1 and 3 denoted $\wh_1$ and $\wh_2$ respectively.}
\end{table}

{\setcounter{equation}{0}
\renewcommand{\theequation}{A.\arabic{equation}}
{\setcounter{section}{0}
\renewcommand{\thesection}{\Alph{section}}

\section{Appendix}{\label{proof}}

\subsection{Proof of Theorem 1.}
 a) For any $\varepsilon>0$, let $\itX_0$ be a compact set such that $P(\bx\notin \itX_0)<\varepsilon $. Then, we have that
\begin{eqnarray*}
\dst\sup_{\bbech,\bb \in \itS_1; a \in \itK}\left|\Delta_n(\bbe,\weta_{\bb,a},a)-\Delta_n(\bbe,\eta_{\bb,a},a)\right|&\leq&\sup_{\bb \in \itS_1,a \in \itK}\|\weta_{\bb,a}-\eta_{\bb,a}\|_{0,\infty} \|\tau \|_{\infty}  \|\phi^{\prime}\|_{\infty}\\
&+& 2    \|\phi\|_{\infty}\frac{1}{n}\sum_{i=1}^n \indica_{(\bx_i\notin \itX_0)} \tau(\bx_i)
\end{eqnarray*}
and so, using (\ref{condicion1}), the fact that $P(\bx\notin \itX_0)<\varepsilon $ and the Strong Law of Large Numbers,  we get that
$$
\dst\sup_{\bbech,\bb \in \itS_1; a \in \itK}\left|\Delta_n(\bbe,\weta_{\bb,a},a)-\Delta_n(\bbe,\eta_{\bb,a},a)\right| \convpp 0\;.
$$
Therefore, it remains to show that $\dst \sup_{\bbech,\bb \in \itS_1; a \in \itK}\left|\Delta_n(\bbe,\eta_{\bb,a},a)-\Delta(\bbe,\eta_{\bb,a},a)\right|  \convpp 0$. Define the following class of functions 
$\itH=\{f_{\bbech}(y,\bx)=\phi(y,\eta_{\bb,a}(\bbe\trasp \bx), a) \tau(\bx) \,,\, \bbe,\bb \in \itS_1, a \in \itK \}$. Using Theorem 3 from Chapter 2 in  Pollard (1984), the compactness of   $\itK$, \textbf{A1}, the continuity of $\eta_{\bbech, \alpha}(u)$ given in \textbf{A6} and  analogous arguments to those considered in  Lemma 1 from Bianco and Boente (2002), we get that  $\dst \sup_{\bbech,\bb \in \itS_1; a \in \itK}\left|\Delta_n(\bbe,\weta_{\bb,a},a)-\Delta(\bbe,\eta_{\bb,a},a)\right|  \convpp 0$ and a) follows.

\noi b) Let $\wbbe_k$ be a subsequence of $\wbbe$ such that $\wbbe_k\rightarrow \bbe^*$, where 
$\bbe^*$ lies in the compact set $\itS_1$. Let us assume, without loss of generality, that  $\wbbe\convpp\bbe^*$. Then, \textbf{A7}, the continuity of $\eta_{\bbech, \alpha}$, the consistency of $\walfa_{\rob}$ and  a)  entail that
 $\Delta_n(\wbbe,\weta_{\wbbech,\walfach_{\rob}},\walfa_{\rob})-\Delta(\bbe^*,\eta_0,\alpha_0)\convpp 0$ 
 and 
 $\Delta_n( \bbe_0,\weta_{\wbbech,\walfach_{\rob}},\walfa_{\rob} )-\Delta(\bbe_0,\eta_{0},\alpha_0)\convpp 0$,
 since $\eta_{\bbech_0,\alpha_0}=\eta_0$. Now, using that
$\Delta_n( \bbe_0,\weta_{\wbbech,\walfach_{\rob}},\walfa_{\rob} )\geq
\Delta_n(\wbbe,\weta_{\wbbech,\walfach_{\rob}},\walfa_{\rob})$ and
$\Delta(\bbe,\eta_0,\alpha_0)$ has a unique minimum at
$\bbe_0$,  we conclude the proof. \square

\subsection{Proof of Proposition  1.}
a) The single index parameter estimation related to \textbf{Step LG2} is obtained by means of the minimization with respect to $\bbe$ of  
$$\sum_{i=1}^n \rho\left(\frac{\sqrt{d\left(y_i, \weta_{\bbech} \left(\bbe\trasp \bx_i\right)\right)}}{c}\right)\tau(\bx_i) \, ,$$
among the vectors of length one, where, at the same time, $\weta_{\bbech}(u)$ is defined as
\begin{eqnarray*}
\weta_{\bbech}(u)&=&\dst\argmin_{a\in \real}
\sum_{i=1}^n \rho\left(\frac{\sqrt{d(y_i,a)}}{c}\right)   W_{h}(u,\bbe\trasp \bx_i ).
\end{eqnarray*} 
Hence, if we denote $\itB(\bthe)=\bthe/\|\bthe\|$, we have that $\wbbe_{\varepsilon}=\wbthe_{\varepsilon}/\|\wbthe_{\varepsilon}\|=\itB(\wbthe_{\varepsilon})$ where $\wbthe_{\varepsilon}$ is the solution of 
$$\argmin_{\bthech} \frac {1-\varepsilon}{n}  \sum_{i=1}^n \rho\left(\frac{\sqrt{d\left(y_i, \weta^{\varepsilon}_{\itB(\bthech)} \left(\itB(\bthe)\trasp \bx_i\right)\right)}}{c}\right)\tau(\bx_i) 
+ \varepsilon \, \rho\left(\frac{\sqrt{d\left(y_0, \weta^{\varepsilon}_{\itB(\bthech)} \left(\itB(\bthe)\trasp \bx_0\right)\right)}}{c}\right)\tau(\bx_0).$$
Then, $\wbthe_{\varepsilon}$ satisfies
\begin{eqnarray*}
\bcero &=& \left(\identidad- \itB\left(\wbthe_{\varepsilon}\right) \itB\left(\wbthe_{\varepsilon}\right)\trasp\right)
\left[ \frac{(1-\varepsilon)}{n}\sum_{i=1}^n \psi \left(y_i,\weta^{\varepsilon}_{\itB(\wbthech_{\varepsilon})}(\itB(\wbthe_{\varepsilon})\trasp  \bx_i),c\right) 
\wbnu_i^{\epsilon}\left(\itB(\wbthe_{\varepsilon}),\itB(\wbthe_{\varepsilon}) \bx_i\right) \tau(\bx_i) \right.\\
&+&   \varepsilon \; \psi \left(y_0,\weta^{\varepsilon}_{\itB(\wbthech_{\varepsilon})}(\itB(\wbthe_{\varepsilon})\trasp  \bx_0),c\right) 
\wbnu_0^{\epsilon}\left(\itB(\wbthe_{\varepsilon}),\itB(\wbthe_{\varepsilon}) \bx_0\right) \tau(\bx_0) \Bigg]
 \; ,
\end{eqnarray*}
where 
$$\psi(y,a,c)= \frac{\partial}{\partial a} \phi(y,a,c)= \frac{1}{2c}\Psi\left(\frac{\sqrt{d(y,a)}}{c}\right)  \frac{1-  \exp(y-a)}{\sqrt{d(y,a)}} $$
as defined in (\ref{phi1}), $\Psi$ stands for the derivative of $\rho$  and $\wbnu_i^{\epsilon}(\bb,t)$ are given by 
$$\wbnu_i^{\epsilon}(\bb,t)={ \frac{\partial}{\partial\bbech}
\weta_{\bbech}^{\epsilon}(s)|_{(\bbech,s)=(\bb,t)}+
\frac{\partial}{\partial s}
\weta_{\bbech}^{\epsilon}(s)|_{(\bbech,s)=(\bb,t)}\;\bx_i} \; . 
$$
Using that  $\wbbe_{\varepsilon}=\itB(\wbthe_{\varepsilon})$,  we get that the estimator $\wbbe_{\varepsilon}$ verifies
\begin{eqnarray*}
\bcero &=& \left(\identidad- {\wbbe_{\varepsilon}} {\wbbe_{\varepsilon}\trasp}\right)
\left[ \frac{(1-\varepsilon)}{n}\sum_{i=1}^n \psi \left(y_i,\weta^{\varepsilon}_{\bbech_{\varepsilon}}({\wbbe_{\varepsilon}\trasp} \bx_i),c\right) 
\wbnu_i^{\epsilon}({\wbbe_{\varepsilon}},{\wbbe_{\varepsilon}\trasp} \bx_i) \tau(\bx_i) \right.\\
&&  +
 \varepsilon \; \psi \left(y_0,\weta^{\varepsilon}_{\bbech_{\varepsilon}}({\wbbe_{\varepsilon}\trasp} \bx_0),c\right) 
 \wbnu_0^{\epsilon}({\wbbe_{\varepsilon}},{\wbbe_{\varepsilon}\trasp} \bx_0) \tau(\bx_0)  \Bigg]
 \; 
\end{eqnarray*}
and   $\weta^{\varepsilon}_{\bbech}(u)$ is the solution of
\begin{equation}
 \frac{(1-\varepsilon)}{n}\sum_{i=1}^n \psi \left(y_i,\weta^{\varepsilon}_{\bbech}(u),c\right) 
  W_{h}(u,\bbe_\varepsilon\trasp \bx_i )  +\varepsilon \; \psi \left(y_0,\weta^{\varepsilon}_{\bbech}(u),c\right)  W_{h}(u,\bbe_\varepsilon\trasp \bx_0 )  
 = 0 \; .
 \label{eqeta}
\end{equation}
Then, if we call 
\begin{eqnarray*}
\bla (\varepsilon)=  \frac{(1-\varepsilon)}{n}\sum_{i=1}^n \psi \left(y_i,\weta^{\varepsilon}_{\bbech_{\varepsilon}}({\wbbe_{\varepsilon}\trasp} \bx_i),c\right) 
\wbnu_i^{\epsilon}({\wbbe_{\varepsilon}},{\wbbe_{\varepsilon}\trasp} \bx_i) \tau(\bx_i) +
 \varepsilon \; \psi \left(y_0,\weta^{\varepsilon}_{\bbech_{\varepsilon}}({\wbbe_{\varepsilon}\trasp} \bx_0),c\right) 
 \wbnu_0^{\epsilon}({\wbbe_{\varepsilon}},{\wbbe_{\varepsilon}\trasp} \bx_0) \tau(\bx_0)
\end{eqnarray*}
we get that, for any $0\le \epsilon <1$, $\wbbe_{\varepsilon}$ satisfies
$
\bcero = \left(\identidad- {\wbbe_{\varepsilon}} {\wbbe_{\varepsilon}\trasp}\right) \; \bla(\varepsilon) \, .
$
Therefore, differentiating  with respect to $\varepsilon$ and evaluating at $\varepsilon=0$ and using that $\bla(0)=\bcero$, we obtain that
\begin{eqnarray}
\bcero&=&\frac{\partial}{\partial \varepsilon}\left.\left[
\left(\identidad- {\wbbe_{\varepsilon}} {\wbbe_{\varepsilon}\trasp}\right) \bla(\varepsilon)
\right]\right|_{\varepsilon=0}
=\frac{\partial}{\partial \varepsilon}\left.\left[
\left(\identidad- {\wbbe_{\varepsilon}} {\wbbe_{\varepsilon}\trasp}\right) 
\right]\right|_{\varepsilon=0} \bla(0)
+ \left(\identidad- {\wbbe} {\wbbe\trasp}\right)\frac{\partial}{\partial \varepsilon}\left.
 \bla(\varepsilon) \right|_{\varepsilon=0}
 \nonumber\\
&=& \left(\identidad- {\wbbe} {\wbbe\trasp}\right)\frac{\partial}{\partial \varepsilon}\left.
 \bla(\varepsilon) \right|_{\varepsilon=0} \,. \label{eq0}
\end{eqnarray}
Henceforth, in order to compute $\left.  ({\partial  \bla(\varepsilon)}/{\partial \varepsilon})
 \right|_{\varepsilon=0}$ and to simplify the presentation, we consider the following functions:
$$
h(\varepsilon,\bbe,u) =\weta^{\varepsilon}_{\bbech}(u)\;,\qquad
h_{\bbech}(\varepsilon,\bbe,u) = \frac{\partial}{\partial \bbech} \weta^{\varepsilon}_{\bbech}(u)\;,\qquad
h_{u}(\varepsilon,\bbe,u) = \frac{\partial}{\partial u} \weta^{\varepsilon}_{\bbech}(u)
$$
and their corresponding derivatives with respect to $\varepsilon$
$$
H_i= \left.\frac{\partial}{\partial \varepsilon} h(\varepsilon,\wbbe_{\varepsilon},{\wbbe_{\varepsilon}\trasp} \bx_i)\right|_{\varepsilon=0}\;,\qquad
H_{\bbech,i}=  \left.\frac{\partial}{\partial \varepsilon} h_{\bbech}(\varepsilon,\wbbe_{\varepsilon},{\wbbe_{\varepsilon}\trasp} \bx_i)\right|_{\varepsilon=0}\;,\qquad
H_{u,i}=  \left.\frac{\partial}{\partial \varepsilon} h_u(\varepsilon,\wbbe_{\varepsilon},{\wbbe_{\varepsilon}\trasp} \bx_i)\right|_{\varepsilon=0}\;.
$$
Thus, we have that
\begin{eqnarray*}
\left. \frac{\partial}{\partial \varepsilon} \bla(\varepsilon) \right|_{\varepsilon=0}
&=& 
-\frac 1n \sum_{i=1}^n \psi \left(y_i,\weta_{\bbech}({\wbbe\trasp} \bx_i),c\right) 
\wbnu_i({\wbbe},{\wbbe\trasp} \bx_i) \tau(\bx_i) \\
&+& \frac 1n \sum_{i=1}^n \left\{\chi \left(y_i,\weta_{\bbech}({\wbbe\trasp} \bx_i),c\right) \; H_i 
\; \wbnu_i({\wbbe},{\wbbe\trasp} \bx_i)  +
\psi \left(y_i,\weta_{\bbech}({\wbbe\trasp} \bx_i),c\right) \;  (H_{\bbech,i}+ \bx_i H_{u,i})\right\}
\tau(\bx_i)\\
&+&  \; \psi \left(y_0,\weta_{\bbech}({\wbbe\trasp} \bx_0),c\right) 
\wbnu_0({\wbbe},{\wbbe\trasp} \bx_0) \tau(\bx_0)\, .
\end{eqnarray*}
Since $\bla(0)=\bcero$, we obtain that
\begin{eqnarray}
\left. \frac{\partial}{\partial \varepsilon} \bla(\varepsilon) \right|_{\varepsilon=0}
&=& 
\frac 1n \sum_{i=1}^n \left\{\chi \left(y_i,\weta_{\bbech}({\wbbe\trasp} \bx_i),c\right) \; H_i 
\; \wbnu_i({\wbbe},{\wbbe\trasp} \bx_i)  +
\psi \left(y_i,\weta_{\bbech}({\wbbe\trasp} \bx_i),c\right) \;  (H_{\bbech,i}+ \bx_i H_{u,i})\right\}
\tau(\bx_i) \nonumber\\
&+&  \; \psi \left(y_0,\weta_{\bbech}({\wbbe\trasp} \bx_0),c\right) 
\wbnu_0({\wbbe},{\wbbe\trasp} \bx_0) \tau(\bx_0)\, . \label{lambda0}
\end{eqnarray}
It remains to compute the functions $H_i$, $H_{\bbech,i}$ and $ H_{u,i}$. Straightforward arguments lead to
\begin{eqnarray*}
H_i&=&\left.\frac{\partial}{\partial\varepsilon}
h(\varepsilon,\wbbe_{\varepsilon},\wbbe_{\varepsilon}\trasp\bx_i)\right|_{\varepsilon=0}\\
&=&\left.\frac{\partial}{\partial\varepsilon}
h(\varepsilon,\bbe,u)\right|_{(\varepsilon,\bese)=(0,\widehat{\bese}_i)}
+\left.\frac{\partial}{\partial\bbech}  h(\varepsilon,\bbe,u)\right|_{(\varepsilon,\bese)=(0,\widehat{\bese}_i)}\left.\frac{\partial}{\partial\varepsilon}\wbbe_{\varepsilon}\right|_{\varepsilon=0}
+\left.\frac{\partial}{\partial u} h(\varepsilon,\bbe,u)\right|_{(\varepsilon,\bese)=(0,\widehat{\bese}_i)}\left.\frac{\partial}{\partial\varepsilon}\wbbe_{\varepsilon}\right|_{\varepsilon=0} \bx_i\; ,
\end{eqnarray*}
where $\widehat{\bese}_i=(\wbbe,\wbbe\trasp \bx_i)$.  Then, we get that
\begin{eqnarray*}
H_i&=&\left.\EIF(\weta_{\bbech}(u))\right|_{(\bbech, u)=\widehat{\bese}_i}+\left.\frac{\partial \weta_{\bbech}(u)}{\partial\bbech}  \right|_{(\bbech, u)=\widehat{\bese}_i}\EIF(\wbbe)
+\left.\frac{\partial \weta_{\bbech}(u)}{\partial u}  \right|_{(\bbech, u)=\widehat{\bese}_i}\EIF(\wbbe)\bx_i\\
&=& \left.\EIF(\weta_{\bbech}(u))\right|_{(\bbech,u)=\widehat{\bese}_i} + \wbnu_i(\widehat{\bese}_i)\;.
\end{eqnarray*}
Analogously, we have that
\begin{eqnarray*}
H_{\bbech,i}&=&\left.\frac{\partial}{\partial\varepsilon}
h{\bbech}(\varepsilon,\wbbe_{\varepsilon},\wbbe_{\varepsilon}\trasp\bx_i)\right|_{\varepsilon=0}\\
&=&\left.\frac{\partial}{\partial\varepsilon}\frac{\partial}{\partial\bbech}
h(\varepsilon,\bbe,u)\right|_{(\varepsilon,\bese)=(0,\widehat{\bese}_i)}
+\left.\frac{\partial}{\partial\bbech} \frac{\partial}{\partial\bbech} h(\varepsilon,\bbe,u)\right|_{(\varepsilon,\bese)=(0,\widehat{\bese}_i)}\left.\frac{\partial}{\partial\varepsilon}\wbbe_{\varepsilon}\right|_{\varepsilon=0} 
\\
&&
+\left.\frac{\partial}{\partial u} \frac{\partial}{\partial\bbech} h(\varepsilon,\bbe,u)\right|_{(\varepsilon,\bese)=(0,\widehat{\bese}_i)} \left.\frac{\partial}{\partial\varepsilon}\wbbe_{\varepsilon}\right|_{\varepsilon=0}  \bx_i\; ,
\end{eqnarray*}
so
\begin{eqnarray*}
H_{\bbech,i}&=&\left.\EIF(\frac{\partial}{\partial\bbech} \weta_{\bbech}(u))\right|_{(\bbech,u)=\widehat{\bese}_i}+\left.\frac{\partial^2 \weta_{\bbech}(u)}{\partial^2\bbech} \right|_{(\bbech,u)=\widehat{\bese}_i}\EIF(\wbbe)
+\left.\frac{\partial^2 \weta_{\bbech}(u)}{\partial u \partial\bbech}\right|_{(\bbech,u)=\widehat{\bese}_i}\EIF(\wbbe)\bx_i \; .
\end{eqnarray*}
Finally,  in a similar way, we obtain that
\begin{eqnarray*}
H_{u,i}&=&\left.\frac{\partial}{\partial\varepsilon}
h_{u}(\varepsilon,\wbbe_{\varepsilon},\wbbe_{\varepsilon}\trasp\bx_i)\right|_{\varepsilon=0}\\
&=&\left.\frac{\partial}{\partial\varepsilon}\frac{\partial}{\partial u}
h(\varepsilon,\bbe,u)\right|_{(\varepsilon,\bese)=(0,\widehat{\bese}_i)}
+\left.\frac{\partial}{\partial\bbech} \frac{\partial}{\partial u} h(\varepsilon,\bbe,u)\right|_{(\varepsilon,\bese)=(0,\widehat{\bese}_i)}\left.\frac{\partial}{\partial\varepsilon}\wbbe_{\varepsilon}\right|_{\varepsilon=0} 
\\
&&+\left.\frac{\partial}{\partial u } \frac{\partial}{\partial u} h(\varepsilon,\bbe,u)\right|_{(\varepsilon,\bese)=(0,\widehat{\bese}_i)} \left.\frac{\partial}{\partial\varepsilon}\wbbe_{\varepsilon}\right|_{\varepsilon=0}  \bx_i\; ,
\end{eqnarray*}
which implies that
\begin{eqnarray*}
H_{u,i}&=&\left.\EIF(\frac{\partial}{\partial u} \weta_{\bbech}(u))\right|_{(\bbech,u)=\widehat{\bese}_i}+\left.\frac{\partial^2 \weta_{\bbech}(u)}{\partial\bbech\partial u} \right|_{(\bbech,u)=\widehat{\bese}_i}\EIF(\wbbe)
+\left.\frac{\partial^2 \weta_{\bbech}(u)}{\partial^2
u }\right|_{(\bbech,u)=\widehat{\bese}_i}\EIF(\wbbe)\bx_i\; .
\end{eqnarray*}
Using the previous expressions, we deduce that
\begin{eqnarray*}
H_{\bbech,i}+ \bx_i H_{u,i}&=& \left.\EIF(\frac{\partial}{\partial\bbech} \weta_{\bbech}(u))\right|_{(\bbech,u)=\widehat{\bese}_i} + \left.\EIF(\frac{\partial}{\partial u} \weta_{\bbech}(u))\right|_{(\bbech,u)=\widehat{\bese}_i} \bx_i\; \\
&+& \left[\left.\frac{\partial^2 \weta_{\bbech}(u)}{\partial^2\bbech}   \right|_{(\bbech,u)=\widehat{\bese}_i} +
\left.\frac{\partial^2 \weta_{\bbech}(u)}{\partial^2 u}   \right|_{(\bbech,u)=\widehat{\bese}_i}
\bx_i \bx_i\trasp 
\right.
\\
&& 
\left. 
+\left.\frac{\partial^2 \weta_{\bbech}(u)}{\partial u \partial \bbech}   \right|_{(\bbech,u)=\widehat{\bese}_i} \bx_i\trasp
+ \left.\frac{\partial^2 \weta_{\bbech}(u)}{\partial \bbech \partial u}   \right|_{(\bbech,u)=\widehat{\bese}_i}
\bx_i\trasp \right] \EIF(\wbbe)\; .
\end{eqnarray*}
Now, replacing in (\ref{lambda0}) $H_i$, $H_{\bbech,i}$ and $H_{u,i}$ with the obtained expression, we have that 
\begin{eqnarray*}
\left. \frac{\partial}{\partial \varepsilon} \bla(\varepsilon) \right|_{\varepsilon=0}
&=& 
\frac 1n \sum_{i=1}^n \chi \left(y_i,\weta_{\bbech}({\wbbe\trasp} \bx_i),c\right) \; \tau(\bx_i) \; {\left.\EIF(\weta_{\bbech}(u))\right|_{(\bbech,u)=\widehat{\bese}_i}}
\; \wbnu_i({\wbbe},{\wbbe\trasp} \bx_i) \nonumber\\
&+&
\frac 1n \sum_{i=1}^n \chi \left(y_i,\weta_{\bbech}({\wbbe\trasp} \bx_i),c\right) \;  
\; \tau(\bx_i) \; \wbnu_i({\wbbe},{\wbbe\trasp} \bx_i) \wbnu_i({\wbbe},{\wbbe\trasp} \bx_i)\trasp  \EIF(\wbbe) 
\\
&+&  \frac 1n \sum_{i=1}^n 
\psi \left(y_i,\weta_{\bbech}({\wbbe\trasp} \bx_i),c\right) \;  \tau(\bx_i)  \left\{
 \left.\EIF(\frac{\partial}{\partial\bbech} \weta_{\bbech}(u))\right|_{(\bbech,u)=\widehat{\bese}_i} + \left.\EIF(\frac{\partial}{\partial u} \weta_{\bbech}(u))\right|_{(\bbech,u)=\widehat{\bese}_i} \bx_i\; \right. 
 \\
&+&\left. \left[\left.\frac{\partial^2 \weta_{\bbech}(u) }{\partial^2\bbech}  \right|_{(\bbech,u)=\widehat{\bese}_i} +
\left.\frac{\partial^2 \weta_{\bbech}(u) }{\partial^2 u} \right|_{(\bbech,u)=\widehat{\bese}_i}
\bx_i \bx_i\trasp 
+\left.\frac{\partial^2\weta_{\bbech}(u)  }{\partial
u \partial \bbech} \right|_{(\bbech,u)=\widehat{\bese}_i} \bx_i\trasp
\right.\right. 
\\
&&
\left.\left.
+ \left.\frac{\partial^2 \weta_{\bbech}(u) }{\partial \bbech \partial u}  \right|_{(\bbech,u)=\widehat{\bese}_i}
\bx_i\trasp \right] \EIF(\wbbe)\right\}+  \psi \left(y_0,\weta_{\bbech}({\wbbe\trasp} \bx_0),c\right) 
\wbnu_0({\wbbe},{\wbbe\trasp} \bx_0) \tau(\bx_0)\, . 
\end{eqnarray*}
Recall that
$$\bV(\widehat{\bese}_i)= \left[\left.\frac{\partial^2 \weta_{\bbech}(u) }{\partial^2\bbech}  \right|_{(\bbech,u)=\widehat{\bese}_i} +
\left.\frac{\partial^2 \weta_{\bbech}(u) }{\partial^2 u} \right|_{(\bbech,u)=\widehat{\bese}_i}
\bx_i \bx_i\trasp 
+\left.\frac{\partial^2\weta_{\bbech}(u)  }{\partial
u \partial \bbech} \right|_{(\bbech,u)=\widehat{\bese}_i} \bx_i\trasp
+ \left.\frac{\partial^2 \weta_{\bbech}(u) }{\partial \bbech \partial u}  \right|_{(\bbech,u)=\widehat{\bese}_i}
\bx_i\trasp \right] \; .
$$
Then, we get that
$$
\left. \frac{\partial}{\partial \varepsilon} \bla(\varepsilon) \right|_{\varepsilon=0} 
= \Bell_n+ \bM_n \EIF(\wbbe)\, ,  $$
where $\Bell_n\in \real^q$   and $\bM_n\in \real^{q\times q}$ are defined in (\ref{Bell}) and (\ref{bMn}).
Replacing in (\ref{eq0}), we have that
\begin{eqnarray*}
\bcero &=& \left(\bI- {\wbbe} {\wbbe\trasp}\right) (\Bell_n+ \bM_n \EIF(\wbbe) ) \, .
\end{eqnarray*}
It is worth noticing that since $\| \wbbe_{\varepsilon}\|^2=1$, differentiating with respect to $\varepsilon$ and evaluating at $\varepsilon=0$, we have that  
$$0=\left. \frac{\partial}{\partial \varepsilon}\wbbe_{\varepsilon}\trasp  \wbbe_{\varepsilon} \right|_{\varepsilon=0}= 2 \wbbe \trasp \EIF(\wbbe) \; $$
which, taking into account that $\wbbe=\be_q$, implies that $\EIF(\wbbe)_q=0$. Therefore, we only have to compute $\EIF(\wbbe)_j$ for $j=1,\dots,q-1$.

Using again that $\wbbe=\be_q$, we obtain that
$$\left(\bI- {\wbbe} {\wbbe\trasp}\right) = 
\left(\begin{array}{cc} \bI_{q-1}& \bf{0}\\
\bf{0}& 0\end{array}\right)\,.$$
Hence, we have that  the left superior matrix of $\left(\bI- {\wbbe} {\wbbe\trasp}\right) \bM_n$ equals the matrix $\bM_{n,1}\in \real^{(q-1)\times (q-1)}$, so that $ \bcero = \left(\bI- {\wbbe} {\wbbe\trasp}\right) (\Bell_n+ \bM_n \EIF(\wbbe) ) $ implies
\begin{equation}
\bcero =   \Bell_n^{(q-1)}+ \bM_{n,1} \EIF(\wbbe^{(q-1)})   \, .\label{IFbell}
\end{equation}
Therefore, from (\ref{IFbell}) we get that 
$\EIF(\wbbe^{(q-1)}) = - \bM_{n,1}^{-1} \Bell_n^{(q-1)}  $. 

It is worth noticing that $\Bell_n$ and $\bM_n$ involve $\left.\EIF(\weta_{\bbech}(u))\right|_{(\bbech,u)=\widehat{\bese}_i}$, $ \left.\EIF( {\partial  \weta_{\bbech}(u)}/{\partial\bbech})\right|_{(\bbech,u)=\widehat{\bese}_i}$ and \linebreak $ \left.\EIF( {\partial  \weta_{\bbech}(u)}/{\partial u})\right|_{(\bbech,u)=\widehat{\bese}_i}$.

b) Let us derive $\left.\EIF(\weta_{\bbech}(u))\right|_{(\bbech,u)=\widehat{\bese}_i}$. Since $\weta^{\varepsilon}_{\bbech_{\varepsilon}}(u)$ is the solution of (\ref{eqeta}), we have that 
\begin{eqnarray*}
 \frac{(1-\varepsilon)}{n}\sum_{i=1}^n  K_h(\bbe\trasp \bx_i-u) \psi \left(y_i,\weta^{\varepsilon}_{\bbech_{\varepsilon}}(u),c\right) 
 +\varepsilon \;  K_h(\bbe\trasp \bx_0-u) \psi \left(y_0,\weta^{\varepsilon}_{\bbech_{\varepsilon}}(u),c\right)   
= 0 \; .
\end{eqnarray*}
  Differentiating  with respect to $\varepsilon$ and evaluating at $\varepsilon=0$, we obtain that 
\begin{equation}
\EIF(\weta_{\bbech}(u))=
-\frac{K_h(\bbe\trasp \bx_0-u) \psi \left(y_0,\weta_{\bbech}(u),c\right)}{\dst\frac 1n\sum_{i=1}^n K_h(\bbe\trasp \bx_i-u) \psi \left(y_i,\weta_{\bbech}(u),c\right)} \,.
\label{eifweta}
\end{equation}
Analogously, differentiating first with respect to $\bbe$  on both sides of   equation (\ref{eqeta}) and then,  with respect to $\varepsilon$ and evaluating at $\varepsilon=0$, we can obtain an expression for $ \left.\EIF({\partial \weta_{\bbech}(u)}/{\partial\bbe})\right|_{(\bbech,u)=\widehat{\bese}_i}$.  Alternatively, we may differentiate (\ref{eifweta}) with respect to $\bbe$ to obtain 
\begin{eqnarray*}
\EIF\left(\frac{\partial \weta_{\bbech}(u)}{\partial\bbech}\right) &=& 
-\frac{\frac{1}{h} K_h^{\prime}(\bbe\trasp \bx_0-u) \psi \left(y_0,\weta_{\bbech}(u),c\right) \bx_0+ K_h (\bbe\trasp \bx_0-u) \chi \left(y_0,\weta_{\bbech}(u),c\right)\dst\frac{\partial}{\partial \bbech} \weta_{\bbech}(u)}{\dst \frac 1n\sum_{i=1}^n K_h(\bbe\trasp \bx_i-u) \psi \left(y_i,\weta_{\bbech}(u),c\right)}\\
& + & \frac{K_h(\bbe\trasp \bx_0-u) \psi \left(y_0,\weta_{\bbech}(u),c\right)}{\left\{\dst\frac 1n\sum_{i=1}^n K_h(\bbe\trasp \bx_i-u) \psi \left(y_i,\weta_{\bbech}(u),c\right)\right\}^2}\,  \left[\frac 1n \sum_{i=1}^n \frac 1h K_h^{\prime}(\bbe\trasp \bx_i-u) \psi \left(y_i,\weta_{\bbech}(u),c\right) \bx_i\right.\\
&& \left.+ \frac 1n \sum_{i=1}^n  K_h (\bbe\trasp \bx_i-u) \chi \left(y_i,\weta_{\bbech}(u),c\right)\dst\frac{\partial}{\partial \bbech} \weta_{\bbech}(u)\right] \,.
\end{eqnarray*}
 Similar arguments lead to the expression for $ \left.\EIF({\partial \weta_{\bbech}(u)}/{\partial u})\right|_{(\bbech,u)=\widehat{\bese}_i}$.

Finally, note that   $\weta_{\bbech}(u)$, satisfies
\begin{eqnarray}
 \sum_{i=1}^n K\left(\frac{\bbe\trasp \bx_i-u}{h}\right) \psi\left(y_i,\weta_{\bbech}(u),\alpha\right) &=&0 \, . \label{implicita}
\end{eqnarray}
Hence, differentiating with respect to $\bbe$   equation (\ref{implicita}), we get that
\begin{eqnarray*}
0 &=& \frac 1h  \sum_{i=1}^n K^{\prime}\left(\frac{\bbe\trasp \bx_i-u}{h}\right) \psi\left(y_i,\weta_{\bbech}(u),\alpha\right) \bx_i + \sum_{i=1}^n K\left(\frac{\bbe\trasp \bx_i-u}{h}\right) \chi\left(y_i,\weta_{\bbech}(u),\alpha\right) \times \frac{\partial}{\partial \bbech} \weta_{\bbech}(u)\, ,
\end{eqnarray*} 
which implies that
\begin{eqnarray*}
\frac{\partial}{\partial \bbech} \weta_{\bbech}(u) &=& -\frac 1h \left[
 \sum_{i=1}^n K\left(\frac{\bbe\trasp \bx_i-u}{h}\right) \chi\left(y_i,\weta_{\bbech}(u),\alpha\right)\right]^{-1}  \, \sum_{i=1}^n K^{\prime}\left(\frac{\bbe\trasp \bx_i-u}{h}\right) \psi\left(y_i,\weta_{\bbech}(u),\alpha\right) \bx_i  
\end{eqnarray*} 
On the other hand, differentiating (\ref{implicita}) with respect to $u$, we obtain that
\begin{eqnarray*}
0 &=&  -\frac{1}{h} \sum_{i=1}^n K^{\prime}\left(\frac{\bbe\trasp \bx_i-u}{h}\right) \psi\left(y_i,\weta_{\bbech}(u),\alpha\right)  + \sum_{i=1}^n K\left(\frac{\bbe\trasp \bx_i-u}{h}\right) \chi\left(y_i,\weta_{\bbech}(u),\alpha\right) \times \frac{\partial}{\partial u} \weta_{\bbech}(u)
\end{eqnarray*} 
which entails that
\begin{eqnarray*}
\frac{\partial}{\partial u} \weta_{\bbech}(u) &=& \frac 1h \left[
 \sum_{i=1}^n K\left(\frac{\bbe\trasp \bx_i-u}{h}\right) \chi\left(y_i,\weta_{\bbech}(u),\alpha\right)\right]^{-1}  \, \sum_{i=1}^n K^{\prime}\left(\frac{\bbe\trasp \bx_i-u}{h}\right) \psi\left(y_i,\weta_{\bbech}(u),\alpha\right) \,. \,\square
\end{eqnarray*}   


\vskip0.1in
\small
\noi \textbf{Acknowledgements.}  This research was partially supported by Grants   \textsc{pict} 2014-0351 from \textsc{anpcyt} and  Grant 20120130100279BA from the Universidad de Buenos Aires  at Buenos Aires, Argentina. It was also supported by the Italian-Argentinian project
\textsl{Metodi robusti per la previsione del costo e della durata della degenza ospedaliera}
funded by the joint collaboration program MINCYT-MAE AR14MO6 (IT1306) between  \textsc{mincyt} from Argentina and \textsc{mae} from Italy.

\normalsize
\section*{References}
\small
\begin{description}
 
\item A\"it Sahalia, Y. (1995). The delta method for nonaparmetric kernel functionals. PhD. dissertation, University of Chicago.

\item Bianco, A., Boente, G. (2002) On the asymptotic behavior of one-step estimation. \textsl{Stat. Probab. Lett.}, \textbf{60}, 33-47.

\item Bianco, A. and Boente, G. (2007). Robust estimators under a semiparametric partly linear autoregression model:
asymptotic behavior and bandwidth selection. \textsl{J. of Time Series Anal.}, \textbf{28}, 274-306.

\item Bianco, A., Garc\'\i a Ben, M. and Yohai, V. (2005). Robust estimation for linear regression with asymmetric errors. \textsl{Canad. J. Statist.} \textbf{33}, 511-528.

\item
Boente, G., Fraiman, R. and Meloche, J. (1997). Robust plug-in bandwidth estimators in nonparametric regression. \textsl{J. Statist. Plann. Inf.}, \textbf{57}, 109-142.

\item Boente, G. and Rodriguez, D. (2008). Robust bandwidth selection in semiparametric partly linear regression models: Monte Carlo study and influential analysis. \textsl{Comput. Stat. Data Anal.}, \textbf{52}, 2808-2828.

\item  Boente, G. and Rodriguez, D. (2010). Robust inference in generalized partially linear models.\textsl{Comput. Stat. Data Anal.}, \textbf{54}, 2942-2966.

\item Boente, G. and Rodriguez, D. (2012).  Robust estimates in generalized partially linear single-index models.  \textsl{TEST}, \textbf{21}, 386-411.

\item Cantoni, E. and Ronchetti, E. (2001). Resistant selection of the smoothing parameter for smoothing splines. \textsl{Statistics and Computing}, \textbf{11(2)}, 141-146. 
 
\item
Carroll, R., Fan, J., Gijbels, I. and Wand, M. (1997). Generalized partially linear single-index models. \textsl{J. Amer. Statis. Assoc.}, \textbf{92}, 477-489.

\item Chang, Z. Q., Xue, L. G. and Zhu, L. X. (2010). On an asymptotically more efficient estimation
of the single--index model. \textsl{J. Multivariate Anal.}, 101, 1898-1901.

\item  Croux, C. and Ruiz--Gazen, A.  (2005). High Breakdown   Estimators for Principal Components: the Projection--Pursuit
Approach   Revisited. \textsl{J.  Multivariate Anal.}, \textbf{95}, 206-226.

\item Delecroix, M., H\"ardle, W. and Hristache, M. (2003). Efficient estimation in conditional single--index regression. \textsl{J.   Multivariate Anal.}, \textbf{86}, 213-226.

\item Delecroix, M., Hristache, M. and Patilea, V. (2006). On semiparametric $M-$estimation in single-index
regression. \textsl{J. Statist. Plann. Inf.}, \textbf{136}, 730-769.

\item Hampel, F.R (1974). The influence curve and its role in robust estimation. \textsl{J. Amer. Statist. Assoc.}, \textbf{69}, 383-394.

\item H\"ardle, W. and Stoker, T. M. (1989). Investigating smooth multiple regression by method of
average derivatives. \textsl{J. Am. Statist. Assoc.}, \textbf{84}, 986-95.

\item H\"ardle, W., Hall, P. and Ichimura,H. (1993). (1993). Optimal smoothing in single-index models. \textsl{Ann. Statist.}, \textbf{21}, 157-178.

\item Leung, D. (2005). Cross-validation in nonparametric regression with outliers. \textsl{Annals of Statistics}, \textbf{33}, 2291-2310.

\item Leung, D., Marriott, F. and Wu, E. (1993). Bandwidth selection in robust smoothing. \textsl{J. Nonparametric Statist.}, \textbf{4}, 333-339.

\item Li, W. and Patilea, W. (2017). A new inference approach for single-index models. \textsl{J.  Multivariate Anal.}, \textbf{158}, 47-59.

\item Liu, J., Zhang, R., Zhao, W. and Lv, Y. (2013). A robust and efficient estimation method for single index models. \textsl{J.  Multivariate Anal.}, \textbf{122}, 226-238.

\item  Mallows, C. (1974). On some topics in robustness. \textsl{Memorandum, Bell Laboratories, Murray Hill}, N.J.

\item Manchester, L. (1996). Empirical influence for robust smoothing. \textsl{Austral. J. Statist.}, \textbf{38}, 275-296.

\item Maronna, R., Martin, D. and Yohai, V. (2006). Robust statistics: Theory and methods. John Wiley \& Sons, New York.

\item Pollard. D. (1984). \textsl{Convergence of stochastic processes}. Springer Series in Statistics. Springer-Verlag, New York.

\item Powell, J. L., Stock, J. H. and Stoker, T. M. (1989). Semiparametric estimation of index coefficients. \textsl{Econometrica}, \textbf{57}, 1403-30.

\item Rodriguez, D. (2007). \textsl{Estimaci\'on robusta en modelos parcialmente lineales generalizados}. PhD. Thesis (in spanish), Universidad de Buenos Aires. \newline Available at \url{http://cms.dm.uba.ar/academico/carreras/doctorado/tesisdanielarodriguez.pdf}

\item Severini, T. and Staniswalis, J. (1994). Quasi-likelihood estimation in semiparametric models. \textsl{J. Amer. Statist. Assoc.}, \textbf{89}, 501-511.

\item Severini, T. and Wong, W. (1992).  Profile likelihood and conditionally parametric models. \textsl{Ann. Statist.},
\textbf{20}, 4, 1768-1802.

\item Sherman, R. (1994). Maximal inequalities for degenerate $U-$processes with applications to optimization estimators. \textsl{Ann. Statist.},
\textbf{22}, 439-459.

\item Tamine, J. (2002). Smoothed influence function: another view at robust nonparametric regression. Discussion paper 62, Sonderforschungsbereich 373, Humboldt-Universit¨at zu Berlin.

\item Tukey, J. (1977). \textsl{Exploratory Data Analysis}. Reading, MA: Addison--Wesley.

\item van der Vaart, A. (1988).  Estimating a real parameter in a class of semiparametric models. \textsl{Ann. Statist.}, \textbf{16}, 4, 1450-1474.


\item Wang, F. and Scott, D. (1994). The L1 method for robust nonparametric regression. \textsl{J. Amer. Statist. Assoc.}, \textbf{89}, 65-76.


\item Wang, Q., Zhang, T. and H\"adle, W (2014). An Extended Single Index Model with Missing Response at Random, SFB 649 Discussion Paper 2014-003.

\item Wu, T. Z., Yu, K., and Yu, Y. (2010). Single index quantile regression. \textsl{J. Multivariate Anal.}, \textbf{101}, 1607-1621.

\item Xia, Y. and H\"ardle, W. (2006) Semi-parametric estimation of partially linear single-index models. \textsl{J.   Multivariate  Anal.}, \textbf{97}, 1162-1184.

\item Xia, Y., H\"ardle, W, and Linton, O. (2012). Optimal smoothing for a computationally and
statistically efficient single index estimator. In \textsl{Exploring Research Frontiers in Contemporary Statistics and Econometrics: A Festschrift for L\'eopold Simar}, 229-261.

\item Xia, Y., Tong, H., Li, W. K. and Zhu, L. (2002) An adaptive estimation of dimension reduction space (with discussion). \textsl{J. Royal Statist. Soc. Series B}, \textbf{64}, 363-410. 

\item Xue, L.G. and  Zhu, L.X. (2006). Empirical likelihood for single-index model. \textsl{J. Multivariate Anal.}, \textbf{97}, 1295-1312.
  
\item Zhang,R., Huang, R. and Lv, Z. (2010). Statistical inference for the index parameter in single-index models. \textsl{J. Multivariate Anal.}, \textbf{101}, 1026-1041.

\end{description}

\end{document}